\documentclass[a4paper,12pt]{article}
\usepackage{jheppub,esint,psfrag}
\usepackage[utf8]{inputenc}

\usepackage{epsfig,amssymb,amsmath,psfrag,subfigure,rotate,color,wasysym,xcolor}
\usepackage[bbgreekl]{mathbbol}

\allowdisplaybreaks 
\newcommand{\insertfig}[2]{\includegraphics[width=#1cm]{#2}}

\DeclareSymbolFontAlphabet{\mathbbm}{bbold}
\DeclareSymbolFontAlphabet{\mathbb}{AMSb}%

\def\XXint#1#2#3{{\setbox0=\hbox{$#1{#2#3}{\int}$ }
\vcenter{\hbox{$#2#3$ }}\kern-.6\wd0}}

\def \be  {\begin{equation}}
\def \ee  {\end{equation}}
\def \ba  {\begin{eqnarray}}
\def \ea  {\end{eqnarray}}
\def \baa {\begin{eqnarray*}}
\def \eaa {\end{eqnarray*}}

\def \lab #1 {\label{#1}}

\newcommand\re[1]{(\ref{#1})}
\def\d{\hbox{{d}\kern-.20em\hbox{l}}}

\def \matrix #1 {\left(\begin{array}{cc} #1 \end{array}\right)}

\def \tr {\mathop{\rm tr}\nolimits}

\def \e  {\mathop{\rm e}\nolimits}

\newcommand \vev [1] {\langle{#1}\rangle}

\newcommand \ket [1] {|{#1}\rangle}
\newcommand \bra [1] {\langle {#1}|}
\newcommand{\bit}[1]{\mbox{\boldmath$#1$}}
\def\1{\hbox{{1}\kern-.25em\hbox{l}}}

\newcommand{\ft}[2]{{\textstyle\frac{#1}{#2}}}

\makeatletter

\input pdf-trans
\newbox\qbox
\def\usecolor#1{\csname\string\color@#1\endcsname\space}
\newcommand\bordercolor[1]{\colsplit{1}{#1}}
\newcommand\fillcolor[1]{\colsplit{0}{#1}}
\newcommand\outline[1]{\leavevmode%
  \def\maltext{#1}%
  \setbox\qbox=\hbox{\maltext}%
  \boxgs{Q q 2 Tr \thickness\space w \fillcol\space \bordercol\space}{}%
  \copy\qbox%
}
\makeatother
\newcommand\colsplit[2]{\colorlet{tmpcolor}{#2}\edef\tmp{\usecolor{tmpcolor}}%
  \def\tmpB{}\expandafter\colsplithelp\tmp\relax%
  \ifnum0=#1\relax\edef\fillcol{\tmpB}\else\edef\bordercol{\tmpC}\fi}
\def\colsplithelp#1#2 #3\relax{%
  \edef\tmpB{\tmpB#1#2 }%
  \ifnum `#1>`9\relax\def\tmpC{#3}\else\colsplithelp#3\relax\fi
}
\bordercolor{black}
\fillcolor{white}
\def\thickness{.3}

\def\1{\mathbbm{1}}

\title{Collinear anatomy}
 
\author{A.V.~Belitsky}
\affiliation
{Department of Physics, Arizona State University,  Tempe, AZ 85287-1504, USA}

 \abstract
{We study the collinear factorization of off-shell scattering amplitudes in maximally supersymmetric Yang-Mills (sYM) theory. These 
are constructed starting from six-dimensional $\mathcal{N} = (1,1)$ sYM, taking advantage of an available unconstrained spinor-helicity 
formalism combined with a unitarity-cut sewing procedure. After generalized dimensional reduction, their collinear behavior is dissected 
with assistance from the Method of Regions. We then construct off-shell splitting amplitudes directly using the same techniques, 
establishing equivalence to the amplitude analysis. The calculations are performed at one-loop order.}

\begin{document}

\maketitle
\flushbottom
\setcounter{footnote} 0

\section{Introduction}

The decomposition of on-shell scattering amplitudes $A_n$ for generic values of momentum invariants in terms of incoherent 
momentum modes responsible for short- and long-wave physics is well established via factorization theorems 
\cite{Collins:1989gx,Agarwal:2021ais}. They form the foundation for the applicability of perturbative calculations for quantitative 
predictions of hadronic cross sections starting from QCD partonic physics. When some invariants vanish, momentum and/or 
energy configurations become exceptional. Since the dependence on these is non-analytic, the limits are generally singular. In 
these situations, the amplitudes, nevertheless, enjoy factorization properties in terms of lower point amplitudes accompanied 
by process-independent, universal functions. These are instrumental in practical computations of higher-order effects and, thus, 
for precision QCD physics. 

In this work, we are interested in the so-called collinear behavior, when the momenta of two external particles `coalesce'
\begin{align}
\label{CollLimit}
p_{n-1} \to z (p_{n-1} + p_n)
\, , \qquad
p_n \to \bar{z} (p_{n-1} + p_n)
\, ,
\end{align}
(here and below $\bar{z} \equiv 1 - z$). A universal function of $z$ that governs this limit is known as the splitting amplitude\footnote{Here, 
we ignore the helicity dependence for now as well as do not display the dependence on the vanishing invariant mass $(p_{n-1}+p_n)^2$
in ${\rm Split}$.},
\begin{align}
\label{SplitDef} 
A_n (\dots, p_{n-1}, p_n) \xrightarrow{\scriptscriptstyle (n-1)||n} {\rm Split} (z) A_{n - 1} (\dots, p_{n-1} + p_n)
\, .
\end{align}
At the tree level, these functions have been known since the mid-eighties \cite{Parke:1986gb,Berends:1987me,Mangano:1990by}, 
while at one loop they were calculated in Refs.\ \cite{Bern:1994zx,Kosower:1999xi,Kosower:1999rx} followed up by two-loop 
\cite{Bern:2004cz,Badger:2004uk} and quite recently three-loop \cite{Guan:2024hlf} computations.

There is a solid understanding of massless QCD amplitudes to high perturbative orders \cite{Dixon:1996wi,Bern:2007dw} as well as their 
resummed all-order structure \cite{Sen:1982bt,Catani:1998bh,Sterman:2002qn,Aybat:2006mz,Dixon:2008gr,Gardi:2009qi,Becher:2009qa,Agarwal:2021ais}. 
Within the context of QCD's distant cousin, the planar maximally supersymmetric Yang-Mills theory or $\mathcal{N} = 4$ sYM for short, 
this knowledge goes even further, into the nonperturbative realm. The progress in this endeavor was made in part due to the dual 
description of gluon scattering amplitudes in terms of vacuum expectation values of Wilson loops $W_n$ on null polygons formed by 
external massless gluon momenta\footnote{In retrospect, this does not seem to be significant, however, \cite{Basso:2018tif}.} 
\cite{Alday:2007hr,Drummond:2007cf,Brandhuber:2007yx}. The Wilson links source a flux of the non-Abelian gauge field which is 
stretched between the contour's sides and forms a two-dimensional world sheet as it sweeps the contour. As one takes the collinear 
limit of a pair of adjacent links, i.e., external particle momenta \re{CollLimit}, the flux tube gets excited. This can be put in a systematic 
plaquette expansion in increasing powers of the gluon field-strength tensor \cite{Corrigan:1978zg,Durand:1979sw}, i.e., the operator 
product expansion \cite{Alday:2010ku,Belitsky:2011nn,Sever:2011da}. These operator insertions create flux-tube excitations that 
propagate on top of the world sheet. Their dynamics is integrable with known dispersion relations and scattering matrices exactly in 
't Hooft coupling. This served as a foundation for the so-called pentagon operator product expansion \cite{Basso:2013vsa} which 
became an effective nonperturbative framework for the calculation of massless  $\mathcal{N}=4$ amplitudes in multi-collinear 
limits and even resummations of their entire series.

Off-shell partonic subprocesses are equally important for practical QCD applications. However, their knowledge is not on par with 
the above on-shell physics and is presently only in its infancy. This is due, to a large extent, to difficulties in maintaining gauge 
invariance when imposing off-shell kinematics on available massless gauge amplitudes. One way out of this predicament is to rely 
on the background-field-gauge formalism \cite{Abbott:1980hw,Abbott:1981ke,Gates:1983nr} to perform calculations. This implies using 
traditional Feynman diagrams, which are quite impractical in higher loops. Recently, a complimentary viewpoint \cite{Caron-Huot:2021usw}  
was offered within the $\mathcal{N} = 4$ sYM by looking at its ten-dimensional progenitor, i.e., the $\mathcal{N} = 1$ sYM and a 
generalized dimensional reduction down to four space-time dimensions \cite{Selivanov:1999ie,Boels:2010mj}. In this setup, the 
off-shellness of external gluons is embedded as out-of-four-dimensional components of their ten-dimensional momenta such that 
the masslessness condition in higher dimensions is equivalent to the four-dimensional momenta being taken off their mass shell 
\cite{Selivanov:1999ie,Boels:2010mj}. This 
allowed one to obtain several higher-loop predictions for off-shell scattering amplitudes \cite{Caron-Huot:2021usw,Bork:2022vat} and 
form factors \cite{Belitsky:2022itf,Belitsky:2023ssv,Belitsky:2024agy,Belitsky:2024dcf} in the $\mathcal{N} = 4$ sYM with an uplift of 
four-dimensional integrands by merely promoting all Lorentz invariants in loop integrands to higher dimensions. While for a low number 
of external states, one does not anticipate a problem with this naive approach, taking it for granted is quite unsatisfactory and calls for 
a direct demonstration of this fact. This will be one of the goals of the current consideration.

With an eye on QCD applications, the focus of the current paper is on the planar $\mathcal{N} = 4$ sYM and the use of the 
higher-dimensional vantage point to construct higher loop off-shell amplitudes. This model will allow us `to sharpen the sword' in 
the off-shell toolbox by having better control over various ingredients due to the high level of symmetry of the former. Presently, 
we will rely however on the six-dimensional $\mathcal{N} = (1,1)$ sYM rather than its ten-dimensional $\mathcal{N} = 1$ 
counterpart. The reason is rather obvious: in the six-dimensional spinor-helicity formalism \cite{Cheung:2009dc,Dennen:2009vk}, 
the Weyl spinors are arbitrary since an antisymmetric tensor product of any two contains the same number of degrees of freedom 
as a null vector. In ten dimensions, on the other hand, the spinors obey nonlinear constraints \cite{Caron-Huot:2010nes}. Armed 
with this formalism, we will be able to apply the unitarity-cut sewing \cite{Bern:1994zx,Bern:1994cg,Bern:2004cz} to construct one- 
and two-loop integrands of a six-dimensional five-leg amplitude from available trees \cite{Cheung:2009dc,Dennen:2009vk,Bern:2010qa}. 
An appropriate dimensional reduction will then yield its off-shell four-dimensional limit. The off-shellness will serve as a regulator of 
infrared divergences, and we will be interested in the near-mass-shell kinematics. Superficially this looks like just a different way 
to regulate the theory by moving it to its Coulomb branch \cite{Alday:2009zm}. The reality is however trickier.

It appeared as a surprise from loop calculations of near-mass-shell scattering amplitudes \cite{Caron-Huot:2021usw,Bork:2022vat} and 
form factors \cite{Belitsky:2022itf,Belitsky:2023ssv,Belitsky:2024agy,Belitsky:2024dcf} that their infrared behavior is governed by an
anomalous dimension, known as octagon \cite{Coronado:2018cxj,Belitsky:2019fan,Belitsky:2020qzm}, which is different from the cusp 
\cite{Polyakov:1980ca,Korchemsky:1987wg}. The reason for this is attributed to the presence of the so-called ultrasoft modes in loop 
integrals that are absent in the strictly on-shell case. While this fact has been known for ages at one loop 
\cite{Fishbane:1971jz,Mueller:1981sg,Korchemsky:1988hd} from studies of the Sudakov form factor 
\cite{Sudakov:1954sw}, a higher loop analysis was performed only recently \cite{Belitsky:2024yag}. The Method of Regions (MofR)
\cite{Beneke:1997zp} was used as a discovery tool. In light of this, how does the collinear behavior \re{CollLimit} change compared 
to the massless case? This is the main question that we address below.

Our subsequent presentation is organized as follows. In the next section, we set up the stage by recalling the superspace formulation 
of six-dimensional scattering amplitudes in the $\mathcal{N} = (1,1)$ sYM. Then in Sect.~\ref{6D5LegSection}, we use these results in 
tandem with unitarity techniques to construct integrands of a five-leg amplitude which will be sufficient to address the question of the
collinear behavior \re{SplitDef}. Then after dimensional reduction discussed in Sect.~\ref{DimRedSection}, we turn in 
Sect.~\ref{1LoopOffShellAmplitude} to the infrared factorization of the obtained amplitude in terms of incoherent momentum components 
with MofR and then study it in the limit \re{CollLimit} in Sect.\ \ref{ColFactSection}. The off-shell analogue of the on-shell splitting functions 
is derived in that section to one-loop order. In Sect.~\ref{6DSplitSection}, the splitting amplitudes are defined in terms of three-particle 
vertices and their one-loop integrands are derived from unitarity cuts. Making use of the MofR, the anatomy of these integrals is studied, 
and the off-shell splitting amplitude is extracted in agreement with the earlier amplitude analysis. Finally, we conclude. Several appendices 
are added, providing details of the calculations involved. First, in Appendix~\ref{DiracAppendix}, we construct representations of Clifford 
algebras in four and six dimensions using our conventions, which differ from the literature. This is followed in Appendix~\ref{SpinorHelicityAppendix} 
by a refresher of the spinor-helicity formalisms making use of the previosuly established  notations. Then in 
Appendix~\ref{4D5LegAppendix}, we re-derive one-loop expressions for five-leg amplitudes in four dimensions, which is followed by a 
reminder of its collinear behavior in Appendix \ref{CollinearLimitAppendix}.  We then provide a concise recollection of splitting amplitudes 
in four dimensions in Appendix \ref{4DSplittingAmplitudeAppendix}. Finally, in Appendix~\ref{6DSplitComponentsAppendix}, we give 
a detailed derivation of the one-loop integrand for the six-dimensional gluon splitting amplitude. 

\section{$\mathcal{N} = (1,1)$ superspace}
\label{N11superspaceSection}

Let us begin by setting up the stage for six-dimensional calculations. As reviewed at length in Appendices~\ref{DiracAppendix} and 
\ref{SpinorHelicityAppendix}, the six-dimensional Lorentz group is $\mbox{SO}(5,1)$ with the $\mbox{SU}(2) \times \mbox{SU}(2)$ 
being its little subgroup \cite{Cheung:2009dc,Dennen:2009vk}. Representations of the latter are determined by tensors with two sets 
of indices, dotted and undotted. These classify on-shell polarization states in the theory. The notion of helicity does not exist. A 
six-dimensional gauge field transforms in the $(\ft12, \ft12)$ representation. As a consequence, the superspace formulation of the 
$\mathcal{N} = (1,1)$ sYM is non-chiral\footnote{In light of our use of the six-dimensional sYM for off-shell four-dimensional physics, 
this suggests that the off-shell superspace in four dimensions has to be non-chiral as well. The on-shell superspace is typically chosen 
as chiral \cite{Nair:1988bq}, see, Eq.\ \re{NairField}.} \cite{Dennen:2009vk}. The $R$-symmetry of the model is $\mbox{SU}_{R} (2) 
\times \mbox{SU}_{R} (2)$, and thus the superspace's Grassmann variables possess two indices transforming with respect to the little 
and $R$ groups. However, due to the CPT-self-conjugated nature of the spectrum of physical states in the theory, only half of the fermionic degrees
of freedom are independent. The superspace requires a truncation, which is done with a set of harmonic variables \cite{Galperin:2001seg}. 
Choosing the little group indices to label states, the projection is implemented on the $R$-indices yielding $\eta_a$ and $\bar\eta_{\dot{a}}$
\cite{Dennen:2009vk}. One can package all on-shell states into a single superfield since one possesses enough supersymmetry to 
reach all of them. However, it is different from its four-dimensional counterpart \re{NairField} and reads
\cite{Dennen:2009vk}
\begin{align}
\label{6Dsuperfield}
{\mit\Phi} = \phi + \chi^a \eta_a + \bar{\chi}_{\dot{a}} \bar\eta^{\dot{a}} + \eta^2 \phi^\prime + \bar\eta^2 \phi^{\prime\prime}
+
g^{a}{}_{\dot{a}} \eta_a \bar\eta^{\dot{a}} 
+ 
\psi^a \eta_a \bar\eta^2
+
\bar\psi_{\dot{a}} \bar\eta^{\dot{a}} \eta^2
+
\eta^2 \bar\eta^2 \phi^{\prime\prime\prime}
\, .
\end{align}
Here, we adopted the shorthand notations $\eta^2 \equiv \ft12 \eta^a \eta_a$ and $\bar\eta^2 \equiv \ft12 \bar\eta_{\dot{a}} \bar\eta^{\dot{a}}$.
The coefficients $\phi, \dots, \phi^{\prime\prime\prime}$ are the four six-dimensional scalars, $\chi^a, \dots, \bar\psi_{\dot{a}}$ are 
the eight six-dimensional fer\-mi\-ons and finally $g^{a}{}_{\dot{a}}$ is the six-dimensional gluon. Upon dimensional reduction down to four 
space-time dimensions, the two gluon modes with $a = 1, \dot{a} = \dot{1}$ and $a = 2, \dot{a} = \dot{2}$ account for the `missing' 
two scalars of the $\mathcal{N} = 4$ sYM, while $a = 2, \dot{a} = \dot{1}$ and $a = 1$, $\dot{a} = \dot{2}$ correspond to the positive
and negative helicity gluons, respectively. To further make contact with the field content of the conformal theory reviewed in Appendix 
\ref{4D5LegAppendix}, one can identify the four-dimensional limits of the scalars $\phi$ and $\phi^{\prime\prime\prime}$ with their 
four-dimensional counterparts $\phi^{34}$ and $\phi^{12}$, respectively.

A six-dimensional massless state created by the superfield \re{6Dsuperfield} from the vacuum is characterized by the super-Poincar\'e 
quantum numbers, the momentum $P^{AB}$ and super-charges $Q^A$ and $\bar{Q}_A$. The power of the spinor-helicity formalism is
to represent these using the unconstrained Weyl spinors ${\mit\Lambda}^{A, a}$ and $\bar{\mit\Lambda}_{A, \dot{a}}$,
\begin{align}
\label{6Dsupercharges}
P^{AB} = {\mit\Lambda}^{a,A} {\mit\Lambda}^B_{a}
\, , \qquad
Q^A = {\mit\Lambda}^{A, a} \eta_a
\, , \qquad
\bar{Q}_A = \bar{\mit\Lambda}_{A, \dot{a}} \bar\eta^{\dot{a}}
\, .
\end{align}
The resulting spinor-helicity formalism was instrumental in obtaining a concise form of the color-ordered gluon scattering 
amplitudes \cite{Cheung:2009dc} and their superspace uplift \cite{Dennen:2009vk},
\begin{align}
\label{6DColorOrderedAmp}
\mathcal{A}_n 
&\equiv  
\mathcal{A}_n (1,2, \dots, n)
=
\vev{{\mit\Phi}_1 \dots {\mit\Phi}_n} 
\\
&=
i (2 \pi)^6 \delta^{(6)} \left(\sum\nolimits_{i = 1}^n P_i\right) 
\delta^{(4)} \left(\sum\nolimits_{i = 1}^n Q_i\right)  
\delta^{(4)} \left(\sum\nolimits_{i = 1}^n \bar{Q}_i\right)
\widehat{\mathcal{A}}_n
\, , \nonumber
\end{align}
where mimicking the four-dimensional nomenclature in Eq.\ \re{4DgenericAmplitude}, we extracted the bosonic and fermionic delta functions
imposing the supermomentum conservation condition. The color-ordered reduced amplitudes $\widehat{\mathcal{A}}_n$ are of Grassmann 
degree $n - 4$. They were found explicitly at tree level for the four- and five legs to be \cite{Cheung:2009dc,Dennen:2009vk}
\begin{align}
\widehat{\mathcal{A}}_4^{(0)} 
&
= \frac{1}{S_{12} S_{23}}
\, ,
\\
\label{5legTree6D}
\widehat{\mathcal{A}}_5^{(0)} 
&
= \frac{1}{\prod_{i = 1}^5 S_{i, i+1}}
\Big(
Q_1 \bar{P}_2 P_3 \bar{P}_4 P_5 \bar{Q}_1 + \mbox{cyclic}
\\
&
+
\ft12
\left(
Q_1 \bar{P}_2 [P_3,\bar{P}_4,P_5] \bar{Q}_2
+
Q_3 \bar{P}_4 [P_5,\bar{P}_1,P_2] \bar{Q}_4
+
(Q_3 + Q_4) \bar{P}_5 [P_1,\bar{P}_2,P_3] \bar{Q}_5 + \mbox{cc}
\right)
\Big)
\, , \nonumber
\end{align}
respectively. The cc stands for the chiral conjugate, i.e., replacing all symbols without bars with bars and vice versa.
Here, as in the four-dimensional case, 
\begin{align}
S_{ij} = (P_i + P_j)^2
\end{align}
are the six-dimensional Mandelstam variables\footnote{Below, we will also encounter $S_{ijk} = (P_i + P_j + P_k)^2$.}. In writing 
$\widehat{\mathcal{A}}_5^{(0)}$, the expression in the second line uses the notation
\begin{align}
[P_i, \bar{P}_j,P_k] \equiv P_i \bar{P}_j P_k - P_k \bar{P}_j P_i
\, .
\end{align}
In this form, $\widehat{\mathcal{A}}_5^{(0)}$ lacks explicit cyclic symmetry, though it implicitly obeys this property on the shell of the 
supermomentum conservation condition. Compared to the four-dimensional MHV case, the five-leg amplitude is not as simple in 
structure (it contains multiple terms compared to just one) and, moreover, it is a polynomial in the Grassmann variables as well. 

\section{Five-leg amplitude in six dimensions}
\label{6D5LegSection}

The calculation of loop corrections to tree amplitudes was streamlined in recent years, bypassing the use of the cumbersome Feynman 
diagram technique which is highly inefficient in their computation. Its Lorentz-covariant nature requires dealing with a large number of 
unphysical degrees of freedom propagating in quantum loops, which in turn proliferates the amount of contributing terms making them 
practically intractable at relatively low orders of perturbation theory. Unitarity-based techniques filled the void of competing, superseding 
methods for a highly proficient evaluation of scattering amplitudes. Unitarity \cite{Bern:1994zx,Bern:1994cg,Bern:2004cz} and generalized 
unitarity \cite{Bern:1997sc,Britto:2004nc,Cachazo:2008vp} can be used in parallel for these purposes 
\cite{Bern:2007dw,Bern:2011qt,Carrasco:2011hw}. With the help of these methods, an amplitude is found from its cuts. In the case of 
the maximally supersymmetric theories, there are no further rational contributions to them and this framework can be employed for their 
unique reconstruction in integer dimensions without further amendments. 

Below, we rely on iterated unitarity cuts rather than generalized unitarity since the latter would force us to use three-leg amplitudes, 
which can only be written with the help of auxiliary spinors \cite{Cheung:2009dc}, thus unnecessarily complicating intermediate 
calculations. However, even so, the construction of loop integrands in six dimensions is a bit more involved compared to their 
four-dimensional counterparts due to the complexity of five- and higher-leg amplitudes. To date, only the four-leg amplitude was 
studied in detail at loop order higher than one\footnote{One-loop five-leg amplitudes were studied in Ref.\ \cite{Brandhuber:2010mm}
but their representation is not particularly practical for our purposes.} \cite{Bern:2010qa}. For five legs and beyond the construction 
of loop integrands benefits from a proper choice of external states, such that, on the one hand, it has a well-defined four-dimensional 
massless limit, and, on the other, contains the fewest number of terms (hopefully, just one). Then, making use of the supersymmetry, 
one can restore the entire amplitude from just one component, see, however, Ref.\ \cite{Plefka:2014fta}. 

\begin{figure}[t]
\begin{center}
\mbox{
\begin{picture}(0,85)(180,0)
\put(0,0){\insertfig{13}{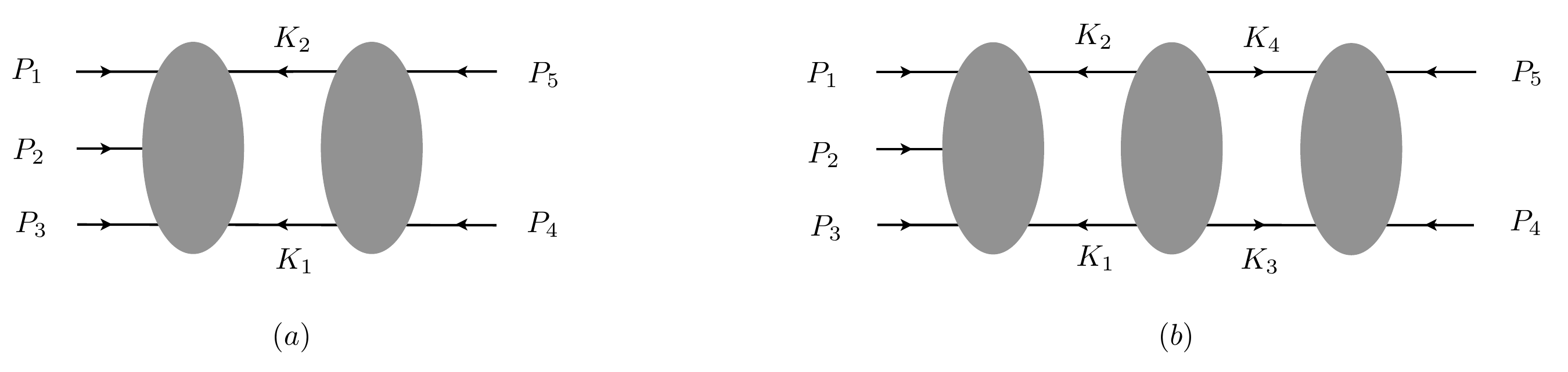}}
\end{picture}
}
\end{center}
\caption{\label{fig12loops} Iterated unitarity cuts for the five-leg amplitude.}
\end{figure}

\subsection{Tree amplitude}

To simplify the construction of loop integrands, we better select a proper component. The four-scalar--gluon amplitude,  
$\vev{\phi^{12}_1 \phi^{12}_2 g^+_3 \phi^{34}_4 \phi^{34}_5}^{(0)}$, with its four-dimensional tree limit \re{4Dscalargluon},
stands up to the challenge of the five-leg case. It is extracted by setting the supercharges $Q_4 = Q_5 = \bar{Q}_4 = \bar{Q}_5 = 0$, 
which will be cumulatively denoted for brevity as the ${Q}_{45} = 0$, and integrating over the Grassmann measure
\begin{align}
\int d^2 \eta_1 d^2 \bar\eta_1 d^2 \eta_2 d^2 \bar\eta_2
\end{align}
making use of the normalization conditions
\begin{align}
d^2 \eta \equiv \ft12 d \eta^a d \eta_a =  d \eta_2 d \eta_1
\, , \qquad
d^2 \bar\eta \equiv \ft12 d \bar\eta_{\dot{a}} d \bar\eta^{\dot{a}} =  d \bar\eta_{\dot{2}} d \bar\eta_{\dot{1}}
\, ,
\end{align}
such that
\begin{align}
\int d^2 \eta\, \eta_a \eta_b = \varepsilon_{ab}
\, , \qquad
\int d^2 \bar\eta \, \bar\eta^{\dot{a}} \bar\eta^{\dot{b}} =  \varepsilon^{\dot{a}\dot{b}}
\, .
\end{align}
These Grassmann integrations can be streamlined with the analogue of the Grassmann identity \re{GrassmannDeltaID4D} generalized
to the six-dimensional setup \cite{Bern:2010qa}
\begin{align}
\label{GrassmannDeltaID6D}
\delta^{(4)} (Q_1 + Q_2 + Q) = - S_{12} 
\delta^{(2)} \left( \eta_1+ \frac{\bra{1} \bar{P}_2 Q}{S_{12}} \right)
\delta^{(2)} \left( \eta_2 + \frac{\bra{2} \bar{P}_1 Q}{S_{12}} \right)
\, .
\end{align}
and the same relation for the barred variables. Dropping the energy-momentum delta function along with the overall numerical 
prefactor\footnote{This will be done systematically in this section for presentation clarity.} in Eq.\ \re{6DColorOrderedAmp}, we get
\begin{align}
\eta_{3, a} \bar\eta_3^{\dot{a}}
\vev{\phi_1^{\prime\prime\prime} \phi_2^{\prime\prime\prime} g_3^{a}{}_{\dot{a}} \phi_4 \phi_5}^{(0)}
=
\int d^2 \eta_1 d^2 \bar\eta_1 d^2 \eta_2 d^2 \bar\eta_2 \mathcal{A}_5^{(0)} |_{{Q}_{45} = 0}
\, ,
\end{align}
where the tree amplitude (before simplifications)
\begin{align}
\vev{\phi_1^{\prime\prime\prime} \phi_2^{\prime\prime\prime} g_3^{a}{}_{\dot{a}} \phi_4 \phi_5}^{(0)}
=
\frac{S_{12} \bra{3^a}{\mit\Pi}_{P_1, P_2, P_3, P_4, P_5} |3_{\dot{a}}]}{\prod_{i = 1}^5 S_{i,i+1}}
\, ,
\end{align}
is encoded in a polynomial of particle momenta,
\begin{align}
{\mit\Pi}_{P_1, P_2, P_3, P_4, P_5} 
=
 \bar{P}_2 P_3 \bar{P}_4 P_5 \bar{P}_1 P_2 
&
+
\bar{P}_1 P_2 \bar{P}_3 P_4 \bar{P}_5 P_1
+
S_{12}
\bar{P}_4 P_5 \bar{P}_1 P_2
\\
&
+
\ft12 
\bar{P}_2 [P_3, \bar{P}_4, P_5] \bar{P}_2 P_1 
+
\ft12 
\bar{P}_1 P_2 [\bar{P}_3, P_4, \bar{P}_5] P_2
\, . \nonumber
\end{align}
To simplify it, let us look at its four-dimensional massless limit first. Making use of Eqs.\ \re{LambdaFromlambdas}, we  then find
for the component with the little group indices $a = 2$, $\dot{a} = \dot{1}$
\begin{align}
\vev{\phi_1^{\prime\prime\prime} \phi_2^{\prime\prime\prime} g_3^{a}{}_{\dot{a}} \phi_4 \phi_5}^{(0)}_{a = 2, \dot{a} = \dot{1}}
=
\frac{s_{12} [ 3 | {\mit\Pi}_{p_1,p_2,p_3,p_4,p_5} | 3 ]}{\prod_{i = 1}^5 s_{i,i+1}}
\, ,
\end{align}
where the numerator can be simplified to the expression
\begin{align}
[ 3 | {\mit\Pi}_{p_1,p_2,p_3,p_4,p_5}  | 3 ] = s_{45} [3| p_4 p_5 p_1 p_2 |3]
\, ,
\end{align}
by projecting both sides with $\langle 3 | p_2 p_1 p_5 p_4 | 3 \rangle$ and evaluating emerging  two-dimensional traces with, say,
{\tt FeynCalc} \cite{Mertig:1990an,Shtabovenko:2023idz}. Further, making use of the four-dimensional spinor-helicity for four-momenta
\re{FourMomentumSpinors}, one immediately confirms that this is indeed the MHV amplitude in question \re{4Dscalargluon}.

Returning to six dimensions, we will only be after the component amplitude, which has the above four-dimensional limit.
Thus, we need to find the overlap of $\bra{3^{a}} {\mit\Pi} |3_{\dot{a}}]$ with the structure 
\begin{align}
( T_3 )^{a}{}_{\dot{a}} = \bra{3^a} \bar{P}_4 P_5 \bar{P}_1 P_2 |3_{\dot{a}}]
\, .
\end{align}
Projecting the five-leg amplitude on $T_3$, we get
\begin{align}
\bra{3^{a}} {\mit\Pi}_{P_1, P_2, P_3, P_4, P_5} |3_{\dot{a}}] ( T_3^{-1} )^{\dot{a}}{}_{a} 
=
- 
\frac{
\tr_4 [{\mit\Pi}_{P_1, P_2, P_3, P_4, P_5} \bar{P}_3 P_2 \bar{P}_1 P_5 \bar{P}_4 P_3] 
}{S_{12} S_{23} S_{34} S_{45} S_{51}}
= 2 S_{45}
\, .
\end{align}
So that the sought-after six-dimensional tree amplitude reads
\begin{align}
\label{6DA5component}
\vev{\phi_1^{\prime\prime\prime} \phi_2^{\prime\prime\prime} g_3^{a}{}_{\dot{a}} \phi_4 \phi_5}^{(0)}
=
\frac{S_{12} S_{45}}{\prod_{i = 1}^5 S_{i,i+1}}  (T_3)^{a}{}_{\dot{a}}
\, .
\end{align}
As we already established above for $a = 2$, $\dot{a} = \dot{1}$, it correctly reduces to the MHV amplitude 
$\vev{\phi^{12}_1 \phi^{12}_2 g^+_3 \phi^{34}_4 \phi^{34}_5}^{(0)}$, while for $a = 1$, $\dot{a} = \dot{2}$ we uncover 
the ${\rm NMHV} = \overline{\rm MHV}$ amplitude $\vev{\phi^{12}_1 \phi^{12}_2 g^-_3 \phi^{34}_4 \phi^{34}_5}^{(0)}$. 
The other two polarizations will not be of interest.

\subsection{One-loop integrand}

\begin{figure}[t]
\begin{center}
\mbox{
\begin{picture}(0,275)(155,0)
\put(0,0){\insertfig{10}{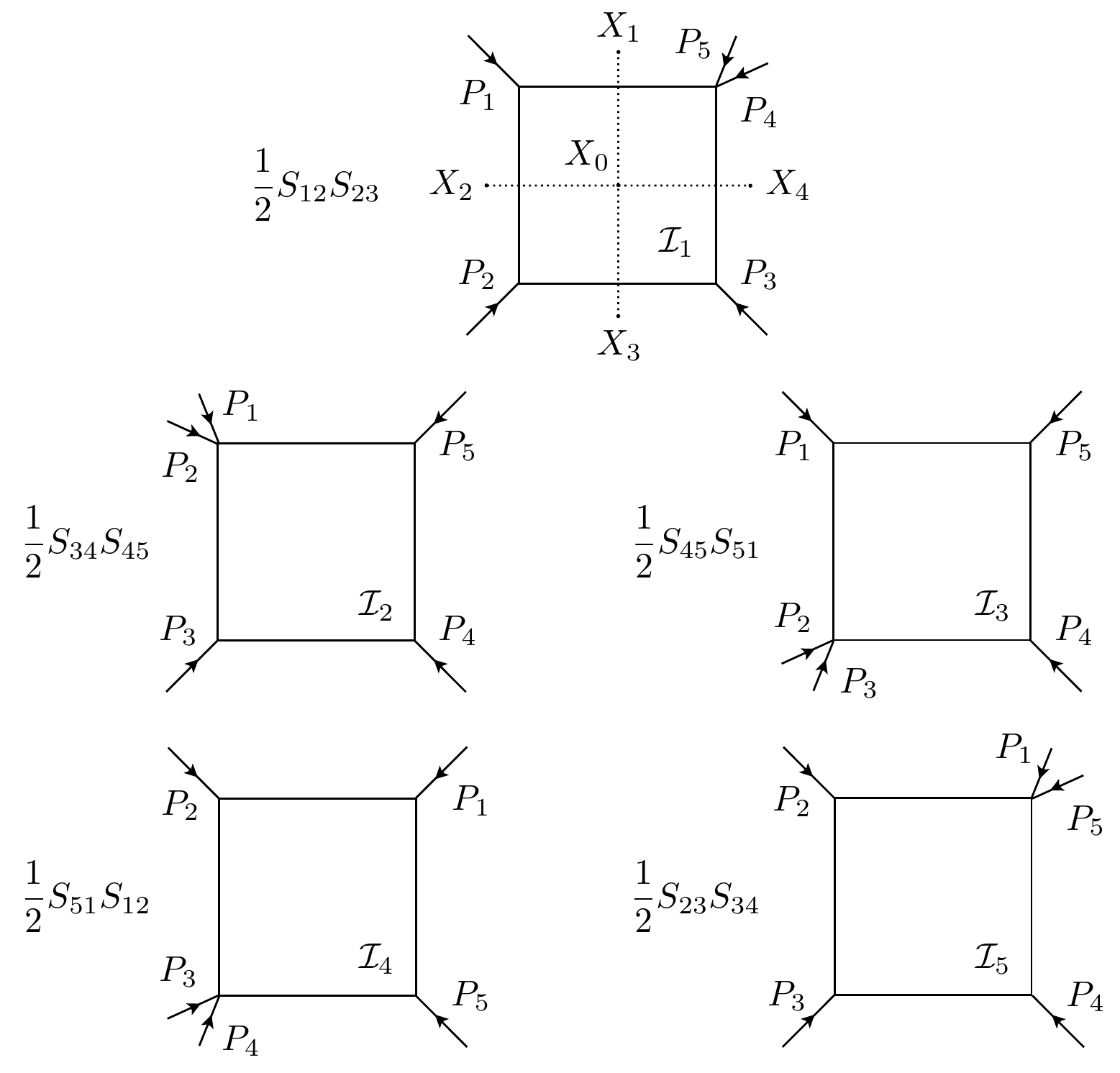}}
\end{picture}
}
\end{center}
\caption{\label{5gColFig} One-loop scalar box expansion of the five-leg amplitude.}
\end{figure}

Let us move on to the one-loop case. We will use two-particle unitarity cuts to find its integrand. The calculation will proceed in 
the same fashion as in four dimensions, as we recall for the reader's convenience in Appendix \ref{4D5LegAppendix}. Namely, 
the starting point is the $S_{123}$ cut
\begin{align}
\mathcal{A}^{(1)}_5 |_{S_{123}-{\rm cut}}
=
\int \prod_{i = 1}^2
d^2 \eta_{K_i} d^2 \bar\eta_{K_i} 
\mathcal{A}^{(0)}_5 (P_1, P_2, P_3, K_1, K_2)
\mathcal{A}^{(0)}_4 (P_4, P_5, - K_2, - K_1)
\, ,
\end{align}
shown in Fig.\ \ref{fig12loops} (a). Here, we showed only the bosonic momenta, with their corresponding chiral charges being
implicit. It is at this point that the choice of the chosen component \re{6DA5component} becomes extremely advantageous for 
subsequent efficient derivation. Namely, extracting the overall fermionic delta function of the (anti-)chiral supercharges, and 
using the condition ${Q}_{45} = 0$, only the $\phi_4 \phi_5 \to \phi^{\prime\prime\prime}_{K_1} \phi^{\prime\prime\prime}_{K_2}$ 
component of the four-leg amplitude contributes to the cut in question. This is akin to the singlet contribution used in the past as 
a simplifying tool \cite{Bern:2007ct} to compute gluon amplitudes. As a consequence, in the equation,
\begin{align}
\label{1Loop5Leg6D}
&
\mathcal{A}^{(1)}_5 |_{S_{123}-{\rm cut}, {Q}_{45} = 0}
=
\delta^{(4)} \left(\sum\nolimits_{i = 1}^3 Q_i\right)  
\delta^{(4)} \left(\sum\nolimits_{i = 1}^3 \bar{Q}_i\right)
\\
&\qquad\qquad
\times
\int \prod_{i = 1}^2
d^2 \eta_{K_i} d^2 \bar\eta_{K_i} 
\widehat{\mathcal{A}}_5^{(0)} (P_1, P_2, P_3, K_1, K_2)
\frac{
\delta^{(4)} (Q_{K_1} + Q_{K_2}) \delta^{(4)} (\bar{Q}_{K_1} + \bar{Q}_{K_2})
}{S_{45} S_{4, -K_1}}
\, , \nonumber
\end{align}
the intermediate-state Grassmann integrations can be performed trivially making use of Eq.\ \re{GrassmannDeltaID6D}.
Here $\widehat{\mathcal{A}}_5^{(0)}$ in the integrand is given by Eq.\ \re{5legTree6D} with the substitutions $P_4 \to K_1$ and $P_5 \to K_2$.
Extracting the scalar states in legs 1 and 2, we obtain
\begin{align}
\int d^2 \eta_1 d^2 \bar\eta_1 d^2 \eta_2 d^2 \bar\eta_2
\mathcal{A}^{(1)}_5 |_{S_{123}-{\rm cut}, {Q}_{45} = 0}
=
\frac{\bra{3^{a}} {\mit\Pi}_{P_1, P_2, P_3, K_1, K_2} |3_{\dot{a}}]}{S_{23} S_{3, K_1} S_{K_1, K_2} S_{K_2, 1}} 
\times
\frac{S_{45}}{S_{4, - K_1}}
\, ,
\end{align}
which is written as a product of the five- and four-leg amplitudes, $\phi^{\prime\prime\prime} \phi^{\prime\prime\prime} g^a{}_{\dot{a}} \to 
\phi \phi$ and $\phi^{\prime\prime\prime} \phi^{\prime\prime\prime} \to  \phi \phi$, respectively. As in the tree case, we are looking at its 
projection on the $T_3$ structure, which yields a single trace
\begin{align}
\bra{3^{a}} {\mit\Pi}_{P_1, P_2, P_3, K_1, K_2} |3_{\dot{a}}]
&
=
(T_3)^a{}_{\dot{a}}
\frac{\tr_4 [{\mit\Pi}_{P_1, P_2, P_3, K_1, K_2} \bar{P}_1 P_2 \bar{P}_3 P_2 \bar{P}_1 P_5 \bar{P}_4 P_3]
}
{2
S_{12} S_{23} S_{34} S_{45} S_{51}}
\nonumber\\
&
=
\frac{(T_3)^a{}_{\dot{a}}}{2 S_{34} S_{51}}
\left[
S_{12} S_{23} S_{4, -K_1} + S_{45} S_{51} S_{3,K_1} + S_{34} S_{45} S_{K_2, 1}
\right]
\, .
\end{align}
Combining these results together gives for the $S_{123}$ cut
\begin{align}
\int d^2 \eta_1 d^2 \bar\eta_1 d^2 \eta_2 d^2 \bar\eta_2
\mathcal{A}^{(1)}_5 |_{S_{123}-{\rm cut}, {Q}_{45} = 0}
=
\frac{1}{2}
\mathcal{A}^{(0)}_5
\left[
\frac{S_{12} S_{23}}{S_{3,K_1} S_{K_2, 1}}
+ 
\frac{S_{45} S_{51}}{S_{K_2, 1} S_{4, -K1}}
 + 
\frac{S_{34} S_{45}}{S_{3,K_1}S_{4, -K1}}
\right]
\, .
\end{align}
This is indeed the uplift of the four-dimensional integrand \re{1Loop5LegMHV} up to six dimensions, which is simply obtained by changing 
all Mandelstam variables from $s_{ij}$ to $S_{ij}$. A similar phenomenon was also observed in Refs.\ \cite{Huang:2011um,Plefka:2014fta} 
for certain components of $\mathcal{N} = (1,1)$ amplitudes. Cuts in other channels are not `singlet' and are more involved. However, since 
the full superamplitude is cyclic, there is no need to analyze these separately. One can restore the rest of the full integrand either by using 
cyclicity or choosing a projection akin to the one adopted above in those channels as well. It is given by the sum of one-loop box
integrals accompanied by Madelstam invariants, as shown in Fig.\ \ref{5gColFig}. This result is in agreement with a generalized cut analysis 
of Ref.\ \cite{Brandhuber:2010mm}.

\subsection{Two-loop integrand}

\begin{figure}[t]
\begin{center}
\mbox{
\begin{picture}(0,95)(220,0)
\put(0,0){\insertfig{15}{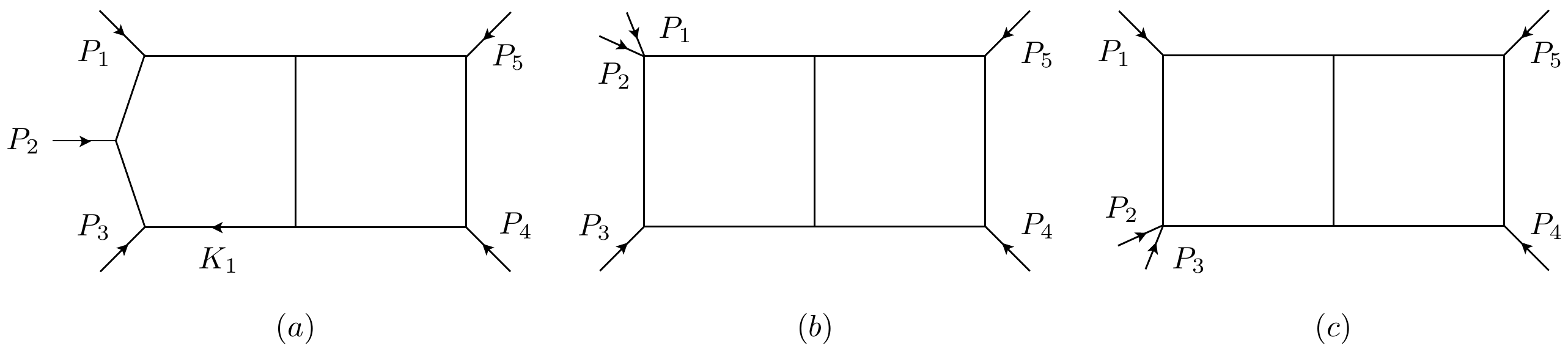}}
\end{picture}
}
\end{center}
\caption{\label{2LoopGraphs} Two-loop integrands defining the parity-even part of the five-leg off-shell amplitude in the $S_{123}$
channel. The pentabox possesses a nontrivial numerator $(K_1 - P_4)^2$.}
\end{figure}

To demonstrate the proficiency of the chosen component, let us carry out a two-loop calculation as a final exercise. The iterative 
$S_{123}$-cut, shown in Fig.\ \ref{fig12loops} (b), will allow us to detect all appearing integrands (up to cyclic contributions). Its reads
\begin{align}
\mathcal{A}^{(2)}_5 |_{S_{123}-{\rm cut}}
=
\int \prod_{i = 1}^4 d^2 \eta_{K_i} d^2 \bar\eta_{K_i} 
&
\mathcal{A}^{(0)}_5 (P_1, P_2, P_3, K_1, K_2)
\\
\times&
\mathcal{A}^{(0)}_4 (- K_2, - K_1, - K_3, - K_4)
\mathcal{A}^{(0)}_4 (P_4, P_5, K_4, K_3)
\, . \nonumber
\end{align}
Choosing the supercomponent in question by imposing the condition ${Q}_{45} = 0$ has the domino effect on the Grassmann
dependence of the integrand by `localizing' them at $\eta_{K_i} = \bar\eta_{K_i} = 0$, namely,
\begin{align}
\label{2Loop5Leg6D}
\mathcal{A}^{(2)}_5 |_{S_{123}-{\rm cut}; {Q}_{25} = 0}
&
=
\delta^{(4)} \left(\sum\nolimits_{i = 1}^3 Q_i\right)  
\delta^{(4)} \left(\sum\nolimits_{i = 1}^3 \bar{Q}_i\right)
\int \prod_{i = 1}^4 d^2 \eta_{K_i} d^2 \bar\eta_{K_i} 
\widehat{\mathcal{A}}^{(0)}_5 (P_1, P_2, P_3, K_1, K_2)
\nonumber\\
&\times
\frac{\delta^{(4)} (Q_{K_1} + Q_{K_2}) \delta^{(4)} (\bar{Q}_{K_1} + \bar{Q}_{K_2})}{S_{K_1,K_2} S_{K_1,K_3}} 
\frac{\delta^{(4)} (Q_{K_3} + Q_{K_4}) \delta^{(4)} (\bar{Q}_{K_3} + \bar{Q}_{K_4})}{S_{45} S_{4,K_3}} 
\, . 
\end{align}
This is just an iteration of the one-loop cut \re{1Loop5Leg6D}. So the subsequent calculation repeats the same steps as above
and can be easily obtained from it by relabeling the cut momenta and an extra factor of the intermediate four-leg amplitude.
The result is
\begin{align}
&
\int d^2 \eta_1 d^2 \bar\eta_1 d^2 \eta_2 d^2 \bar\eta_2
\mathcal{A}^{(2)}_5 |_{S_{123}-{\rm cut}, {Q}_{45} = 0}
\\
&\qquad
=
\frac{1}{2}
\mathcal{A}^{(0)}_5
\left[
\frac{S_{12} S_{23} S_{45} S_{4, - K1}}{S_{3,K_1} S_{K_2, 1} S_{K_1,K_3} S_{4, K_3}}
+ 
\frac{S_{45}^2 S_{51}}{S_{K_2, 1} S_{K_1,K_3} S_{4, K_3}}
 + 
\frac{S_{34} S_{45}^2}{S_{3,K_1} S_{K_1,K_3} S_{4, K_3}}
\right]
\, . \nonumber
\end{align}
The first term in the square bracket is the pentabox with the loop-momentum dependent numerator $S_{4, - K_1}$, while
the rest are the double boxes, as shown in Fig.\ \ref{2LoopGraphs}. Adding the cyclically symmetric contributions yields 
the full amplitude. Again, as in the one-loop case, we observe that the six-dimensional integrand is an uplift $s_{ij} \to S_{ij}$ 
of the parity-even part of the well-known four-dimensional five-leg amplitude \cite{Bern:1997it,Bern:2006vw,Cachazo:2008vp,Carrasco:2011mn}.

The just-demonstrated explicit construction puts the uplift used in the past for construction of loop integrands of amplitudes and form factors 
on the Coulomb branch \cite{Henn:2010ir,Caron-Huot:2021usw,Belitsky:2022itf,Belitsky:2023ssv,Belitsky:2024agy,Belitsky:2024dcf} 
on a firm foundation.

\section{Dimensional reduction}
\label{DimRedSection}

All of our subsequent calculations will be performed at one loop order. In light of our findings in the previous section, the five-leg 
$\mathcal{N} = (1,1)$ ratio function $\mathcal{M}_5 = \mathcal{A}_5/\mathcal{A}^{(0)}_5$ is then given by the perturbative expansion 
in the six-dimensional gauge coupling $g_{6, \rm\scriptscriptstyle YM}$
\begin{align}
\label{RatioM5}
\mathcal{M}_5
=
1 - g^2_{6, \rm\scriptscriptstyle YM} N_c \left[ \frac{1}{2} S_{12} S_{23} \mathcal{I}_1 + \mbox{cyclic} \right]
+
O (g^4_{6, \rm\scriptscriptstyle YM})
\, ,
\end{align}
with the coefficients determined by the cyclic permutation of the box integral
\begin{align}
\mathcal{I}_1 = \int \frac{d^6 K}{i (2 \pi)^6} \frac{1}{K^2 (K + P_1)^2 (K + P_1 + P_2)^2 (K + P_1 + P_2 + P_3)^2}
\, .
\end{align}
To obtain an off-shell amplitude from this expression, we impose a set of kinematical constraints in a dual conformally invariant 
fashion \cite{Caron-Huot:2021usw}. First, we pass to the region momenta $X_i$, $P_i = X_i - X_{i + 1}$ and $K = X_0 - X_1$, 
with the dual graph shown in the top panel of Fig.\ \ref{5gColFig}. Second, we decompose these in terms of their four- and 
out-of-four-dimensional components $X_i = (x_i, y_i)$ and impose a null condition on $y_i$, $y_i^2 = 0$. This can be realized 
with complex values of the extra-dimensional positions. It implies that all internal lines are massless. The external states, on the 
other hand, are massive or off-shell from the four-dimensional perspective, 
\begin{align}
\label{6DtoOffShell}
P_i^2 = 0 \quad \to \quad p_i^2 \equiv x_{i,i+1}^2 = - y_{i,i+1}^2 \equiv - m
\, .
\end{align}
All virtualities are chosen to have the same value (for simplicity). Finally, the coordinate of the integration vertex $X_0$ is dimensionally 
reduced down to four-dimensions, $X_0 = (x_0, 0)$ with $d^6 X_0 = R_0^2 d^4 x_0$. The four-dimensional dimensionless gauge 
coupling $g_{\rm\scriptscriptstyle YM}$ emerges from its six-dimensional counterpart as $g_{\rm\scriptscriptstyle YM} = 
g_{6, \rm\scriptscriptstyle YM} R_0/(2 \pi)$. One can then pass back to the momentum representation but now involving only the 
four-dimensional momenta,
\begin{align}
\label{1loopBox}
I_1 
= 
(8 \pi^2/R_0)^2 \times \mbox{dim.red.} [\mathcal{I}_1]
=
\int \frac{d^4 k}{i \pi^2} \frac{1}{k^2 (k + p_1)^2 (k + p_1 + p_2)^2 (k + p_1 + p_2 + p_3)^2}
\, ,
\end{align}
where we took the factor of $(8 \pi^2)^2$ out of the measure so that the loop integrals comply with standard conventions. Thus, 
the starting point for our subsequent analysis is the off-shell ratio
\begin{align}
\label{1loopAmplitude}
M_5 = 
1 
- 
g^2 \left[ \frac{1}{2} s_{12} s_{23} I_1 + \mbox{cyclic} \right]
+
O (g^4, m^2)
\, ,
\end{align}
to leading order in 't Hooft coupling $g^2 = g^2_{\rm\scriptscriptstyle YM} N_c/(4 \pi)^2$. Here, we neglected $O(m^2)$ terms 
in all numerators since our interest is in the near-mass-shell limit, i.e., $m \to 0$.

\section{Infrared factorization}
\label{1LoopOffShellAmplitude}

To prepare the ground for the collinear analysis \re{SplitDef}, when the invariant mass of the `coalescing' collinear pair can be as small 
as the off-shellness $m$, let us construct the infrared factorized formula for the amplitude first when there are just two groups of widely 
separated scales, namely, the large Madelstam invariants $s_{i,i+1}$, on the one hand, and the small off-shellness, on the other. We will 
work in the Euclidean domain to avoid dealing with discontinuities, i.e., 
\begin{align}
\label{NearMassShellLimit}
- s_{i,i+1} \gg m \to 0_+
\, .
\end{align}

Recently, we studied this asymptotic domain in Ref.\ \cite{Belitsky:2024yag} for the off-shell Sudakov form factor, the simplest
prototypical example of an infrared-sensitive quantity akin to amplitudes. Instead of a traditional analysis of Landau equations
to determine when some propagators go on-shell and induce singularities, we employed the Method of Region (MofR) 
\cite{Beneke:1997zp}. The latter, formulated in the momentum space, determines all scalings of loop momenta in terms of external 
kinematics and unambiguously establishes leading contributing regions to the limit in question. Each of these corresponds to a 
pinch surface of the Landau equations. However, it goes even further and endows them with an integral representation 
of reduced integrands, however, with unrestricted integration over the entire space. An operator definition can then be 
reversed-engineered from these. This setup provided a factorized expression for the form factor in terms of incoherent momentum 
components that do not `talk' to each other.

\begin{figure}[t]
\begin{center}
\mbox{
\begin{picture}(0,150)(235,0)
\put(0,0){\insertfig{16.5}{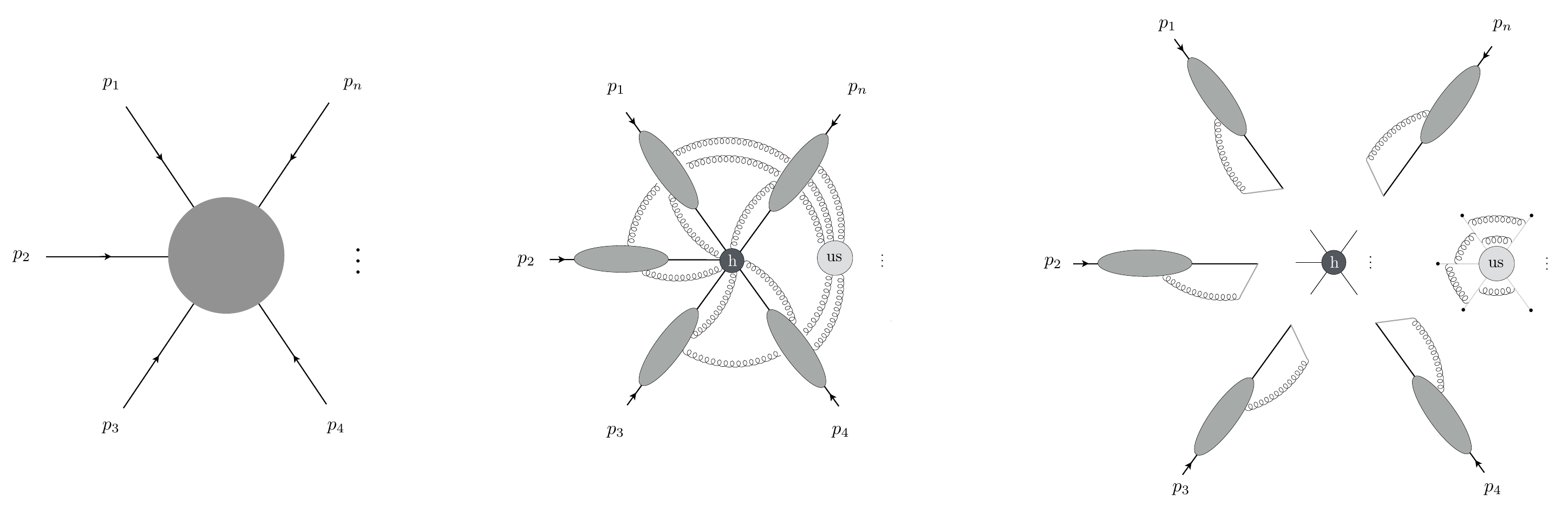}}
\end{picture}
}
\end{center}
\caption{\label{PinchFig} Leading reduced graph (middle) for an $n$-leg amplitude (left panel) near its mass shell and its
factorized form (right panel) in terms of matrix elements of Wilson-line operators. As the shade of gray gets more intense, it
represents the harder momentum: from ultrasoft to collinear and then to hard, see Eq.\ \re{LeadingRegions}.}
\end{figure}

The scattering amplitudes in the asymptotic limit \re{NearMassShellLimit} will enjoy a similar factorization. Three distinct scales
emerge from the analysis, the hard, collinear, and ultrasoft
\begin{align}
\label{LeadingRegions}
\mu^2_{\rm h} = O (s_{i,i+1})
\, , \qquad
\mu^2_{\rm c} = O (m)
\, , \qquad
\mu^2_{\rm us} = O (m^2/s_{i,i+1})
\, ,
\end{align}
respectively. These reflect the physical processes underlying amplitude's decomposition into independent components. As 
the nearly mass-shell states travel toward the scattering region, they copiously emit collinear gluons forming jets. These do 
not resolve each other and only see the overall color and direction of propagation. Thus, they can only interact by means of 
long-wave ultrasoft gluons. The hard region is a matching `coefficient' that does not depend on any soft scales, i.e., $\mu_{\rm c}$
or $\mu_{\rm us}$. In fact, they are equal to the massless on-shell amplitude in the current MofR-based approach. The pinch 
surfaces corresponding to these regions are demonstrated in the middle panel of Fig.\ \ref{PinchFig} along with a further 
disentanglement of these from each other in the right panel. The clear-cut separation of various momentum modes suggested 
by the right panel in Fig.\ \ref{PinchFig} is not immediately implied by the MofR. As was shown in Ref.\ \cite{Belitsky:2024yag} 
for the case of the Sudakov form factor, additional subtle modifications of factorized components is required beyond one-loop 
order. Currently, it will not be an issue since our consideration is done at leading order in 't Hooft coupling.

\subsection{One loop with MofR}
\label{IRoneloop}

Since the separation of a finite loop-momentum integral in individual regions inevitably introduces singularities for the simple fact that 
propagators are expanded according to the kinematical region in question while the integration is nevertheless performed over 
the entire infinite domain of the loop's energy-momentum, we have to introduce dimensional regularization at intermediate steps. 
This is accomplished with substitutions
\begin{align}
\int \frac{d^4 k}{i \pi^2} \to \int_k \equiv \e^{\varepsilon \gamma_{\rm E}} \mu^{2 \varepsilon} \int \frac{d^D k}{i \pi^{D2}}
\, , \qquad
g^2 \to g^2_\varepsilon = \e^{- \varepsilon \gamma_{\rm E}} \frac{g^2_{\rm\scriptscriptstyle YM} N_c}{(4 \pi)^{D/2}}
\, ,
\end{align}
in Eqs.\ \re{1loopBox} and \re{1loopAmplitude}, with $D = 4 - 2 \varepsilon$.

A MofR analysis with {\tt FIESTA} \cite{Smirnov:2021rhf} of the integral $I_1$ in the limit \re{NearMassShellLimit} identifies six leading 
regions. They can be identified with the above-introduced momentum scalings \re{LeadingRegions} and possess the following region 
vectors
\begin{align}
\label{IRregions}
\bit{v}_{\rm h} 
&
= (0,0,0,0)
\, , \\
\bit{v}_{\rm c1} 
&
= (0,0,1,1)
\, , \qquad
\bit{v}_{\rm c2} = (1,0,0,1)
\, , \qquad
\bit{v}_{\rm c3} = (1,1,0,0)
\, , \nonumber\\
\bit{v}_{\rm us2} 
&= (1,0,1,2)
\, , \qquad
\bit{v}_{\rm us3} = (2,1,0,1)
\, . \nonumber
\end{align}
The first one, $\bit{v}_{\rm h} $, stands for the loop momentum being hard $k^\mu = O (s_{i,i+1})$, the next three $\bit{v}_{{\rm c}i}$
describe the momentum collinear to the external momenta\footnote{This behavior of the loop momentum is sensitive to the choice 
of momentum routing. We imply that a proper one was adopted by effecting shifts with external momenta.} $p^\mu_1$, 
$p^\mu_2$, and $p^\mu_3$, respectively. Finally, the last two correspond to the ultrasoft momenta $k^\mu = O(m^2/s_{12})$ and 
$k^\mu = O(m^2/s_{23})$ when propagators 2 and 3, i.e., between the $p_1$ and $p_2$, and $p_2$ and $p_3$ vertices, respectively, 
blow up. The calculation of parametric Feynman integrals defining these is straightforward and yields the Laurent expansion in 
$\varepsilon$ for the hard
region
\begin{align}
\label{1LoopHard}
I_1^{\rm h}
&
=
\mbox{Box}_1 [s_{12}, s_{23}; s_{45}]
\\
&
\equiv
\frac{c_\varepsilon}{s_{12} s_{23}} \left(- \frac{\mu^2 s_{45} }{s_{12} s_{23}} \right)^\varepsilon
\left[
\frac{2}{\varepsilon^2} + 2 {\rm Li}_2 \left( 1 - \frac{s_{12}}{s_{45}} \right) + 2 {\rm Li}_2 \left( 1 - \frac{s_{23}}{s_{45}} \right)
- 2 \zeta_2
+
O (\varepsilon)
\right]
\, . \nonumber
\end{align}
This expression is nothing else than the well-known one-mass box with the heavy $p_4+p_5$ leg 
\cite{Fabricius:1979tb,Papadopoulos:1981ju,Bern:1993kr} that defines the massless five-leg amplitude 
\cite{Bern:1997it,Bern:2006vw}. Here and below, the overall ratio of Euler gammas is
\begin{align}
c_\varepsilon = 
\frac{\e^{\varepsilon \gamma_{\rm E}} \Gamma^2 (1 - \varepsilon) \Gamma (1 + \varepsilon)}{\Gamma (1 - 2 \varepsilon)}
\, .
\end{align}
The collinear regions are found similarly and read
\begin{align}
I_1^{\rm c1}
&
=
I_1^{\rm c3}|_{s_{12} \leftrightarrow s_{23}}
\\
&
=
- \frac{c_\varepsilon}{s_{12} s_{23}} \left( \frac{\mu^2 s_{45}}{m s_{23}} \right)^\varepsilon
\left[
\frac{1}{\varepsilon^2} + 2 {\rm Li}_2 \left( 1 - \frac{s_{23}}{s_{45}} \right) 
+ O(\varepsilon)
\right]
\, , \nonumber\\
I_1^{\rm c2}
&
=
-
\frac{c_\varepsilon}{s_{12} s_{23}} \left( \frac{\mu^2}{m} \right)^\varepsilon\frac{2}{\varepsilon^2}
\, .
\end{align}
The $I_1^{\rm c2}$ is exact in $\varepsilon$ and agrees with the known expression for the one-loop jet factor, see Eq.\ (4.6) in 
\cite{Belitsky:2024yag}. Finally, the ultrasoft regions produce
\begin{align}
I_1^{\rm us2}
=
I_1^{\rm us3}|_{s_{12} \leftrightarrow s_{23}}
=
-
\frac{2 c_\varepsilon}{s_{12} s_{23}} \left( - \frac{\mu^2 s_{12}}{m^2} \right)^\varepsilon
\frac{\Gamma (-2 \varepsilon) \Gamma (\varepsilon)}{\Gamma (1-\varepsilon)}
\, .
\end{align}
They coincide with the one in the Sudakov form factor as well, see Eq.\ (5.14) in \cite{Belitsky:2024yag}.

\subsection{Hard, collinear and ultrasoft}
\label{HardColUSsection}

To this order in 't Hooft coupling, one can immediately write a factorized expression for the ratio \re{1loopAmplitude}
\begin{align}
\label{IRfactorizationM5}
M_5 = h_5 \left(\mu^2/\mu_{\rm h}^2, \varepsilon \right) S_5 (\mu^2/\mu_{\rm us}^2, \varepsilon) \prod_{i = 1}^5 J_i (\mu^2/\mu_{\rm c}^2, \varepsilon) 
\, .
\end{align}
Here, the matching coefficient $h_5$ is determined by the sum of all hard regions in the cyclic sum of \re{1LoopHard} with 
accompanying products of Mandelstam invariants. It can be reduced to the form
\begin{align}
h_5 = 1 + g_\varepsilon^2 c_\varepsilon V^{(1)}_5 + O (g_\varepsilon^4)
\, ,
\end{align}
where $V^{(1)}_5$ is the well-known one-loop massless on-shell five-leg gluon amplitude \cite{Bern:1993mq,Bern:1993kr},
\begin{align}
V^{(1)}_5 = - \sum_{i = 1}^5 \left[ \frac{1}{\varepsilon^2}  \left(- \frac{\mu^2}{s_{i,i+1}} \right)^\varepsilon 
- 
\log \frac{s_{i,i+1}}{s_{i+1, i+2}} \log \frac{s_{i+2,i+3}}{s_{i-2, i-1}}
- 
\zeta_2
\right]
\, .
\end{align}

Next, in the sum of all cyclic permutations of collinear regions, the ones possessing non-trivial dependence on the hard
scales nicely combine into an $s_{i,i+1}$-independent expression
\begin{align}
\sum_{\rm cyclic}
s_{12} s_{23} \left( I_1^{\rm c1} + I_1^{\rm c3} \right)
=
\sum_{\rm cyclic}
s_{12} s_{23} I_1^{\rm c2} 
\, ,
\end{align}
given by the one-loop jet. This doubles the contributing integral $I_1^{\rm c2}$, correctly yielding the normalized jet function,
\begin{align}
\label{1LoopJet}
J_i = 1 
+
g_\varepsilon^2 J^{(1)} 
+ 
O (g_\varepsilon^4)
\, , \qquad
J^{(1)} = \frac{2 c_\varepsilon}{\varepsilon^2} \left( \frac{\mu^2}{m} \right)^\varepsilon
\, .
\end{align}
As was demonstrated in \cite{Belitsky:2024yag}, this result arises from the off-shell matrix element of a (scalar) field with an 
attached semi-infinite Wilson line $[\infty, 0]_{\bar{n}_i}$ emanating from it and going in the light-like direction $\bar{n}_i$ 
`opposite' to its largest momentum component $n_i$
\begin{align}
\label{SudakovDecomMom}
p_i = n_i - \ft12 m \, \bar{n}_i
\end{align}
with $n_i \cdot \bar{n}_i = 1$,
\begin{align}
J_i = \bra{0} [\infty, 0]_{\bar{n}_i} \phi(0) \ket{p_i}
\, .
\end{align}
This explains the graphical depiction in Fig.\ \ref{PinchFig}.

Finally, the ultrasoft regions are generated by the vacuum expectation of a product of Wilson line segments in the
adjoin representation of the color group, aligned with the largest component of the parent momentum \re{SudakovDecomMom}, 
meeting at the origin
\begin{align}
\label{USWilonSements}
\mathcal{W}_{n_1, n_2, \dots} (\sigma_1, \sigma_2, \dots)
=
\vev{C_{a_1a_2, \dots} [0, 2 \sigma_1 n_1]_{a_1} [0, 2 \sigma_2 n_2]_{a_2} \dots }
\, ,
\end{align}
with $C$ being a color tensor that contracts SU$(N_c)$ indices at the merger (the ones at the opposite end of intervals are
not displayed for brevity). Its structure is inherited from the Chan-Patton factor of the color-ordered amplitude under study.
In the planar approximation we are currently working with, different components do not mix. This implies that at leading order 
in color, the matrix element \re{USWilonSements} falls part into the product of individual wedges of pairwise nearest-neighbor 
segments in the fundamental representation, in complete analogy with the massless case\footnote{It is interesting to note
that the soft (as opposed to the current ultrasoft) factor of massless scattering amplitudes, which is defined in terms of
semi-infinite Wilson links, enjoys a pair-wise, aka dipole, form even for finite $N_c$ at two loops \cite{Aybat:2006mz}. The 
reason being that non-Abelian three-gluon graphs connecting three different lines vanish. Non-trivial corrections start only 
at three loops \cite{Almelid:2015jia}. It would be instructive to see when the arguments of Ref.\ \cite{Aybat:2006mz} fail for the
finite segmented Wilson lines \re{USWilonSements}.} \cite{Bern:2005iz}. However, since we prefer to work with adjoint lines, 
the fundamental ones can be recast in their terms by taking their square root, up to effects suppressed in $N_c$. The 
off-shellness is then introduced as a reciprocal variable to the length of each interval,
\begin{align}
W_{n_1, n_2} (m)
=
\int_0^\infty d\sigma_1 d\sigma_2 
\e^{- i (\sigma_1 + \sigma_2) m_- }
\mathcal{W}_{n_1, n_2} (\sigma_1, \sigma_2)
\, .
\end{align}
Here, the off-shellness has a small negative imaginary part $m_- = m - i 0$ necessary for convergence. This operator was shown 
to produce the correct ultrasoft behavior of the two-loop Sudakov form factor as $m \to 0$ \cite{Belitsky:2024yag}. Amputating the 
external propagators from $W_{n_1, n_2} (m)$, we define the square of the ultrasoft wedge
\begin{align}
w_{12}^2 = (i m)^2 W_{n_1, n_2} (m)
\, ,
\end{align}
such that the function $S_5$ entering the factorized expression for the amplitude \re{IRfactorizationM5} is
\begin{align}
S_5 = \prod_{i = 1}^5 w_{i,i+1} (\mu^2 s_{i,i+1}/m^2, \varepsilon)
\, .
\end{align}
Here, we explicitly displayed its dependence on the kinematical scales involved. At one loop, it is given by
\begin{align}
w_{i,i+1} 
= 
1 + g_\varepsilon^2 w_{i,i+1}^{(1)}
+
O(g_\varepsilon^4) \, ,
\qquad
w_{i,i+1}^{(1)}
=
-
\left(- \frac{\mu^2 s_{i,i+1}}{m^2} \right)^\varepsilon 
\e^{\varepsilon \gamma_{\rm E}} \Gamma (1- \varepsilon) \Gamma^2 (\varepsilon)
\, .
\end{align}
This concludes our one-loop consideration.

For future reference, let us provide now the infrared factorization formula for the near-mass-shell limit of the four-leg amplitude.
It reads
\begin{align}
\label{IRfactorizationM4}
M_4 = h_4 \left(\mu^2/\mu_{\rm h}^2, \varepsilon \right) S_4 (\mu^2/\mu_{\rm us}^2, \varepsilon) \prod_{i = 1}^4 J_i (\mu^2/\mu_{\rm c}^2, \varepsilon) 
\, ,
\end{align}
Here, the matching coefficient $h_5$ is determined by the sum of all hard regions in the cyclic sum of \re{1LoopHard} with 
accompanying products of Mandelstam invariants. It can be reduced to the form
\begin{align}
h_4 = 1 + g_\varepsilon^2 c_\varepsilon V^{(1)}_4 + O (g_\varepsilon^4)
\, ,
\end{align}
where $V^{(1)}_4$ is the well-known on-shell one-loop four-leg gluon amplitude \cite{Bern:1991aq,Kunszt:1993sd,Bern:1993kr},
\begin{align}
\label{V4}
V^{(1)}_4 
= 
- \frac{2}{\varepsilon^2} \left( - \frac{\mu^2}{s_{12}} \right)^\varepsilon 
- \frac{2}{\varepsilon^2} \left( - \frac{\mu^2}{s_{23}} \right)^\varepsilon
+
\log^2 \frac{s_{12}}{s_{23}} + 6 \zeta_2
\, .
\end{align}

\section{Collinear factorization}
\label{ColFactSection}

Now that we have a clear understanding of various momentum modes defining the near-mass-shell amplitude, we are in a position
to study its collinear behavior \re{CollLimit} for $n = 5$. Of course, having the explicit form of the amplitude, we can merely evaluate 
it by using its integrated expression. However, the latter were derived under the assumption that all inter-particle momentum 
invariants are large in absolute value, i.e., larger than the off-shellness $m$. On the other hand, as we take the invariant mass of the 
coalescing pair to be small, it can be as small as $m$. This kinematics cannot be achieved from the asymptotic analysis of the previous 
section. So we will go back to the drawing board and re-analyze one-loop integrals anew by considering the problem with three scales
\begin{align}
\label{Col45Limit}
\mbox{col-lim}_{45} = \{ - s_{i,i+1} \gg - s_{45} \gtrsim m \to 0 \ | \ i=5,1,2,3 \}
\, .
\end{align}
It is crucial for the subsequent consideration to implement a stronger condition, i.e.,  $- s_{45} \gg m$, on the rational prefactors 
to enforce the cancellation of denominators induced by momentum integration with Mandelstam invariants accompanying graphs. 
Otherwise, the purity of emerging functions is violated by effects of the form $s_{i,i+1}/[s_{45}/m + s_{i,i+1}/s_{k,k+1}]$. A weaker 
inequality $- s_{45} \gtrsim m$ can be used in all arguments of transcendental functions. This is how Eq.\ \re{Col45Limit} will be
understood below.

To make the discussion more comprehensible, we present a summary of a similar analysis but for the on-shell case in 
Appendix \ref{CollinearLimitAppendix}. Another reason to apply MofR from scratch to the one-loop expressions in the collinear 
limit is to clearly identify the types of loop momentum modes that determine it. In the following, we take
\begin{align}
s_{45} = - r m
\, ,
\end{align}
with real $r$, which can vary in a wide interval, and apply MofR as $m \to 0$. We will use the notation
\begin{align}
\mathcal{J}_i = \mbox{col-lim}_{45} [I_i]
\, ,
\end{align}
for the appearing integrals.

\subsection{Collinear limits of graphs}

We start with $I_1$, for which we identify nine leading regions. One is hard
\begin{align}
\mathcal{J}_1^{\rm h} = \mathcal{J}_{0;1}^{\rm h}
\end{align}
and coincides with \re{I1MasslessS45Lim} since one ignores all soft scales in the integrand. Four of them are collinear
\begin{align}
\mathcal{J}_1^{\rm c1} = \mathcal{J}_1^{\rm c2} = \mathcal{J}_1^{\rm c3} 
= 
-
\frac{c_\varepsilon}{s_{12} s_{23}} \left( \frac{\mu^2}{m} \right)^\varepsilon\frac{2}{\varepsilon^2}
\, , \qquad
\mathcal{J}_1^{\rm c45} 
=
-
\frac{c_\varepsilon}{s_{12} s_{23}} \left( \frac{\mu^2}{r m} \right)^\varepsilon \frac{2}{\varepsilon^2}
\, ,
\end{align}
with the first three corresponding to the loop momentum aligned with the external momenta $p_{1,2,3}$ and the last to the 
near-mass-shell momentum $p_4+p_5$. Finally, there are four ultrasoft regions
\begin{align}
\mathcal{J}_1^{\rm us1}
&
=
-
\frac{2 c_\varepsilon}{s_{12} s_{23}} \left( - \frac{\mu^2 s_{23}}{r m^2} \right)^\varepsilon
\frac{\Gamma (-2 \varepsilon) \Gamma (\varepsilon)}{\Gamma (1-\varepsilon)}
\, , \qquad
\mathcal{J}_1^{\rm us2}
=
-
\frac{2 c_\varepsilon}{s_{12} s_{23}} \left( - \frac{\mu^2 s_{12}}{m^2} \right)^\varepsilon
\frac{\Gamma (-2 \varepsilon) \Gamma (\varepsilon)}{\Gamma (1-\varepsilon)}
\, , \\
\mathcal{J}_1^{\rm us3}
&
=
-
\frac{2 c_\varepsilon}{s_{12} s_{23}} \left( - \frac{\mu^2 s_{23}}{m^2} \right)^\varepsilon
\frac{\Gamma (-2 \varepsilon) \Gamma (\varepsilon)}{\Gamma (1-\varepsilon)}
\, , \qquad
\mathcal{J}_1^{\rm us4}
=
-
\frac{2 c_\varepsilon}{s_{12} s_{23}} \left( - \frac{\mu^2 s_{12}}{r m^2} \right)^\varepsilon
\frac{\Gamma (-2 \varepsilon) \Gamma (\varepsilon)}{\Gamma (1-\varepsilon)}
\, ,
\end{align}
when each of the four propagators in the integrand approaches its mass shell.

It is instructive to find a correspondence between the limit \re{Col45Limit} of the original six regions and those found by 
the direct MofR analysis of above. A short consideration gives
\begin{align}
&
\mbox{col-lim}_{45} [I_1^{\rm h}] = \mathcal{J}_1^{\rm h} + \mathcal{J}_1^{\rm c45}
\, , 
\nonumber\\
&
\mbox{col-lim}_{45} [I_1^{\rm c1}] = \mathcal{J}_1^{\rm c1} + \mathcal{J}_1^{\rm us1}
\, , 
\qquad
&&
\mbox{col-lim}_{45} [I_1^{\rm c2}] = \mathcal{J}_1^{\rm c2} 
\, , \qquad
\mbox{col-lim}_{45} [I_1^{\rm c3}] = \mathcal{J}_1^{\rm c3} + \mathcal{J}_1^{\rm us4}
\, , \nonumber\\
&
\mbox{col-lim}_{45} [I_1^{\rm us2}] = \mathcal{J}_1^{\rm us2} 
\, , 
\qquad
&&
\mbox{col-lim}_{45} [I_1^{\rm us3}] = \mathcal{J}_1^{\rm us3}
\, . 
\end{align}
We observe that as in the massless case \re{I1MasslessS45Lim}, the hard contribution $I_1^{\rm h}$ generates $r \to 0$ collinear 
singularity $\mathcal{J}_1^{\rm c45}$. However, in addition, the two collinear regions $I_1^{\rm c1}$ and $I_1^{\rm c3}$ induce 
divergences in $r$ through the ultrasoft regimes $\mathcal{J}_1^{\rm us1}$ and $\mathcal{J}_1^{\rm us4}$.

Turning to the graph $I_2$ and $I_3$, which possess an overall $1/r$-power singularity, MofR finds two regions in each, one collinear
\begin{align}
\mathcal{J}_2^{\rm c45} 
&
= \mathcal{J}_3^{\rm c45}|_{s_{12} \leftrightarrow s_{23}, s_{34} \leftrightarrow s_{51}}
\\
&
=
\frac{1}{s_{34} s_{45}}
\left[
- \frac{1}{\varepsilon} \log \frac{r s_{34}}{s_{12}}
+
\log \frac{r m}{\mu^2}\log \frac{r s_{34}}{s_{12}} + \frac{1}{2} \log^2 \frac{r s_{34}}{s_{12}} + 2 {\rm Li}_2 \left( 1 - \frac{s_{12}}{r s_{34}}\right)
+ O(\varepsilon)
\right]
\, , \nonumber
\end{align}
when the loop momentum is collinear with $p_4 + p_5$, and one ultrasoft
\begin{align}
\mathcal{J}_2^{\rm us} 
= \mathcal{J}_3^{\rm us}|_{s_{12} \leftrightarrow s_{23}, s_{34} \leftrightarrow s_{51}}
=
\frac{2 c_\varepsilon}{s_{34} s_{45}}
\left[
\left( - \frac{\mu^2 s_{12}}{r m^2} \right)^\varepsilon
-
\left( - \frac{\mu^2 s_{34}}{m^2} \right)^\varepsilon
\right]
\frac{\Gamma (-2 \varepsilon) \Gamma (\varepsilon)}{\Gamma (1-\varepsilon)}
\, , 
\end{align}
when the momentum in the propagator opposite to the propagator between the heavy vertex $(p_1 + p_2)$/($p_2+p_3$) and 
light $p_5$/$p_4$ becomes ultrasoft. The limit of these graphs in generic kinematics is a bit more scrambled. We find
\begin{align}
\mbox{col-lim}_{45} [I_2^{\rm h} + I_2^{\rm c2} + I_2^{\rm c3} + I_2^{\rm us3}] = \mathcal{J}_2^{\rm c45} + \delta
\, , \qquad
\mbox{col-lim}_{45} [I_2^{\rm c1} + I_2^{\rm us2}] = \mathcal{J}_2^{\rm us} 
\, ,
\end{align}
and analogous relations for $I_3$. Here, as we already advocated earlier, the deviation $\delta$ accounts for a more accurate treatment
of the domain when $- s_{45} \sim m$,
\begin{align}
\label{delta}
\delta = 2 \zeta_2 - 2 {\rm Li}_2 \left( 1 - \frac{s_{12}}{r s_{34}} \right)
\, .
\end{align}
It vanishes in the limit $r \to \infty$ enforced in the previous section where all interparticle invariants were larger than the off-shellness $m$.

Finally, the last two graphs $I_4$ and $I_5$ do not possess additional collinear singularities as $s_{45} \to 0$, so they
are given by the original MofR expression,
\begin{align}
\mbox{col-lim}_{45} [I_{4,5}^{\rm r}] = I_{4,5}^{\rm r} = \mathcal{J}_{4,5}^{\rm r} 
\, ,
\end{align}
where $r$ runs over the six region vectors \re{IRregions}. Their explicit expressions can be found from the ones given 
in Sect.\ \ref{IRoneloop} by a cyclic permutation of indices (or by a set of replacements)
\begin{align}
\mathcal{J}_{4}^{\rm r} = I_{1}^{\rm r}|_{s_{45} \to s_{34}, s_{12} \to s_{51}, s_{23} \to s_{12}}
\, , \qquad
\mathcal{J}_{5}^{\rm r} = I_{1}^{\rm r}|_{s_{45} \to s_{51}, s_{12} \to s_{23}, s_{23} \to s_{34}}
\, ,
\end{align}
followed by the substitution \re{CollLimit}.

\subsection{Extracting splitting amplitude}

By having the required ingredients in our hands, we can now extract the one-loop splitting amplitude and observe how Eq.\ \re{SplitDef}
gets modified in the off-shell regime. In anticipation of what is coming at us, let us notice that collinear regions for off-shell kinematics,
as opposed to the on-shell cas, contain both $m \to 0$ and $s_{45} \to 0$ singularities of collinear origin. As we observed in Appendix 
\ref{CollinearLimitAppendix}, the use of MofR in the on-shell case eliminates all jet factors from the consideration since they are determined 
by scaleless integrals and, therefore, cannot contaminate the collinear limit of particle coalescence \re{CollLimit}. Only $s_{45} \to 0$
divergencies need to be selected in the off-shell case to properly define the splitting amplitude.

To start with, we confirm that by taking the sum of the hard regions, we reproduce the on-shell situation, i.e., their sum is determined
by the four-leg amplitude $V^{(1)}_4$,
\begin{align}
s_{12} s_{23} \mathcal{J}_1^{\rm h} + s_{51} s_{12} \mathcal{J}_4^{\rm h} + s_{23} s_{34} \mathcal{J}_5^{\rm h} = - 2 c_\varepsilon V^{(1)}_4
\, .
\end{align}
In the sum of all collinear regions when the loop momentum is parallel to $p_1$, $p_2$ and $p_3$ external momenta 
all nontrivial dependence on Mandelstam invariants cancel, and they add up to three times the one-loop jet functions \re{1LoopJet}, 
as expected,
\begin{align}
\label{P123jets}
s_{12} s_{23} [ \mathcal{J}_1^{\rm c1} + \mathcal{J}_1^{\rm c2} + \mathcal{J}_1^{\rm c3} ]
+ 
s_{51} s_{12} [ \mathcal{J}_4^{\rm c1} + \mathcal{J}_4^{\rm c2} ]
+ 
s_{23} s_{34} [ \mathcal{J}_5^{\rm c2} + \mathcal{J}_5^{\rm c3} ]
=
- 6 J^{(1)}
\, .
\end{align}
The remaining collinear regions indeed contain the sought-after $s_{45} \to 0$ collinear amplitude. However, they also have 
residual $m \to 0$ collinear singularities from the two jet factors stemming from legs 4 and 5. While one of these will be 
absorbed as a fourth factor (in addition to the other three from Eq.\ \re{P123jets}) into the four-leg amplitude $M_4$ 
\re{IRfactorizationM4}, there will be one left. The latter is not a part of the splitting amplitude as it is not related to the $s_{45} 
\to 0$ behavior of the scattering amplitude! So, there will be an additional factor in the collinear factorization formula due to 
the mismatch in the number of jet functions in $n$ and $(n-1)$ point amplitudes. We find 
\begin{align}
s_{12} s_{23} \mathcal{J}_1^{\rm c45}
+
s_{34} s_{45} \mathcal{J}_2^{\rm c45}
+
s_{45} s_{51} \mathcal{J}_3^{\rm c45}
+ 
s_{51} s_{12} \mathcal{J}_4^{\rm c5}
+ 
s_{23} s_{34} \mathcal{J}_5^{\rm c4}
=
- 4 J^{(1)} - 2 R^{(1)}_{s_{45}}
\, ,
\end{align}
where the $R^{(1)}_{s_{45}}$ is the one-loop splitting amplitude given as a Laurent expansion
\begin{align} 
\label{1loopSplitAmplObscure}
R^{(1)}_{s_{45}} (z)
=
-
\frac{2}{\varepsilon^2} + \frac{1}{\varepsilon} \log \frac{z \bar{z} m^2}{\mu^4} 
&
- 
\log \left( - \frac{s_{45} z \bar{z}}{\mu^2} \right) \log\left(- \frac{s_{45}}{\mu^2} \right)
+
2 \log \frac{m}{\mu^2} \log \left( - \frac{s_{45}}{m} \right)
\nonumber\\
&
-
\frac{1}{2} \log^2 \frac{z}{\bar{z}} - {\rm Li}_2 \left( 1 + \frac{m}{z s_{45}} \right) - {\rm Li}_2 \left( 1 + \frac{m}{\bar{z} s_{45}} \right)
\, .
\end{align}
For brevity, we do not display the mass scales $m$ and $\mu^2$ in the argument of this function. This expression has an unorthodox 
form, especially compared to the mass-shell case \re{Massless1loopSplittingAmplitude}, and displays no clear structure. It is the 
main goal of the next section to uncover it.

Before we move to this question, there are contributions that we did not address, i.e., the ultrasoft regions. Their sum yields
\begin{align}
s_{12} s_{23} & [ \mathcal{J}_1^{\rm us1} 
+ \mathcal{J}_1^{\rm us2} + \mathcal{J}_1^{\rm us3} + \mathcal{J}_1^{\rm us4} ]
+
s_{34} s_{45} \mathcal{J}_2^{\rm us}
+
s_{45} s_{51} \mathcal{J}_3^{\rm us}
\\
&
+ 
s_{51} s_{12} [ \mathcal{J}_4^{\rm us2} + \mathcal{J}_4^{\rm us3} ]
+ 
s_{23} s_{34} [ \mathcal{J}_5^{\rm us2} + \mathcal{J}_5^{\rm us3} ]
=
- 2 [ w_{12}^{(1)} + w_{23}^{(1)} + w_{34}^{(1)} + w_{41}^{(1)} ] - 2 R^{(1)}_{\rm us}
\, , \nonumber
\end{align}
where on the right-hand side, we get the expected four contributions to the ultrasoft function of the four-point amplitude
but there is also a remainder $ R^{(1)}_{\rm us}$. Notice, however, that $r \to 0$ singularities completely cancel between the graphs 
$I_1$, $I_2$ and $I_3$. Namely, in the sum $\mathcal{J}_2^{\rm us}$ and $\mathcal{J}_1^{\rm us4}$, and $\mathcal{J}_3^{\rm us}$ 
and $\mathcal{J}_1^{\rm us1}$, the $r$-dependence disappears, such that the leftover has a purely ultrasoft origin with corresponding 
$\mu^2_{\rm us}$ scales \re{LeadingRegions}, as can be seen from the explicit form
\begin{align}
R^{(1)}_{\rm us} (z) 
&
= w_{34}^{(1)} - w_{12}^{(1)} + w_{51}^{(1)} - w_{23}^{(1)} 
\nonumber\\
&
=
-
\frac{1}{\varepsilon} \log (z \bar{z}) 
+ 
\log z \log \left( - \frac{m^2}{\mu^2 s_{12} \sqrt{z}} \right) + \log \bar{z} \log \left( - \frac{m^2}{\mu^2 s_{23} \sqrt{\bar{z}}} \right)
+ 
O (\varepsilon)
\, .
\end{align}
Notice that for the four-leg kinematics, we re-wrote $w_{34'}^{(1)} = w_{12}^{(1)}$ and $w_{4'1}^{(1)} = w_{23}^{(1)}$,
where prime on the symbol stands for the merged leg $4$ of $M_4$ from the two legs $4$ and $5$ of $M_5$.

\subsection{Factorization formula}

Summarizing our one-loop analysis, we can propose a general collinear factorization formula when two legs, $n-1$ and $n$, 
coalesce. Namely,
\begin{align}
\label{CollFactOffShell}
M_n \xrightarrow{\scriptscriptstyle (n-1)||n} R_{s_{n-1,n}} (z) R_{\rm us} (z) J M_{n-1}
\, .
\end{align}
Here only the function $R_{s_{n-1,n}} (z)$ is responsible for the singular collinear behavior when $s_{n-1,n} \to 0$. At
one loop, it is
\begin{align}
R_{s_{n-1,n}} (z) = 1 + g_\varepsilon^2 R^{(1)}_{s_{n-1,n}} (z) + O (g_\varepsilon^4)
\, .
\end{align}
The jet factor $J$ accounts for the mismatch of the $m \to 0$ collinear behavior between the left and right-hand sides.
Finally, $R_{\rm us} (z)$ is the ratio of the ultrasoft wedge factors between the initial and final amplitudes in channels 
adjacent to the collinear pair,
\begin{align}
R_{\rm us} (z) = \frac{w_{n-2,n-1} w_{n,1}}{w_{n-2,(n-1)'} w_{(n-1)',1}}
\, ,
\end{align}
where we used the notation $(n-1)'$ for the merged leg from the initial $(n-1)$-st and $n$-th, i.e., $(n-1)+n\to (n-1)'$.

\section{Splitting superamplitudes}
\label{6DSplitSection}

Among the ingredients in the right-hand side of the collinear factorization formula, the one-loop expression for the main quantity, 
the collinear amplitude $R_{s_{n-1,n}} (z)$, is the most obscure. Presently, we will clarify its origin. To do this, we will construct it 
from the six-dimensional three-gluon vertex $V_{123}$ and its uplift to superspace.

\subsection{Gluon splitting amplitude}

$V_{123}$ admits the same form in any number of dimensions, so we take its four-dimensional counterpart \re{4Dvertex} 
in Gervais-Neveu gauge \cite{Gervais:1972tr}, which is very handy for dealing with color-ordered Feynman rules 
\cite{Srednicki:2007qs}, and rewrite all inner products in terms of the (traces of the) six-dimensional Dirac matrices
\begin{align}
\label{6DvertexIni}
V_{123} 
= 
\tr_4[ \bar{\mathcal{E}}_1 \mathcal{E}_2] \tr_4[\bar{\mathcal{E}}_3 P_1]
+
\tr_4[ \bar{\mathcal{E}}_2 \mathcal{E}_3] \tr_4[\bar{\mathcal{E}}_1 P_2]
+
\tr_4[ \bar{\mathcal{E}}_3 \mathcal{E}_1] \tr_4[\bar{\mathcal{E}}_2 P_3]
\, .
\end{align}
Here as before, all momenta are incoming and obey the conservation law $P_1 + P_2 + P_3 = 0$. The polarization vectors carry 
implicit little group indices
\begin{align}
{\mathcal{E}}_j \equiv {\mathcal{E}}_{a\dot{a}} (P_j) 
= 
\frac{|j_a \rangle [ j_{\dot{a}}| \mathcal{Q} +  \mathcal{Q} |j_{\dot{a}} ] \langle j_a]}{\sqrt{2} (P_j \cdot \mathcal{Q})}
\, , \qquad
\bar{\mathcal{E}}_j \equiv \bar{\mathcal{E}}_{a\dot{a}} (P_j) 
= 
\frac{|j_{\dot{a}}] \langle j_a| \bar{\mathcal{Q}} + \bar{\mathcal{Q}} |j_a \rangle [j_{\dot{a}}|}{\sqrt{2} (P_j \cdot \mathcal{Q})}
\, ,
\end{align}
and depend on a light-like six-vector $\mathcal{Q}$, $\tr_4[\mathcal{Q} \bar{\mathcal{Q}}] = 0$. They are normalized as
\begin{align}
\ft14 \tr_4 [ {\mathcal{E}}_{a\dot{a}} (P_j)  \bar{\mathcal{E}}_{b\dot{b}} (P_j) ] = - \varepsilon_{ab} \varepsilon_{\dot{a}\dot{b}}
\, .
\end{align} 
Using these expressions, the `decay' vertex of the gluon with momentum $P_3$ and polarization $\mathcal{E}_3$ into two 
on-shell six-dimensional gluons with labels 1 and 2 then reads
\begin{align}
\label{6Dvertex}
V_{123} = \frac{1}{(P_1 \cdot \mathcal{Q}) (P_2 \cdot \mathcal{Q})}
\big[
\langle 1_a| \bar{\mathcal{Q}} | 2_b \rangle [2_{\dot{b}}| \mathcal{Q} |1_{\dot{a}}] \tr_4 [\bar{\mathcal{E}}_3 P_1]
&
+
\langle 2_b | \bar{\mathcal{E}}_3 \mathcal{Q} | 2_{\dot{b}}] [1_{\dot{a}} | \mathcal{Q} \bar{P}_2 | 1_a\rangle 
\\
&
-
\langle 1_a| \bar{\mathcal{E}}_3 \mathcal{Q} | 1_{\dot{a}}] [2_{\dot{b}} | \mathcal{Q} \bar{P}_1 | 2_b\rangle 
\big]
\, . \nonumber
\end{align}
We then introduce the tree-level splitting amplitude by analogy with its four-dimensional counterpart \re{4DTreeSplit}, discussed at 
length in Appendix \ref{4DSplittingAmplitudeAppendix}, to take the form
\begin{align}
\label{6DBosonicSplittingDef}
{\rm Split}^{(0)} (P_3; P_1, P_2) \equiv \frac{2  V_{123}}{S_{12}} 
\, ,
\end{align}
with coalescing momenta
\begin{align}
\label{6DCollinearity}
P_1 \to - z P_3
\, , \qquad
P_2 \to - \bar{z} P_3 
\, .
\end{align}
The calculation of the one-loop integrand starting from this expression is presented in Appendix \ref{6DSplitComponentsAppendix}.
Below, we provide, however, a more compact and more general derivation based on the $\mathcal{N} = (1,1)$ superspace formulation 
that we reviewed in Sect.\ \ref{N11superspaceSection}.

\subsection{Splitting amplitude in $\mathcal{N} = (1,1)$ superspace}

\begin{figure}[t]
\begin{center}
\mbox{
\begin{picture}(0,100)(160,0)
\put(0,0){\insertfig{10}{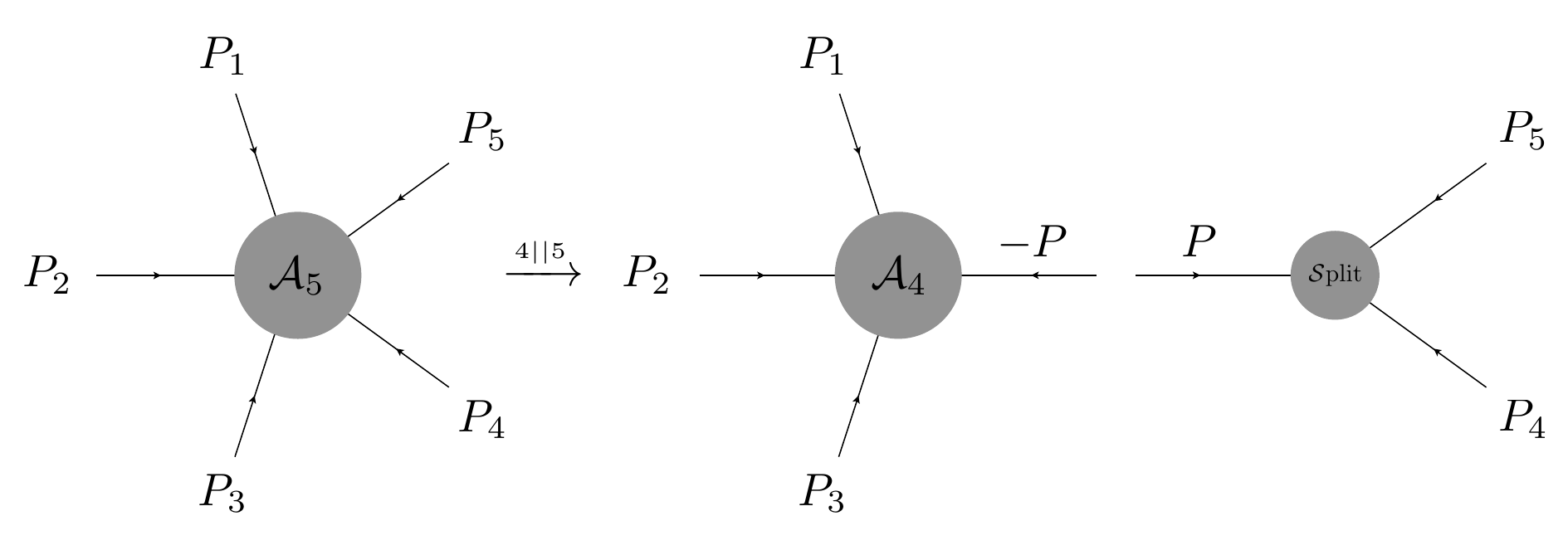}}
\end{picture}
}
\end{center}
\caption{\label{ColFactPicPic} Collinear factorization of the five-leg superamplitude.}
\end{figure}

To determine the $\mathcal{N} = (1,1)$ superspace form of the tree splitting amplitude, we will take advantage of the simplicity of the 
five-leg superamplitude \re{5legTree6D} and extract the latter from there. Since it uses the spinor-helicity variables, we will rewrite
the collinearity condition \re{6DCollinearity} in terms of the corresponding Weyl spinors as
\begin{align}
\bra{1^a} \to i z \bra{3^a}
\, , \qquad
\ket{1^a} \to i z \ket{3^a}
\, , \qquad
[2^{\dot{a}}| \to i \bar{z} [3^{\dot{a}}|
\, , \qquad
|2^{\dot{a}}] \to i \bar{z} |3^{\dot{a}}]
\, ,
\end{align}
with conventional phase assignments. We will not, however, have any collinearity requirements on the corresponding fermionic variables 
$\eta_1$ and $\eta_2$ as we need these to distinguish coalescing states in the splitting superamplitude. The latter is of Grassmann degree 
two in these variables.

We will impose collinearity on legs 4 and 5 of the five-leg super amplitude \re{5legTree6D}, $P_4 + P_5 = - P$ with $P^2 \to 0$.
There is an overall pole in the Mandelstam invariant $S_{45}$, which is absorbed into the definition of the splitting superamplitude 
as in its bosonic counterpart \re{6DBosonicSplittingDef}. The calculation is quite tedious and benefits from a precise parametrization
of the approach to the collinear limit by making use of the light-cone decomposition of the coalescing momenta $P_4$ and $P_5$. 
It is necessary to resolve the ambiguities in terms where the two matrix momenta are adjacent and naively vanish in the limit in
question. This analysis is relegated to a forthcoming publication, where the multi-collinear limits are discussed as well. Here, we merely 
content ourselves with a demonstration of just two terms in the amplitude, which disclose themselves in the limit in question and are 
free from ambiguities. These are the fourth and fifth cyclic permutations in the first line of Eq.\ \re{5legTree6D}. The former gives
\begin{align}
\label{RedStep1}
Q_4 \bar{P}_5 P_1 \bar{P}_2 P_3 \bar{Q}_4 
=
- S_{23} Q_4 \bar{P}_5 P_3 \bar{Q}_4 + O(S_{45})
\, ,
\end{align}
where we used momentum conservation to simplify the string of the Dirac matrices to $\bar{P}_5 P_1 \bar{P}_2 P_3 = - S_{23} \bar{P}_5 P_3$
and ignored $O(S_{45})$ term in the right-hand side in the used kinematical relation $S_{12} + S_{23} + S_{31} = S_{45}$. Further, in the
$4||5$-collinear limit, $P_3$ in Eq.\ \re{RedStep1} is sandwiched between the Weyl spinors $|P]$ such that one can rely on the identity
\begin{align}
[P^{\dot{a}} | P_3 | P^{\dot{b}}] = S_{3P} \varepsilon^{\dot{a} \dot{b}}
\, ,
\end{align}
valid due to the antisymmetry of six-dimensional Dirac matrices, to conclude that
\begin{align}
\label{RedStep2} 
Q_4 \bar{P}_5 P_1 \bar{P}_2 P_3 \bar{Q}_4 
\xrightarrow{\scriptscriptstyle 4||5}
-
\sqrt{z \bar{z}} S_{23} S_{3 P} \langle 4_a | 5_{\dot{a}}] \eta_4^a \bar\eta_4^{\dot{a}}
\, .
\end{align}
Notice that as in four dimensions \re{4DTreeSplit}, the collinear singularity of the amplitude gets softer: it is a `square root' 
$\langle P_a | P_{\dot{a}}]/P^2$ rather than a pole $1/P^2$. Identical steps can be performed for the last, i.e., fifth, cyclic 
permutation as well, yielding a similar result with minor modifications. Namely,
\begin{align}
Q_5 \bar{P}_1 P_2 \bar{P}_3 P_4 \bar{Q}_5
\xrightarrow{\scriptscriptstyle 4||5}
\sqrt{z \bar{z}} S_{12} S_{1 P} \langle 5_a | 4_{\dot{a}}] \eta_5^a \bar\eta_5^{\dot{a}}
\, ,
\end{align}
where the sign difference is due to the reversed position of the momentum matrix $P_4$ in the Dirac string as compared to
$P_5$ in Eq.\ \re{RedStep2}. Adding further contribution from the rest of the terms in the amplitude, we obtain the collinear 
factorization
\begin{align}
\widehat{\mathcal{A}}_5^{(0)} (1,2,3,4,5)
\xrightarrow{\scriptscriptstyle 4||5}
\widehat{\mathcal{A}}_4^{(0)} (1,2,3,- P)\mbox{$\mathcal{S}$plit}^{(0)} (P; P_4, P_5)
\, ,
\end{align}
shown in Fig.\ \ref{ColFactPicPic}, with the color-ordered tree splitting superamplitude
\begin{align}
\label{TreeSuperSplittingAmplitude}
\mbox{$\mathcal{S}$plit}^{(0)} (P; P_4, P_5)
=
\frac{1}{\sqrt{z \bar{z}} S_{45}}
\Big[
&
\langle 5_a | 4_{\dot{a}}]
\eta_5^a \bar\eta_5^{\dot{a}}
-
\langle 4_a | 5_{\dot{a}}]
\eta_4^a \bar\eta_4^{\dot{a}}
\\
-
&
\langle 4_a | 5_{\dot{a}}]
(\sqrt{z} \eta_4^a + \sqrt{\bar{z}} \eta_5^a) (\sqrt{z} \bar\eta_4^{\dot{a}} + \sqrt{\bar{z}} \bar\eta_5^{\dot{a}})
\Big]
\, . \nonumber
\end{align}
This expression possesses the correct four-dimensional limit \re{4DTreeSplit}, as can be easily concluded by taking $a=\dot{a} = 1$ 
values for the little group indices corresponding to the positive helicity gluons $4$ and $5$, together with $\eta_4^{1} = i \sqrt{z} \eta_P$,
$\eta_5^{1} = i \sqrt{\bar{z}} \eta_P$ (and similar relations for the anti-chiral variables) and the explicit form of six-dimensional
Weyl spinors from Eq.\ \re{LambdaFromlambdas}.

To get a full-color\footnote{We introduce it in order to prepare the ground for a two-loop calculation, where the color tensor of the vertex 
is of paramount importance for selecting only planar graphs in the unitarity-based approach \cite{Bern:2004cz}. For instance, for 
a two-leg form factor with the Kronecker delta as the SU$(N_c)$ tensor, non-planar graphs contribute on equal footing with the 
planar ones \cite{Brandhuber:2012vm}.} splitting super-vertex, dubbed the splitting matrix in the nomenclature of Ref.\ \cite{Catani:2011st}, 
we dress the above Eq.\ \re{TreeSuperSplittingAmplitude} with the SU$(N_c)$ structure constant (and changing the labels along the way 
for follow-up uses)
\begin{align}
\mbox{$\mathcal{S}$plit}^{(0); i_3 i_1 i_2} (P_3; P_1, P_2)
=
f^{i_3 i_1 i_2}
\mbox{$\mathcal{S}$plit}^{(0)} (P_3; P_1, P_2)
\, .
\end{align}

\subsection{One-loop integrand}

\begin{figure}[t]
\begin{center}
\mbox{
\begin{picture}(0,65)(90,0)
\put(0,0){\insertfig{5}{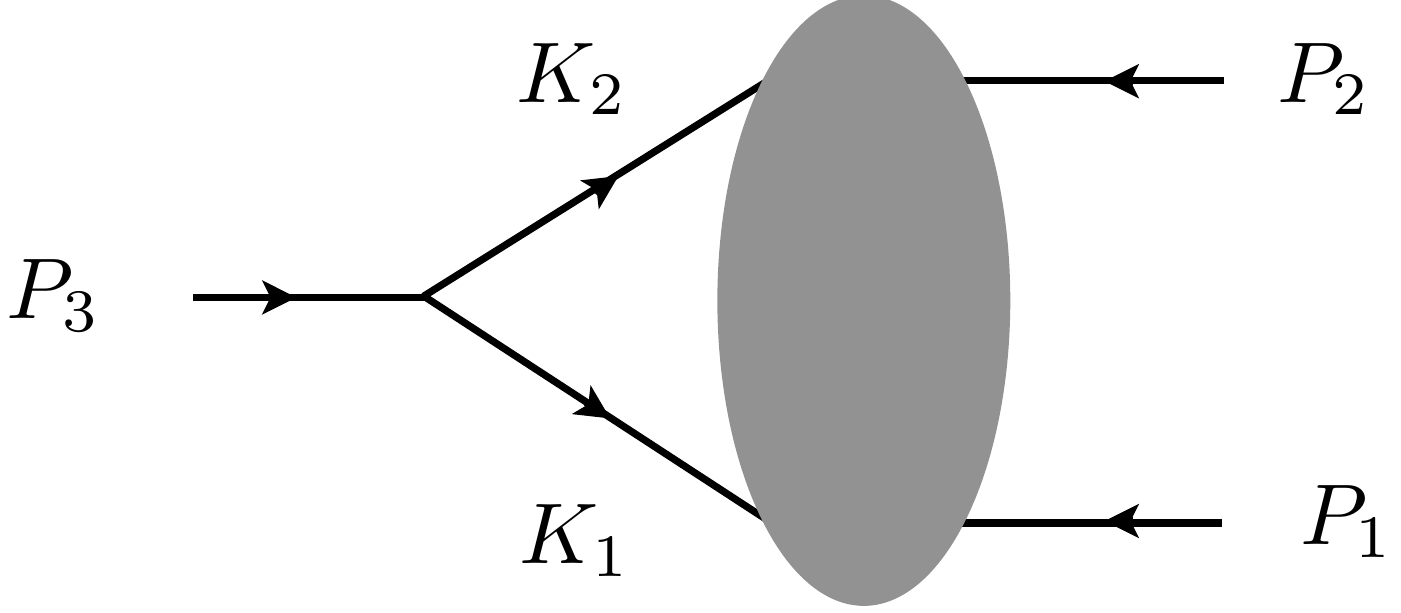}}
\end{picture}
}
\end{center}
\caption{\label{1LoopSplitPic} A two-particle unitarity cut defining the integrand of the one-loop splitting amplitude.}
\end{figure}

Having determined the form of the tree-level splitting superamplitude, we can now easily generate its one-loop integrand.
An equivalent but more practical representation of the tree amplitude for this purpose is
\begin{align}
\label{PracticalTreeSplitting}
\mbox{$\mathcal{S}$plit}^{(0); i_3 i_1 i_2} (P_3; P_1, P_2)
&
=
\frac{f^{i_3 i_1 i_2}}{2 S_{12}} \frac{(P_3 \cdot \mathcal{Q})}{(P_1 \cdot \mathcal{Q}) (P_2 \cdot \mathcal{Q})}
\\
&
\times
\left[
Q_1 \bar{\mathcal{Q}} {P}_2 \bar{Q}_1 + Q_2 \bar{P}_1 \mathcal{Q} \bar{Q}_2
- \frac{( Q_1 + Q_2 ) \bar{\mathcal{Q}} {P}_2 \bar{P}_1 \mathcal{Q} (\bar{Q}_1 + \bar{Q}_1)}{2 (P_3 \cdot \mathcal{Q})}
\right]
\, . \nonumber
\end{align}
We proceed as in Sect.\ \ref{6D5LegSection} and construct the latter from its two-particle unitarity cut shown in Fig.\ \ref{1LoopSplitPic}.
What we need is a full-color four-leg superamplitude. It can be written in the most compact form by using an adjoin basis 
of SU$(N_c)$ generators \cite{DelDuca:1999rs,DelDuca:1999iql}. It contains just two terms and reads
\begin{align}
\mathcal{A}^{i_1i_2 i_3 i_4}_4 (1,2,3,4)
=
i  f^{i_1 i_4 i_5} i  f^{i_5 i_2 i_3} \mathcal{A}_4 (1,2,3,4) 
+
i  f^{i_2 i_4 i_5} i f^{i_5  i_1 i_3} \mathcal{A}_4 (2,1,3,4) 
\, .
\end{align}
Here $\mathcal{A}_4$ is the color-ordered superamplitude from Eq.\ \re{6DColorOrderedAmp} and we explicitly displayed its 
arguments' order. The $S_{12}$-cut is
\begin{align}
\mbox{$\mathcal{S}$plit}^{(1);  i_3 i_1 i_2} |_{S_{12}-{\rm cut}}
=\!
\int \!\prod_{i = 1}^2
d^2 \eta_{K_i} d^2 \bar\eta_{K_i} 
\mbox{$\mathcal{S}$plit}^{(0); i_3 j_1 j_2} (P_3; - K_1, - K_2) \mathcal{A}^{(0); j_1 j_2 i_1 i_2}_4 (K_2, K_1, P_1, P_2)
,
\end{align}
and we show only the bosonic components of the legs' supermomenta. Contracting the color structures, we find
\begin{align}
&
\mbox{$\mathcal{S}$plit}^{(0); i_3 j_1 j_2} (P_3; - K_1, - K_2) \mathcal{A}^{(0); j_1 j_2 i_1 i_2}_4 (K_2, K_1, P_1, P_2)
\nonumber\\
&\qquad
=
\frac{N_c}{2}\, \mbox{$\mathcal{S}$plit}^{(0); i_3 i_1 i_2} (P_3; - K_1, - K_2) 
\left[ \mathcal{A}_4 (K_2, K_1, P_1, P_2) - \mathcal{A}_4 (K_1, K_2, P_1, P_2) \right]
\, .
\end{align}
Using the antisymmetry of the kinematical factor in $\mbox{$\mathcal{S}$plit}^{(0); i_3 i_1 i_2} (P_3; K_1, K_2)$, we
can simply double the first contribution in the square brackets and perform the Grassmann integrations with the help of 
the identity \re{GrassmannDeltaID6D} to obtain
\begin{align}
\mbox{$\mathcal{S}$plit}^{(1); i_3 i_1 i_2} |_{S_{12}-{\rm cut}}
&=
N_c\,
\mbox{$\mathcal{S}$plit}^{(0); i_3 i_1 i_2} (P_3; P_1, P_2) 
\frac{z \bar{z} (P_3 \cdot \mathcal{Q})^2}{(K_1 \cdot \mathcal{Q})(K_2 \cdot \mathcal{Q})}
\frac{\mathcal{N}}{S_{K_1 P_1} S_{K_1 K_2}}
\, .
\end{align}
The numerator $\mathcal{N}$ is determined by a ratio of traces of lengthy strings of six-dimensional Dirac matrices\footnote{To be 
completely accurate, each term in Eq.\ \re{PracticalTreeSplitting} comes accompanied by its own trace since the Dirac structures of
these contributions are different. If we name them $\mathcal{N}_i$ with $i=1,2,3$ for each term in the square brackets in 
\re{PracticalTreeSplitting}, then $\mathcal{N}_1 \equiv \mathcal{N}$ given in the text and $\mathcal{N}_2$ obtained from it by 
exchange of indices such that $\mathcal{N}_2 = \mathcal{N}_1$ in the end. $\mathcal{N}_3$, on the other hand, is determined by the ratio of 
different traces $\mathcal{N}_3 = S_{12}^2 \tr_4 [\bar{\mathcal{Q}} {K}_2 \bar{K}_1 \mathcal{Q} \bar{P}_1 {P}_2]/\tr_4[\bar{\mathcal{Q}} 
{P}_2 \bar{P}_1 \mathcal{Q} \bar{P}_1 {P}_2]$ and it is reassuring that it evaluates to the very same expression in the right-hand side 
of Eq.\ \re{OneLoopNumerator}, i.e., $\mathcal{N}_3 = \mathcal{N}$.}. They are evaluated with the help of {\tt FeynCalc} to yield the 
numerator
\begin{align}
\label{OneLoopNumerator}
\mathcal{N}
&
=
\frac{
\tr_4
[\bar{\mathcal{Q}} {P}_2 \bar{P}_1 K_2 \bar{K}_1 \bar{\mathcal{Q}} K_2 \bar{K}_1 {K}_2 \bar{P}_1]
+
\tr_4
[\bar{\mathcal{Q}} {P}_2 \bar{P}_1 K_1 \bar{K}_2 \bar{K}_1 {\mathcal{Q}}  \bar{K}_2 {K}_1 \bar{P}_1]
}
{\tr_4
[\bar{\mathcal{Q}} {P}_2 \bar{P}_1 \mathcal{Q} \bar{P}_1 P_2]}
\\
&
= 
\frac{S_{12}}{2 (P_3 \cdot \mathcal{Q}) z \bar{z}}
\left[
(P_3 \cdot \mathcal{Q}) (K_2 + P_1)^2 + (P_3 \cdot \mathcal{Q}) S_{12} - (K_2 \cdot \mathcal{Q}) S_{12} z - (K_1 \cdot Q) S_{12} \bar{z}
\right]
\, . \nonumber
\end{align}
We find that it is identical to the gluon component \re{SumGluonComponentSplit6D} calculated in Appendix 
\ref{6DSplitComponentsAppendix} and is just a dimensional uplift of the four-dimensional result \re{4D1loopSplitIntegrand}!

Factoring out the six-dimensional splitting tree superamplitude, 
\begin{align}
\mbox{$\mathcal{S}$plit} = \mbox{$\mathcal{S}$plit}^{(0)} \mathcal{R}
\end{align}
the ratio $ \mathcal{R}$ is determined by the perturbative expansion in the six-dimensional gauge coupling
\begin{align}
\label{6DRatio}
\mathcal{R}
=
1 -
\frac{1}{2}
g^2_{6, \rm\scriptscriptstyle YM} N_c 
[J_{1} + z S_{12} J_{2} + (\bar z - z) S_{12} J_{3} + ({z \leftrightarrow \bar{z}})]
+
O (g^4_{6, \rm\scriptscriptstyle YM})
\, ,
\end{align}
where
\begin{align}
J_{1} 
&= \int \frac{d^6 K_2}{i (2 \pi)^6} \frac{(P_3 \cdot \mathcal{Q})}{K_1^2 K_2^2 (K_2 \cdot \mathcal{Q})}
\, ,\\
J_{2}
&= \int \frac{d^6 K_2}{i (2 \pi)^6} \frac{(P_3 \cdot \mathcal{Q})}{K_1^2 K_2^2 (K_2 + P_1)^2(K_2 \cdot \mathcal{Q})}
\, ,\\
J_{3} 
&= \int \frac{d^6 K_2}{i (2 \pi)^6} \frac{1}{K_1^2 K_2^2 (K_2 + P_1)^2}
\, .
\end{align}
Here one of the $K_i$ momenta is eliminated with the momentum conservation condition $K_1 + K_2 + P_1 + P_2 = 0$.
$J_3$ does not contribute due to the antisymmetry of the $z$-dependent coefficient. This is a straightforward uplift of the 
known four-dimensional result \re{MasslessIntSplitting}.

\subsection{On the choice of auxiliary vector}
\label{QChoiceSection}

Since at the end of the day, we are interested only in the dimensionally reduced form of the six-dimensional splitting 
amplitude and we know how we interpret the extra-dimensional components of the momentum vectors $P_i$ according 
to \re{6DtoOffShell}, but what about the auxiliary null vector $\mathcal{Q}$?

A typical momentum integral defining the splitting amplitude at one loop is
\begin{align}
\mathcal{S}
=
\int \frac{d^6 K}{(2 \pi)^6} \frac{1}{[K^2]^{a_1} [(K+P_1)^2]^{a_2} [(K-P_2)^2]^{a_3}} 
\frac{1}{[(K+P_1) \cdot \mathcal{Q}] [(K-P_2) \cdot \mathcal{Q}]}
\, .
\end{align}
As we perform the dimensional reduction on the dual form of this integral according to the rules of Sect.\ \ref{DimRedSection}, 
we get
\begin{align}
(8 \pi^2/R_0)^2 \mbox{dim.red.}[\mathcal{S}]
=
\int \frac{d^4 x_0}{i \pi^2} 
&
\frac{1}{[(x_0 + x_2)^2]^{a_1} [(x_0 + x_1)^2]^{a_2} [(x_0 + x_3)^2]^{a_3}} 
\nonumber\\
&\times
\frac{1}{[x_0 \cdot q + X_1 \cdot \mathcal{Q} ] [x_0 \cdot q + X_3 \cdot \mathcal{Q}]}
\, ,
\end{align}
where the integration variable $x_0$ projects out only the four-dimensional components $q$ of $\mathcal{Q} = (q, y_q)$. On the
other hand, in the six-dimensional products $X_i \cdot \mathcal{Q} = q \cdot x_i + y_q \cdot y_i$, the extra-dimensional product
$(y_q \cdot y_i)$ is not constrained. The only condition on the six-vector $\mathcal{Q}$ is that it is null $\mathcal{Q} \cdot \mathcal{Q} = 0$.
However, we would like to preserve the ability of its four-dimensional subspace to project the four-dimensional
off-shell momenta $p_1$ and $p_2$ on a common near-light-like direction $p_1 + p_2$, as $s_{12} \to 0$. This 
suggests that we can select $q$ to be light-like, i.e., $q^2 = 0$. Then by default $y_q \cdot y_q = 0$. In order to
treat $x_1$ and $x_3$ region momenta symmetrically, without loss of generality we may choose $y_q =\alpha y_2$
with some unfixed parameter $\alpha$. As a consequence the two `axial' denominators $[x_0 \cdot q + X_i \cdot \mathcal{Q}]
= [(x_0+x_i)\cdot q - \ft12 \alpha m]$ get regularized in four dimensions by the off-shellness $m$. However, there does not seem
to be a place for such denominators in a quantized four-dimensional gauge theory since a gauge-fixing term cannot produce
them.  If these would be possible, they completely wipe out collinear singularities of amplitudes. We are forced to conclude that 
$\alpha = 0$ and, thus, the six-dimensional vector $\mathcal{Q}$ can only reside within the four-dimensional subspace, i.e., our 
Minkowski space-time.

\subsection{Factorized splitting amplitude at one loop}

Dimensionally reducing the six-dimensional representation \re{6DRatio}, and ignoring, as before, negligible effects of $m$ in numerators, 
we obtain the one-loop expression for the off-shell splitting ratio
\begin{align}
\mathcal{R}
=
1 -
\ft{1}{2} g_{\varepsilon}^2 [J_{1} + z s_{12} J_{2} + ({z \leftrightarrow \bar{z}})]
+
O (g_{\varepsilon}^4, m^2)
\, .
\end{align}
We dropped the contribution from $J_3$ since its coefficient vanishes. Here, all integrals are identical to the massless ones
\re{MasslessIntSplitting} except that the external legs are off their mass shell. They can now be evaluated with MofR as
in our earlier consideration of the five-leg amplitude.

The integral $J_1$ coincides with its massless counterpart and is, thus, given by the hard region of the loop momentum alone
\begin{align}
J_1 = J_1^{\rm h} = - \frac{c_\varepsilon}{\varepsilon^2} \left( - \frac{\mu^2}{s_{12}} \right)^\varepsilon
\, .
\end{align}
On the other hand, $J_2$ possesses all three regions \re{LeadingRegions} and receives leading contributions from these,
\begin{align}
J_2^{\rm h}
&
=
\frac{1}{z s_{12}}
\left[
\frac{2 c_\varepsilon}{\varepsilon^2} \left( - \frac{\mu^2}{z s_{12}} \right)^\varepsilon + 2 {\rm Li}_2 (\bar{z}) + O (\varepsilon)
\right]
\, , \\
J_2^{\rm c1}
&
=
-
\frac{1}{z s_{12}} \frac{2 c_\varepsilon}{\varepsilon^2} \left( \frac{\mu^2}{m} \right)^\varepsilon
\, , \\
J_2^{\rm c2}
&
=
-
\frac{1}{z s_{12}} \left[ \frac{c_\varepsilon}{\varepsilon^2} \left( \frac{\mu^2}{z m} \right)^\varepsilon + 2 {\rm Li}_2 (\bar{z}) + O (\varepsilon) \right]
\, , \\
J_2^{\rm us}
&
=
-
\frac{2 c_\varepsilon}{z s_{12}} \left( - \frac{\mu^2 s_{12}}{m^2} \right)^\varepsilon 
\frac{\Gamma (-2 \varepsilon) \Gamma (\varepsilon)}{\Gamma (1-\varepsilon)}
\, .
\end{align}
As before, the regions `c1' and `c2' correspond to the loop momentum being collinear to the external near mass-shell legs 1 and 2,
respectively. These contributions are independent of the invariant mass of the collinear pair $s_{12}$ and are, thus, not relevant for 
the divergent behavior as $s_{12} \to 0$. These need to be subtracted from the definition of the splitting amplitude. Consequently, 
collinear singularities stem only from the hard and ultrasoft regions of the loop momentum.

In light of the above one-loop analysis, we conclude that we have to perform the infrared factorization of the loop integrals 
determining the splitting amplitudes from the three-particle vertex along the lines of Sect.\ \ref{1LoopOffShellAmplitude}. The
true splitting amplitude is then
\begin{align}
\mathcal{R}_{\rm true} (s_{12}, z)= \frac{\mathcal{R} (s_{12}, z)}{J_{\rm ax} (z) J_{\rm ax} (\bar{z})}
\, ,
\end{align}
where $J_{\rm ax}$ is what we will dub the `axial jet' since it is similar to the jet function introduced in Sect.\ \ref{HardColUSsection} 
but its integrand is dressed with additional light-cone denominators akin to ones arizing in calculations of Feynman diagrams in 
axial-like gauges. At one loop, they are given by
\begin{align}
J_{\rm ax} (z)
=
1 + g_\varepsilon^2
\frac{c_\varepsilon}{\varepsilon^2} 
\left[
\left( \frac{\mu^2}{m} \right)^\varepsilon
+
\frac12 \left( \frac{\mu^2}{z m} \right)^\varepsilon + \varepsilon^2 {\rm Li}_2 (\bar{z}) + O (\varepsilon^3)
\right]
+ O (g_\varepsilon^4)
\, .
\end{align}
The one-loop true splitting amplitude is then
\begin{align}
\mathcal{R}_{\rm true}^{(1)} (s_{12}, z)
&
=
-
\ft{1}{2} 
\sum_{r = {\rm h, us}}[J_{1}^{r} + z s_{12} J^{r}_{2} + ({z \leftrightarrow \bar{z}})]
=
r^{(1)}_{\rm h} (s_{12}, z) + r^{(1)}_{\rm us}  (s_{12}, z)
\, ,
\end{align}
where $r^{(1)}_{\rm h}$ is the well-known one-loop massless splitting amplitude \cite{Bern:1994zx,Kosower:1999xi,Kosower:1999rx} 
given in Eq.\ \re{Massless1loopSplittingAmplitude} of Appendix \ref{CollinearLimitAppendix} and 
\begin{align}
r^{(1)}_{\rm us}  (s_{12}, z)
= 
-
\e^{\varepsilon \gamma_{\rm E}}
\Gamma (1 - \varepsilon) \Gamma^2 (\varepsilon)
\left( - \frac{\mu^2 s_{12}}{m^2} \right)^\varepsilon 
\end{align}
is an extra contribution in the off-shell case due to the presence of the ultrasoft regime in loop momentum. It also depends 
on the hard collinear scale $s_{12}$ through $\mu_{\rm us}^2 = - m^2/s_{12}$. It is easy to verify that this expression is identical
to $R^{(1)}_{s_{12}} (z) = \mathcal{R}_{\rm true}^{(1)} (s_{12}, z)$ in \re{1loopSplitAmplObscure}, where, however, one has to set the 
$\delta$-contribution, see Eq.\ \re{delta}, to zero since it exceeds the accuracy of the calculations used to find 
$\mathcal{R}_{\rm true}^{(1)}$. The above representation unravels the obscure structure obtained from the collinear limit of 
off-shell gauge amplitudes.

\section{Conclusions} 

In this paper, we addressed the question of the collinear factorization of near mass-shell scattering amplitudes in maximally supersymmetric
gauge theory. The main tool that allowed us to do it in a clear-cut manner was the Method of Regions, which enables separations of 
momentum modes responsible for different physics. The resulting collinear factorization formula \re{CollFactOffShell} is far more complex 
than its on-shell counterpart \re{SplitDef}. The main source for the difference is the existence of the ultrasoft modes in loop integrations 
which introduce a new scale that encodes hard scales in the problem but in the inverse fashion \re{LeadingRegions}.

Infrared factorization of near mass-shell scattering amplitudes was also studied in this work. Ultrasoft modes again introduced new physics 
into the problem. They were endowed with an operator definition as a vacuum expectation value of a `bursting spray' of Wilson line segments
emanating from the collision point. The length of these segments is inversely proportional to the legs' off-shellness. In the planar limit,
the matrix element factorizes in pair-wise wedge factors and the latter coincides with the ultrasoft factor recently studied within the context
of the off-shell Sudakov form factor \cite{Belitsky:2024yag}. However, naive factorization of the latter in incoherent hard, collinear, and ultrasoft 
momentum components becomes very subtle at higher loop orders. To maintain the product form akin to one for the amplitudes suggested
in Eqs.\ \re{IRfactorizationM5} and \re{IRfactorizationM4}, one has to modify definitions of individual factors at subleading orders in the
$\varepsilon$-expansion and simultaneous finite scheme transformation of the 't Hooft coupling. Undoubtedly, this procedure will have to be
implemented in the context of off-shell scattering amplitudes as well. In the current work, our analysis was performed in the one-loop 
approximation and these complications did not arise. However, they will show themselves up already at two loops. Consequently, the collinear 
factorization properties of amplitudes beyond leading perturbative order will be affected as well. These questions will be addressed in the future.

In principle, the factorization formulas \re{IRfactorizationM5} and \re{IRfactorizationM4} proposed in this paper embody scattering amplitudes 
with a description in terms of matrix elements of operators built (predominantly) from Wilson lines. While this arises naturally for the 
jet and ultrasoft modes, the hard modes admit this form, making use of the duality of massless gauge amplitudes to expectation values 
of null polygonal Wilson loops $h_n = W_n$, as we recalled in the introductory section. Presently, it is unclear, however, whether the entire
off-shell amplitudes can be recast as an expectation value of, potentially generalized, Wilson loops on some contours. A higher-dimensional 
perspective and the dimensional reduction down to four-dimensional Minkowski space-time, similar to the point of view adopted in this paper, 
was taken in Ref.\ \cite{Belitsky:2021huz} to build such a prototype. However, while an agreement was found at one loop order, devitations 
invalidated this conjecture starting from two loops. Can it be remediated? It remains to be seen.

\begin{acknowledgments}
This work was supported by the U.S.\ National Science Foundation under grant No.\ PHY-2207138. 
\end{acknowledgments}

\appendix

\section{Dirac matrices in ...}
\label{DiracAppendix}

In this Appendix, we set up our conventions for Clifford algebras in four and six dimensions.

\subsection{... Four dimensions}
\label{4DGammas}

Let us begin with the more familiar four-dimensional case. We borrow notations from Ref.\ \cite{Belitsky:2003sh}, which were in turn 
adopted from Sohnius \cite{Sohnius:1985qm}. Let us summarize them here for the reader's convenience.

A four-component spinor $\psi$ is composed of two Weyl spinors, 
\begin{align}
\psi
=
\left(
\begin{array}{c}
\lambda_\alpha
\\
\bar\lambda^{\dot\alpha}
\end{array}
\right)
\, .
\end{align}
In the mostly-negative Minkowski-signature metric, $\eta^{\mu\nu} = {\rm diag}(1,-1,-1,-1)$, these transform with respect to conjugate 
factors of the Lorentz group ${\rm SO} (3,1) = {\rm SO} (4, \mathbb{C})_{\downarrow R} \approx \left( {\rm SL} (2, \mathbb{C})  \otimes {\rm SL} 
(2, \mathbb{C}) \right)_{\downarrow R}$ which are labeled by a pair $\left( j_1, j_2 \right)$ of eigenvalues of the ${\rm SL} (2)$ 
quadratic Casimir operators, i.e., $\lambda_\alpha \sim (\ft12, 0)$ and $\bar\lambda^{\dot\alpha} \sim (0 , \ft12)$. A four-dimensional vector
transforms in the  $(\ft12, \ft12)$ representation of ${\rm SL} (2)  \otimes {\rm SL} (2)$. The corresponding Clebsch-Gordan coefficients are
the Dirac matrices which admit the following form
\begin{align}
\label{4Dmatrices}
\gamma^\mu
=
\left(
\begin{array}{cc}
0 & \bar\sigma^\mu{}_{\alpha \dot\beta}
\\
\sigma^\mu{}^{\; \dot\alpha \beta} & 0
\end{array}
\right)
\, , \qquad
\gamma^5
=
i \gamma^0 \gamma^1 \gamma^2 \gamma^3
=
\left(
\begin{array}{cc}
\1_{[2 \times 2]} & 0
\\
0 & - \1_{[2 \times 2]}
\end{array}
\right)
\, ,
\end{align}
where the two-dimensional blocks $\sigma^\mu = (1 , \bit{\sigma})$ and $\bar\sigma^\mu = (1 , - \bit{\sigma})$ are built from the identity 
and Pauli matrices for the temporal and spacial components, respectively. The Clifford algebra $\{ \gamma^\mu , \gamma^\nu \} = 2 
\eta^{\mu\nu} \1_{[4 \times 4]}$ can be rewritten for their two-dimensional off-diagonal elements 
\begin{align}
\label{Clifford2Delements}
\bar\sigma^\mu{}_{\alpha\dot\gamma} \sigma^{\nu \; \dot\gamma\beta}
+
\bar\sigma^\nu{}_{\alpha\dot\gamma} \sigma^{\mu \; \dot\gamma\beta}
= 2 \eta^{\mu\nu} \delta_\alpha{}^\beta
\, ,
\qquad
\sigma^{\mu \; \dot\alpha\gamma} \bar\sigma^\nu{}_{\gamma\dot\beta}
+
\sigma^{\nu \; \dot\alpha\gamma} \bar\sigma^\mu{}_{\gamma\dot\beta}
= 2 \eta^{\mu\nu} \delta^{\dot\alpha}{}_{\dot\beta}
\, .
\end{align}
The raising and lowering of Weyl spinor indices
\begin{align}
\label{RaisingLowering4D}
\lambda^\alpha = \varepsilon^{\alpha \beta} \lambda_\beta
\, , \qquad
\bar\lambda_{\dot\alpha}
= \varepsilon_{\dot\alpha \dot\beta} \bar\lambda^{\dot\beta}
\, , \qquad
\lambda_\alpha = \lambda^\beta \varepsilon_{\beta\alpha}
\, , \qquad
\bar\lambda^{\dot\alpha}
= \bar\lambda_{\dot\beta} \varepsilon^{\dot\beta \dot\alpha}
\, ,
\end{align}
are accomplished with fully antisymmetric tensors $\varepsilon^{\alpha\beta}$ and $\varepsilon_{\dot\alpha \dot\beta}$ 
which are normalized as $\varepsilon^{\alpha\beta} \equiv i \sigma_2^{\alpha\beta}$ such that $\varepsilon^{12} = 
\varepsilon_{12} = - \varepsilon_{\dot 1 \dot 2} = - \varepsilon^{\dot 1 \dot 2} = 1$. While for the $\sigma$-matrices, we have
\begin{align}
\sigma^{\mu \; \dot\alpha\beta}
=
\varepsilon^{\beta\gamma}
\bar\sigma^\mu{}_{\gamma\dot\delta}
\varepsilon^{\dot\delta\dot\alpha}
\, , \qquad
\bar\sigma^\mu{}_{\alpha\dot\beta}
=
\varepsilon_{\dot\beta\dot\gamma}
\sigma^{\mu \; \dot\gamma\delta}
\varepsilon_{\delta\alpha}
\, .
\end{align}
Note that $\left( \varepsilon_{\dot\alpha\dot\beta} \right)^\ast = - \varepsilon_{\alpha\beta}$ (and this is the only inconvenience of these 
conventions), however, in applications we typically complexify all spinors and thus never use complex conjugate nature of the Weyl 
spinors $\lambda$ and $\bar\lambda$.

Finally the traces are
\begin{align}
\label{2Dtraces}
&
\ft12 {\rm tr}_2
\left[
\,
\bar\sigma^\mu \sigma^\nu
\right]
=
\eta^{\mu\nu}
\, , \\
&
\ft12 {\rm tr}_2
\left[
\,
\bar\sigma^\mu \sigma^\nu \bar\sigma^\rho \sigma^\sigma
\right]
=
\eta^{\mu \nu} \eta^{\rho \sigma}
+
\eta^{\mu \sigma} \eta^{\nu \rho}
-
\eta^{\mu \rho} \eta^{\nu \sigma}
-
i \varepsilon^{\mu\nu\rho\sigma}
\, , \nonumber\\
&
\ft12 {\rm tr}_2
\left[
\,
\sigma^\mu \bar\sigma^\nu \sigma^\rho \bar\sigma^\sigma
\right]
=
\eta^{\mu \nu} \eta^{\rho \sigma}
+
\eta^{\mu \sigma} \eta^{\nu \rho}
-
\eta^{\mu \rho} \eta^{\nu \sigma}
+
i \varepsilon^{\mu\nu\rho\sigma}
\, , \nonumber
\end{align}
where the normal position of spinor indices, as in Eq.\ (\ref{4Dmatrices}), is implied in their left-hand sides. The four-dimensional
Levi-Civita tensor is normalized as $\varepsilon^{0123} = 1$.

\subsection{... Six dimensions}
\label{6DGammas}

In six dimensions, the Lorentz group ${\rm SO}(5,1)$ has ${\rm SU}(4)$ as its cover\footnote{It is ${\rm SU}^\ast(4)$ to be precise, 
but the distinction between pseudoreal and real representations will not be important for our discussion.} and so their algebras are 
isomorphic ${\rm so}(5,1) \simeq {\rm su} (4)$. Thus, a six-dimensional vector transforms in the antisymmetric representation of the 
${\rm SU}(4)$. A particularly convenient form for corresponding Clebsch-Gordan coefficients making use of the 't Hooft 
symbols $\eta$ and $\bar\eta$ \cite{tHooft:1976snw} was constructed in Refs.\ \cite{Belitsky:2000ii,Belitsky:2003sh}. These
Dirac matrices admit the off-diagonal form 
\begin{align}
\label{tHooftGammas}
{\mit\Gamma}^M
=
\left(
\begin{array}{cc}
 0 & \bar{\mit\Sigma}^M{}_{AB}  \\
{\mit\Sigma}^{M,AB} & 0 
\end{array}
\right)
\end{align}
just like the four-dimensional ones, with
\begin{align}
{\mit\Sigma}^{M,AB} 
&
= (i \eta_{1,AB}, \eta_{2,AB}, \eta_{3,AB}, +i \bar\eta_{1,AB}, +i \bar\eta_{2,AB}, +i \bar\eta_{3,AB})
\, , \\
\bar{\mit\Sigma}^{M}{}_{AB} 
&
= (i \eta_{1,AB}, \eta_{2,AB}, \eta_{3,AB}, -i \bar\eta_{1,AB}, -i \bar\eta_{2,AB}, -i \bar\eta_{3,AB})
\, ,
\end{align}
in the Minkowski space-time with the mostly negative metric tensor $\eta^{MN} = {\rm diag}(+1,-1,-1,$ $-1,-1,-1)$. However, literally assuming 
these conventions in the current application would result in a scrambled four-dimensional limit, see Eq.\ \re{6DmatrixID} below. To properly 
embed the four-dimensional subspace into the six-dimensional one, we proceed in a step-wise fashion.

To start with, we convert the left indices of the Dirac matrices \re{4Dmatrices} to be in the same positions as the right ones. This is 
accomplished by acting with the two-dimensional Levi-Civita tensors from the right
\begin{align}
\label{gamma2Sigma}
\Sigma^\mu =  \gamma^\mu \cdot ( \1_{[2 \times 2]} \otimes i \sigma_2 )
\, ,
\end{align}
and from the left,
\begin{align}
\label{gamma2SigmaBar}
\bar\Sigma^\mu = ( - \1_{[2 \times 2]} \otimes i \sigma_2 ) \cdot \gamma^\mu
\, ,
\end{align}
on the four-dimensional matrices \re{4Dmatrices}. These become skew symmetric, as can be seen from their explicit expressions,
\begin{align}
\label{First4Sigma}
\Sigma^{\mu, AB}
&=
{
\footnotesize
\left\{
\left(
\begin{array}{cccc}
 0 & 0 & 0 & +1 \\
 0 & 0 & -1 & 0 \\
 0 & +1 & 0 & 0 \\
 -1 & 0 & 0 & 0 
\end{array}
\right)
,
\left(
\begin{array}{cccc}
 0 & 0 & +1 & 0 \\
 0 & 0 & 0 & -1 \\
 -1 & 0 & 0 & 0 \\
 0 & +1 & 0 & 0 
\end{array}
\right)
,
\left(
\begin{array}{cccc}
 0 & 0 & -i & 0 \\
 0 & 0 & 0 & -i \\
 +i & 0 & 0 & 0 \\
 0 & +i & 0 & 0 
\end{array}
\right)
,
\left(
\begin{array}{cccc}
 0 & 0 & 0 & -1 \\
 0 & 0 & -1 & 0 \\
 0 & +1 & 0 & 0 \\
 +1 & 0 & 0 & 0 
\end{array}
\right)
\right\}
}
\, , \\
\label{First4SigmaBar}
\bar\Sigma^\mu_{AB}
&=
{
\footnotesize
\left\{
\left(
\begin{array}{cccc}
 0 & 0 & 0 & -1 \\
 0 & 0 & +1 & 0 \\
 0 & -1 & 0 & 0 \\
 +1 & 0 & 0 & 0 
\end{array}
\right)
,
\left(
\begin{array}{cccc}
 0 & 0 & +1 & 0 \\
 0 & 0 & 0 & -1 \\
 -1 & 0 & 0 & 0 \\
 0 & +1 & 0 & 0 
\end{array}
\right)
,
\left(
\begin{array}{cccc}
 0 & 0 & +i & 0 \\
 0 & 0 & 0 & +i \\
 -i & 0 & 0 & 0 \\
 0 & -i & 0 & 0 
\end{array}
\right)
,
\left(
\begin{array}{cccc}
 0 & 0 & 0 & -1 \\
 0 & 0 & -1 & 0 \\
 0 & +1 & 0 & 0 \\
 +1 & 0 & 0 & 0 
\end{array}
\right)
\right\}
}
\, .
\end{align}
The two-dimensional Weyl indices $\alpha$ and $\dot\alpha$ are embedded as $\ast^A = \left( \ast_\alpha, \ast^{\dot\alpha}\right)^T$, 
$\ast_A = \left( \ast^\alpha, \ast_{\dot\alpha}\right)^T$ into the four-dimensional one $A$.

The above $\Sigma$ and $\bar\Sigma$ matrices define the four-dimensional blocks of the (first four) six-dimensional Dirac matrices. One can 
immediately observe that they are indeed related to the ${\mit\Gamma}$-matrices introduced in Refs.\ \cite{Belitsky:2000ii,Belitsky:2003sh} 
and spelled out at the beginning of this section, see, Eq.\ \re{tHooftGammas}, as
\begin{align}
\label{6DmatrixID}
\Sigma^0 = i {\mit\Sigma}^3
\, , \quad
\Sigma^1 = - {\mit\Sigma}^1
\, , \quad
\Sigma^2 = {\mit\Sigma}^4
\, , \quad
\Sigma^3 = i {\mit\Sigma}^0
\, , 
\end{align}
(we omitted the matrix indices to avoid cluttering) and the same for the barred elements. So the two sets are indeed related by reshuffling of 
the elements in the space-time index.

Finally, we complement the set of the matrices \re{First4Sigma} and \re{First4SigmaBar} with the fifth and sixth elements of the six-dimensional 
space
\begin{align}
\Sigma^4 = {\mit\Sigma}^5 = - \sigma_3 \otimes \sigma_2
\, , \quad
\Sigma^5 = {\mit\Sigma}^2 =  i \sigma_2 \otimes \sigma_2
\, , \quad
\bar\Sigma^4 = \bar{\mit\Sigma}^5 = \sigma_3 \otimes \sigma_2
\, , \quad
\bar\Sigma^5 = \bar{\mit\Sigma}^2 =  i \sigma_2 \otimes \sigma_2
\, .
\end{align}
This is the representation which we will be using throughout this work. A slightly different form was introduced in the early literature on the 
six-dimensional spinor-helicity formalism \cite{Cheung:2009dc}.

Thus introduced Dirac matrices
\begin{align}
\label{6DGammasDef}
\Gamma^M
=
\left(
\begin{array}{cc}
 0 & \bar{\Sigma}^M{}_{AB}  \\
{\Sigma}^{M,AB} & 0 
\end{array}
\right)
\end{align}
obey the Clifford algebra $\{ \Gamma^M , \Gamma^N \} = 2 \eta^{MN} \1_{[8 \times 8]}$, which can be recast in terms of their off-diagonal
blocks, akin to its four-dimensional counterpart \re{Clifford2Delements},
\begin{align}
{\Sigma}^{M,AB} \bar{\Sigma}^N{}_{BC}
+
{\Sigma}^{N,AB} \bar{\Sigma}^M{}_{BC}
=
2 \eta^{MN} \delta^A{}_C
\, , \quad
\bar{\Sigma}^M{}_{AB} {\Sigma}^{N,BC} 
+
\bar{\Sigma}^N{}_{AB} {\Sigma}^{M,BC} 
=
2 \eta^{MN} \delta_A{}^C
\, .
\end{align}
The raising and lowering of pairs of ${\rm SU}(4)$ indices is accomplished with the totally antisymmetric tensors $\varepsilon^{ABCD}$
and $\varepsilon_{ABCD}$ normalized with $\varepsilon^{1234} = \varepsilon_{1234} = 1$,
\begin{align}
\label{6Dduality}
{\Sigma}^{M,AB} = \frac{1}{2} \varepsilon^{ABCD} \bar{\Sigma}^{M}_{CD}
\, , \quad  
\bar{\Sigma}^M{}_{AB} = \frac{1}{2} \varepsilon_{ABCD} {\Sigma}^{M,CD}
\, .
\end{align}
Contractions 
\begin{align}
&
{\Sigma}^{M,AB} {\Sigma}_M{}^{CD} = - 2 \varepsilon^{ABCD}
\, , \qquad
\bar{\Sigma}^M{}_{AB} \bar{\Sigma}_{M, CD} = - 2 \varepsilon_{ABCD}
\, , \\
&
{\Sigma}^{M,AB} \bar{\Sigma}_{M,CD} = - 2 \left( \delta^A{}_C \delta^B{}_D -  \delta^A{}_D \delta^B{}_C \right)
\, ,
\end{align}
and traces 
\begin{align}
&
\ft14 {\rm tr}_4 \left[ \Sigma^M \bar\Sigma^N \right] = \eta^{MN}
\, , \\
&
\ft14 {\rm tr}_4 \left[ \Sigma^M \bar\Sigma^N \Sigma^K \bar\Sigma^L \right] = \eta^{MN} \eta^{KL} + \eta^{ML} \eta^{NK} - \eta^{MK} \eta^{NL}
\, .
\end{align}
of the four-dimensional blocks ${\Sigma}$ and $\bar{\Sigma}$ inherit all of the properties of the 't Hooft symbols.

\section{Spinor-helicity formalism in ...}
\label{SpinorHelicityAppendix}

In four and six dimensions, a light-like vector can be represented in terms of unconstrained spinors, i.e., the Dirac equations for
the latter are automatically satisfied. This makes the spinor-helicity technique particularly powerful. In this Appendix, we merely 
recast the known formalisms in the above-introduced conventions.

\subsection{... Four dimensions}

A massless particle is four-dimensions is defined by its null momentum vector $p^\mu$, $p^2 \equiv p^\mu p_\mu = 0$. This vector is 
left invariant by the ${\rm SO}(2)$ rotations, which becomes obvious in a frame where it admits the form $p^\mu = (p^0,0,0,p^3)$. 
This ${\rm SO}(2) \simeq {\rm U}(1)$ defines the little group of the Lorentz group and yields two values of the helicity for a given spin. 
Dotting $p^\mu$ with the two-dimensional blocks of the Dirac matrices, one obtains its spinor representation\footnote{We do not place 
the bar on top of $p$ in the second relation, since which Clebsh-Gordan coefficients are used to define it is obvious from the position 
of the spinor indices. In all matrix products, the natural position of indices is always assumed.}
\begin{align}
p^\mu \sigma_\mu^{\dot\alpha \beta} = p^{\dot\alpha \beta}
\, , \qquad
p^\mu \bar\sigma_{\mu, \alpha \dot\beta} = p_{\alpha \dot\beta}
\, .
\end{align}
Since $p^2 \equiv {\rm det} p^{\dot\alpha \beta} = 0$, $p^{\dot\alpha \beta}$ is a rank one matrix and it can be written as a tensor product
of spinors \cite{Gastmans:1990xh,Xu:1986xb} (see \cite{Dixon:2013uaa}, for a review)
\begin{align}
\label{FourMomentumSpinors}
p^{\dot\alpha \beta} = \bar\lambda_{p}^{\dot\alpha} \lambda_p^{\beta}
\, ,
\end{align}
which we will treat as independent, thus implying that we are dealing with complexified momenta. We use the bra and ket notation for
these spinors
\begin{align}
\lambda_{p_i, \alpha} \equiv | i \rangle
\, , \quad
\lambda_{p_i}^{\alpha} \equiv \langle i |
\, , \quad
\bar\lambda_{p_i}^{\dot\alpha} \equiv | i ]
\, , \quad
\bar\lambda_{p_i, \dot\alpha} \equiv [ i |
\, ,
\end{align}
and use a uniform way of contracting their ${\rm SL}(2)$ indices: un-dotted indices from upper left to lower right and dotted one 
lower left to upper right,
\begin{align}
\langle i j \rangle  = \lambda_{p_i}^\alpha \lambda_{p_j, \alpha}
\, , \qquad
[ i j ]  = \bar\lambda_{p_i, \dot\alpha} \bar\lambda_{p_j}^{\dot\alpha}
\, .
\end{align}
This was already implied in Eq.\ \re{RaisingLowering4D}. With these conventions in hand, the inner product of four-vectors reads
\begin{align}
p_i \cdot p_j = \ft12 {\rm tr}_2 [p_i p_j] = - \ft12 \langle i j \rangle [ i j ]
\, .
\end{align}

\subsection{... Six dimensions}

A six-dimensional null vector $P^M$, $P^2 \equiv P^M P_M = 0$, is left invariant by the ${\rm SO} (4)$ rotations, as made apparent in 
the frame $P^M = (P^0, 0,0,0,0, P^5)$. This is the little group of the six-dimensional Lorentz group ${\rm SO} (5,1)$. All spin states of 
a particle with the momentum $P$ are then classified with respect to it. ${\rm SO} (4)$ is isomorphic to ${\rm SU}(2) \otimes 
{\rm SU}(2)$ and thus helicity states can be labeled by up to two ${\rm SU}(2)$ indices. Echoing the four-dimensional
notations of Sect.\ \ref{4DGammas}, we will denote the latter by un-dotted and dotted Latin (as opposed to the used above 
Greek) letters, i.e., $a,b,\dots$ and $\dot{a}, \dot{b}, \dots$. We pass to the spinor representation for six-dimensional null vectors 
with the help of the Dirac matrices \re{6DGammasDef}. These read explicitly
\begin{align}
\label{6DLambdaSpinors}
P^{AB} \equiv P^{M} \Sigma_M{}^{AB} 
&
= 
- i
\left(
\begin{array}{cc}
w \sigma_2 & - \bar{p} \cdot \sigma_2  \\
- p \cdot \sigma_2 & \bar{w} \sigma_2
\end{array}
\right)
\, , \\
\label{6DLambdaBarSpinors}
\bar{P}_{AB} \equiv P^{M} \bar\Sigma_{M, AB} 
&
=
- i
\left(
\begin{array}{cc}
\bar{w} \sigma_2 & \sigma_2 \cdot \bar{p}  \\
\sigma_2 \cdot p & w \sigma_2
\end{array}
\right)
\, ,
\end{align}
where we introduced the notations $w = P^5 + i P^4$, $\bar{w} = P^5 - i P^4$ as well as $\bar{p} = P_\mu \bar\sigma^\mu$,
$p = P_\mu \sigma^\mu$ with the summation running over the first four $\mu$ values only of the six-dimensional index $M$. Since 
the four-dimensional momentum $p^\mu$ is off-shell, i.e., $p^2 \equiv w \bar{w} \neq 0$. One then introduces two sets of the Weyl 
spinors \cite{Cheung:2009dc}
\begin{align}
\label{PtoLambda}
P^{AB} = {\mit\Lambda}_p^{A,a} {\mit\Lambda}^B_{p,a}
\, , \qquad
\bar{P}_{AB} = \bar{\mit\Lambda}_{p, A,\dot{a}} \bar{\mit\Lambda}_{p,B}^{\dot{a}}
\, ,
\end{align}
such that
\begin{align}
{\mit\Lambda}^{A,a} \bar{\mit\Lambda}_{A, \dot{a}} = 0
\, .
\end{align}
Notice that the fundamental and anti-fundamental representations of ${\rm SU} (4)$ are inequivalent since there is 
no two-index tensor that can raise/lower the indices, as opposed to the four-dimensional case. As it is obvious from our 
definitions \re{PtoLambda}, we will adopt the same rules for raising/lowering and contractions of the little group indices 
as for the ${\rm SL}(2)$ indices of the four-dimensional Lorentz group, namely,
\begin{align}
{\mit\Lambda}^{A,a} \varepsilon_{ab} = {\mit\Lambda}^{A}_b
\, , \quad
\varepsilon^{ab} {\mit\Lambda}^A_b  = {\mit\Lambda}^{A,a}
\, , \quad
\varepsilon_{\dot{a}\dot{b}} \bar{\mit\Lambda}_A^{\dot{b}} = \bar{\mit\Lambda}_{A, \dot{a}}
\, , \quad
\bar{\mit\Lambda}_{A,\dot{a}} \varepsilon^{\dot{a}\dot{b}} = \bar{\mit\Lambda}_A^{\dot{b}}
\, ,
\end{align}
with the Levi-Civita tensors $\varepsilon^{ab}$ and $\varepsilon_{\dot{a} \dot{b}}$ normalized as $\varepsilon^{12}
= \varepsilon_{12} = - \varepsilon_{\dot{1} \dot{2}} = - \varepsilon^{\dot{1} \dot{2}} = 1$.

One can further introduce bracket notations for the six-dimensional spinors, though the bra and ket are interchangeable
in this case, 
\begin{align}
{\mit\Lambda}_{p_i}^{A,a} = | i^a \rangle = \langle i^a |
\, , \quad
\bar{\mit\Lambda}_{p_i,A,\dot{a}} = | i_{\dot{a}} ] = [ i_{\dot{a}} |
\, ,
\end{align}
and therefore there will be no particular contraction rules assumed for the ${\rm SU} (4)$ indices, i.e., 
\begin{align}
\langle i_a | j_{\dot{b}} ] = [ j_{\dot{b}} | i_a \rangle
\, .
\end{align}

The inner products of six-vectors can be represented in a variety of ways in the spinor formalism,
\begin{align}
P_i \cdot P_j 
&
= \ft{1}{4} {\rm tr}_4 [ P_i \bar{P}_j ] 
= 
\ft{1}{4} {\rm tr}_4 [ \bar{P}_i {P}_j ]
= 
- \ft{1}{8} \varepsilon_{ABCD} P_i^{AB} P_j^{CD}
=
- \ft{1}{8} \varepsilon^{ABCD} \bar{P}_{i, AB} \bar{P}_{j, CD}
\\
&
=
-
\ft{1}{4} \langle i^a | j_{\dot{a}} ] [ j^{\dot{a}} | i_a \rangle
=
-
\ft{1}{4} \langle j^a | i_{\dot{a}} ] [ i^{\dot{a}} | j_a \rangle
=
\ft{1}{2} \det \langle i_a | j_{\dot{a}} ]
=
\ft{1}{2} \det \langle j_a | i_{\dot{a}} ] 
\, . \nonumber
\end{align}

Finally, since we will be interested in the dimensional reduction of the six-dimensional theory down to four and treating the
extra-dimensional components of momenta as masses, we will close this section with an explicit solution to the Dirac equations
for the Weyl spinors,
\begin{align}
\label{6DDiracEqs}
P^{AB} \bar{\mit\Lambda}_{A,\dot{a}} = 0
\, , \quad
\bar{P}_{AB} {\mit\Lambda}^A_{a} = 0
\, .
\end{align}
It is more instructive to work with matrices with natural position of the four-dimensional blocks so we will transform $P$ and $\bar{P}$
by acting with the two-dimensional Levi-Civita tensor, basically undoing the transformation in Eq.\ \re{gamma2Sigma} and
\re{gamma2SigmaBar}. Namely
\begin{align}
&
(i \sigma_2 \otimes \1_{[2 \times 2]}) \cdot {P}_{AB} 
=
\left(
\begin{array}{cc}
w \ \1_{[2 \times 2]} & - p^T \\
- \bar{p}^T & \bar{w} \ \1_{[2 \times 2]}
\end{array}
\right)
\, , \\
&
(i \sigma_2 \otimes \1_{[2 \times 2]}) \cdot \bar{P}_{AB} 
=
\left(
\begin{array}{cc}
\bar{w} \ \1_{[2 \times 2]} & \bar{p} \\
p & w \ \1_{[2 \times 2]}
\end{array}
\right)
\, ,
\end{align}
Following Ref.\ \cite{Boels:2009bv}, we can further decompose the off-diagonal blocks in terms of their light-like components
\begin{align}
\label{OffShellMomentum}
p^{\dot{\alpha} \alpha} 
= 
\bar\lambda_p^{\dot\alpha} \lambda_p^\alpha - \frac{p^2}{\langle p q \rangle [p q]} \bar\lambda_q^{\dot\alpha} \lambda_q^\alpha
\, ,
\end{align}
which are encoded by the two-dimensional Weyl spinors $\lambda_{p, q}$ and $\bar\lambda_{p, q}$, with $\lambda_{q}, \bar\lambda_{q}$ 
being auxiliary. The solutions to the Dirac equations \re{6DDiracEqs} are then \cite{Boels:2009bv}
\begin{align}
\label{LambdaFromlambdas}
{\mit\Lambda}^A_{1}
&
=
\left(
\begin{array}{c}
- \frac{w}{\langle p q \rangle} \lambda_{q, \alpha} 
\\ 
\bar{\mit\lambda}_p^{\dot\alpha}
\end{array}
\right)
\, , \quad
{\mit\Lambda}^A_{2}
&
=
\left(
\begin{array}{c}
\lambda_{p, \alpha}
\\
- \frac{\bar{w}}{[ p q ]} \bar\lambda_q^{\dot\alpha} 
\end{array}
\right)
\, , \quad
\bar{\mit\Lambda}_{A,\dot{1}}
&
=
\left(
\begin{array}{c}
\frac{\bar{w}}{\langle p q \rangle} \lambda_q^{\alpha}
\\
- \bar\lambda_{p, \dot\alpha}
\end{array}
\right)
\, , \quad
\bar\Lambda_{A,\dot{2}}
&
=
\left(
\begin{array}{c}
\lambda_p^\alpha
\\
- \frac{w}{[ p q ]} \bar\lambda_{q, \dot\alpha}  
\end{array}
\right)
\, .
\end{align}

\section{Five-leg amplitudes in four dimensions}
\label{4D5LegAppendix}

In this Appendix, we give a brief recollection of the four-dimensional on-shell amplitudes and unitarity construction of the loop-level 
five-leg MHV superamplitude since the latter will serve as a benchmark for its six-dimensional generalization. All on-shell states in 
the $\mathcal{N} = 4$ theory form a CPT self-conjugate multiplet. These can be assembled into a single chiral superfield $\Phi$ 
with states depending on the supermomentum $(p^{\dot\alpha\alpha}, q^{A,\alpha})$. The fermionic component of the latter has 
the ${\rm SU} (4)$ R-symmetry index $A$ and the Weyl index $\alpha$, and it can be written in the spinor-helicity formalism as 
$q^{A,\alpha} = \eta^A \lambda_p^\alpha$ such that $\Phi$ admits a terminating expansion in $\eta$ \cite{Nair:1988bq}
\begin{align}
\label{NairField}
\Phi (p, \eta) = g^+ + \eta^A \psi_A + \ft12 \eta^A \eta^B \phi_{AB} + \ft{1}{3!} \varepsilon_{ABCD} \eta^A \eta^B \eta^C \bar\psi^D
+ \ft{1}{4!} \varepsilon_{ABCD} \eta^A \eta^B \eta^C \eta^D g^-
\, .
\end{align}
All amplitudes can be combined into a single super-amplitude
\begin{align}
\label{4DgenericAmplitude}
\mathcal{A}_n &
\equiv \vev{\Phi_1 \dots \Phi_n}
\nonumber\\
&
=
i (2 \pi)^4 \delta^{(4)}  \left(\sum\nolimits_{i = 1}^n p_i \right) \delta^{(8)}  \left(\sum\nolimits_{i = 1}^n q_i \right)
\left[
\widehat{A}_n^{\rm MHV} (p_i; g) + \widehat{A}_n^{\rm NMHV} (p_i, \eta_i; g) + \dots
\right]
, 
\end{align}
where the overall delta functions enforce the four-momentum and chiral charge conservation. The sum over the reduced amplitudes in 
the square brackets accounts for a possible scattering of particles with helicities $-(n-4)$, $-(n-6)$, etc. The first term, known as the
maximal helicity violating (reduced) amplitude, is purely bosonic. It develops (like the rest of them) an infinite series expansion in 't Hooft 
coupling
\begin{align}
\widehat{A}_n^{\rm MHV} =  \widehat{A}_n^{(0) \rm MHV}  + g^2 \widehat{A}_n^{(1) \rm MHV} + \dots
\, .
\end{align}
with the tree term being simply given by Parke-Taylor denominator \cite{Parke:1986gb}
\begin{align}
\widehat{A}_n^{(0) \rm MHV} = \frac{1}{\vev{12} \vev{23} \dots \vev{n1}}
\end{align}

Let us use these ingredients together with the unitarity technique of Ref.\ \cite{Bern:1994zx,Bern:1994cg,Bern:2004cz} at one-loop 
order for the five-leg MHV amplitude. This way, we will have an anchor point to compare with in the six-dimensional case to be dealt 
with in Sect.\ \ref{6D5LegSection}. We will not discuss a spanning set of cuts but instead just a single $s_{123}$-channel cut shown 
in Fig.~\ref{fig12loops} (left panel) since it is known to contain all possible structures that can appear in the integrand. It reads
\begin{align}
\mathcal{A}^{(1) {\rm MHV}}_5 |_{s_{123}-{\rm cut}}
=
\int d^4 \eta_{k_1} d^4 \eta_{k_2}
\mathcal{A}^{(0) {\rm MHV}}_5 (p_1, p_2, p_3, k_1, k_2)
\mathcal{A}^{(0) {\rm MHV}}_4 (p_4, p_5, - k_2, - k_1)
\, .
\end{align}
Here, we dropped everywhere the energy-momentum delta functions (with accompanying prefactor of $i (2 \pi)^4$) and Cutkosky 
propagators. Its right-hand side takes the form
\begin{align}
\frac{\delta^{(8)} \left( \sum_{i = 1}^5 q_i\right)}{\vev{12}\vev{23}\vev{3,k_1}\vev{k_1,k_2}\vev{k_2,1} \vev{k_1, 4}\vev{45} \vev{5,k_2} \vev{k_2,k_1}}
\int d^4 \eta_{k_1} d^4 \eta_{k_2}
\delta^{(8)} \left( q_{k_1} + q_{k_2} + Q_{45} \right)
\end{align}
where the Grassmann delta functions naturally combine into the overall chiral-charge conservation condition. Here, we used standard 
assignments, $\lambda_{- k} = i \lambda_k$, $\eta_{-k} = i \eta_k$, etc., to preserve proper signs in their relations to super-momenta. The 
final fermionic integrations can be easily accomplished by making use of the following general identity 
\cite{Drummond:2008bq}
\begin{align}
\label{GrassmannDeltaID4D}
\delta^{(8)} \left( q_{k_1} + q_{k_2} + q_{45} \right)
=
\vev{k_1,k_2}^4 
\delta^{(4)} \left( \eta_{k_1} + \frac{\vev{q_{45}, k_2}}{\vev{k_1, k_2}} \right)
\delta^{(4)} \left( \eta_{k_2} + \frac{\vev{q_{45}, k_1}}{\vev{k_2, k_1}} \right)
\, .
\end{align}
In this manner, we obtain the $s_{123}$-cut integrand of the one-loop amplitude
\begin{align}
\mathcal{A}^{(1) {\rm MHV}}_5 |_{s_{123}-{\rm cut}}
=
\mathcal{A}_5^{(0) {\rm MHV}} \frac{\vev{k_1,k_2}^4 \vev{34} \vev{51}}{\vev{3, k_1} \vev{4, k_1} \vev{5, k_2} \vev{1, k_2}}
\, .
\end{align}
Complementing the second factor with appropriate square-bracket products, it can be converted into the trace form
\begin{align}
\frac{\vev{k_1,k_2}^4 \vev{34} \vev{51}}{\vev{3, k_1} \vev{4, k_1} \vev{5, k_2} \vev{1, k_2}}
=
\frac{\tr_2 [k_1 k_2 p_5 p_1 k_2 k_1 p_3 p_4]}{(k_1 + p_3)^2 (k_2 + p_1)^2 (k_1 - p_4)^2 (k_2 - p_5)^2}
\, .
\end{align}
This is the well-known representation in terms of the hexagon \cite{Bern:1994zx} (accounting for the dropped two cut propagators). To observe 
the emerging structure, we will not calculate the trace right away, but use the Dirac algebra to reduce it to the sum of four boxes with nontrivial 
numerators
\begin{align}
\tr_2 
[k_1 k_2 & p_5 p_1 k_2 k_1 p_3 p_4]
\\
=
&
- (k_1 - p_4)^2 (k_2 - p_5)^2 \tr_2 [p_1 k_2 p_3 k_1]
- (k_1 + p_3)^2 (k_2 - p_5)^2 \tr_2 [p_1 k_2 p_4 k_1]
\nonumber\\
&
- (k_2 + p_1)^2 (k_1 + p_3)^2 \tr_2 [p_5 k_2 p_4 k_1]
- (k_2 + p_1)^2 (k_1 - p_4)^2 \tr_2 [p_5 k_2 p_3 k_1]
\, . \nonumber
\end{align}
The remaining traces are computed by making use of Eq.\ \re{2Dtraces} and decomposed via the Passarino-Veltman reduction
\cite{Passarino:1978jh} into the linear combination of remaining propagators. One immediately observes at this step that all
triangles and bubbles cancel among each other leaving just scalar boxes and a parity-odd pentagon. The latter will eventually 
cancel in the cyclic sum \cite{Bern:1994zx,Bern:1995db,Cachazo:2008vp} from the spanning set of cuts in the integrand
(or individually after the momentum loop integration since the initial amplitude does not possess a sufficient number of 
independent momenta to return a nonvanishing contribution). Thus, the $s_{123}$-cut yields the following integrand of the 
one-loop five-leg MHV amplitude
\begin{align}
\label{1Loop5LegMHV}
\widehat{A}^{(1) {\rm MHV}}_{5, {\rm integrand}} 
=
\frac{1}{2} s_{12} s_{23}
\parbox[c][18mm][t]{20mm}{
\insertfig{2}{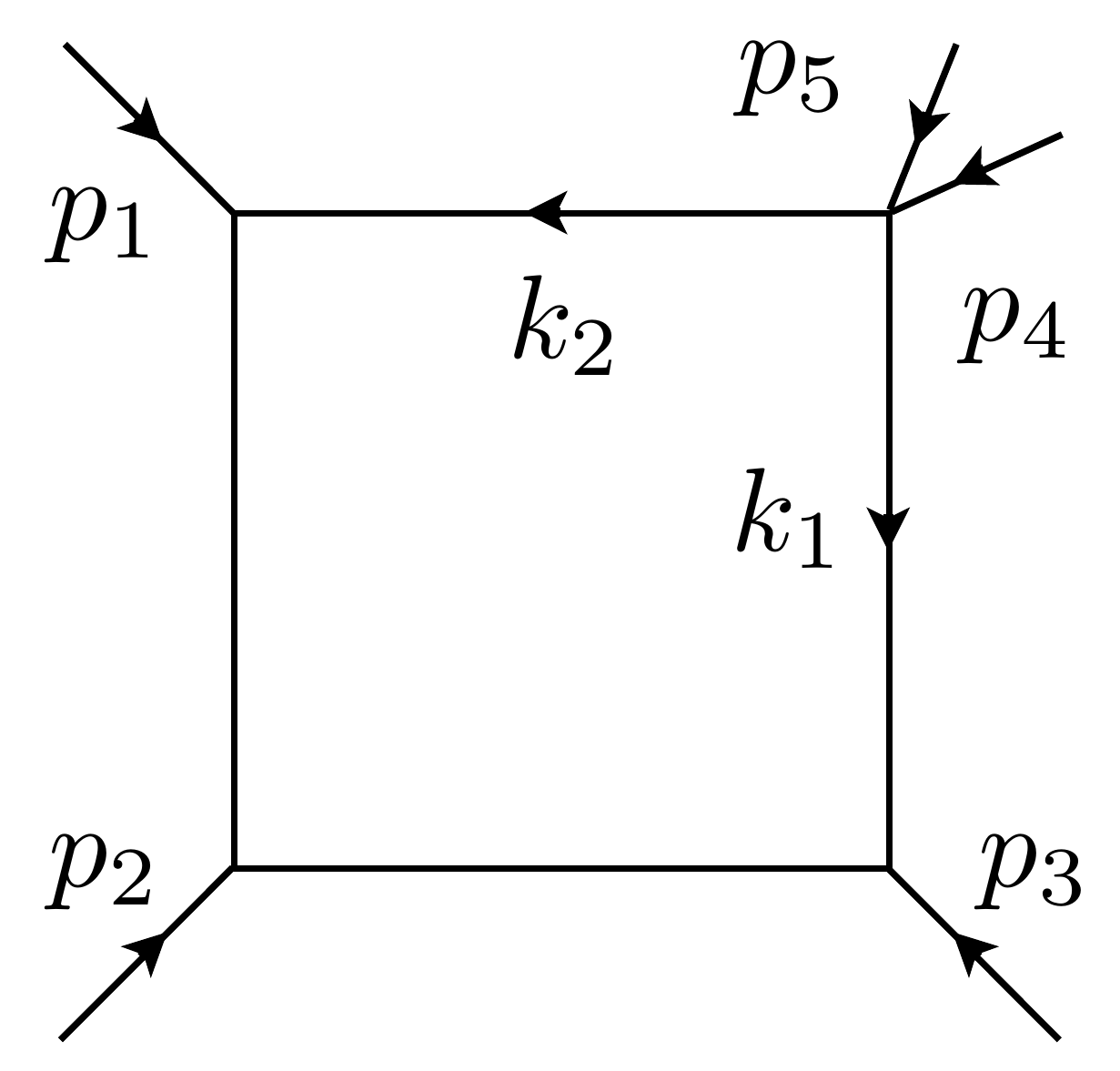}
}
+
\frac{1}{2} s_{45} s_{51}
\parbox[c][18mm][t]{20mm}{
\insertfig{2}{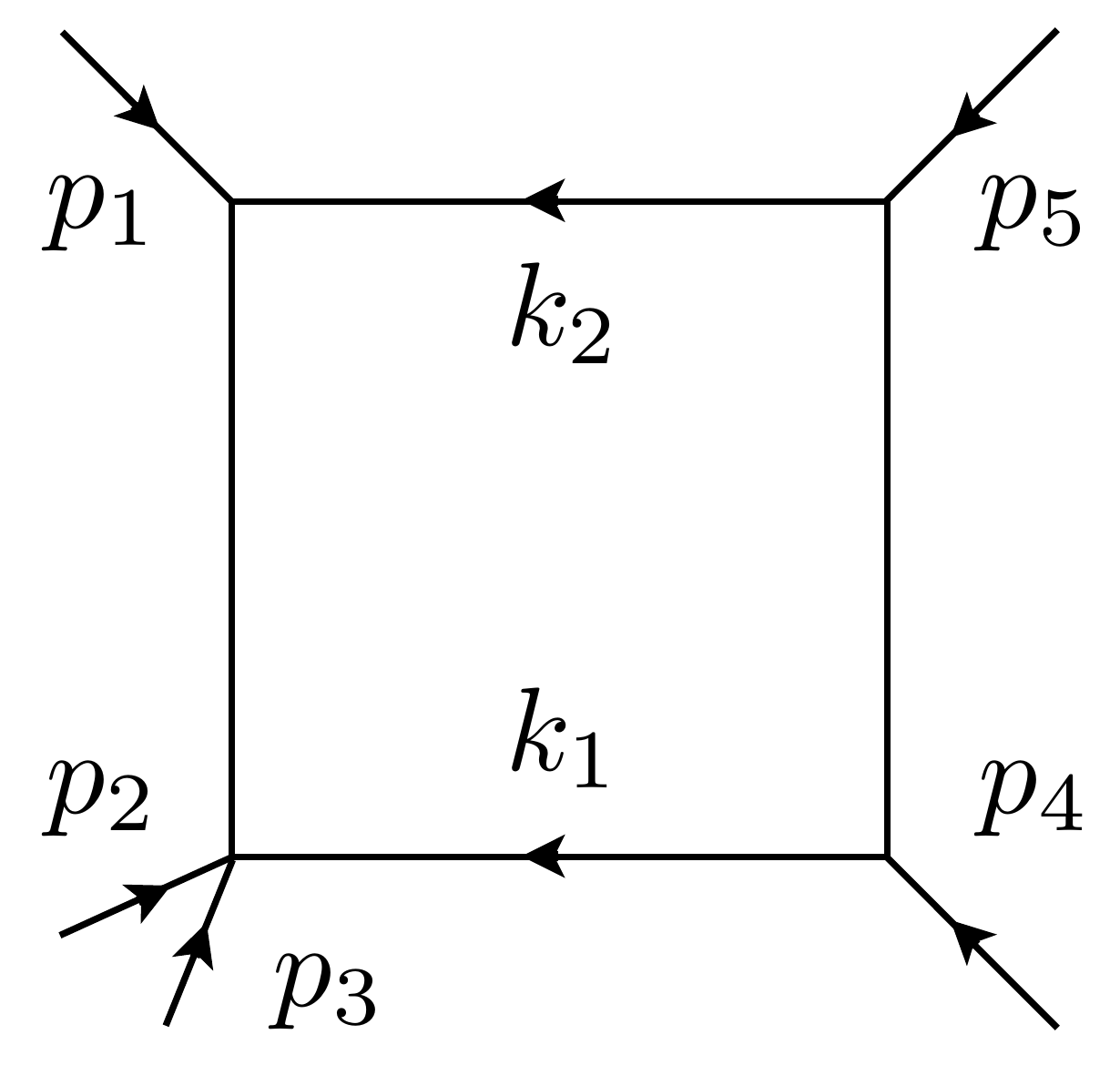}
}
+
\frac{1}{2} s_{34} s_{45}
\parbox[c][18mm][t]{20mm}{
\insertfig{2}{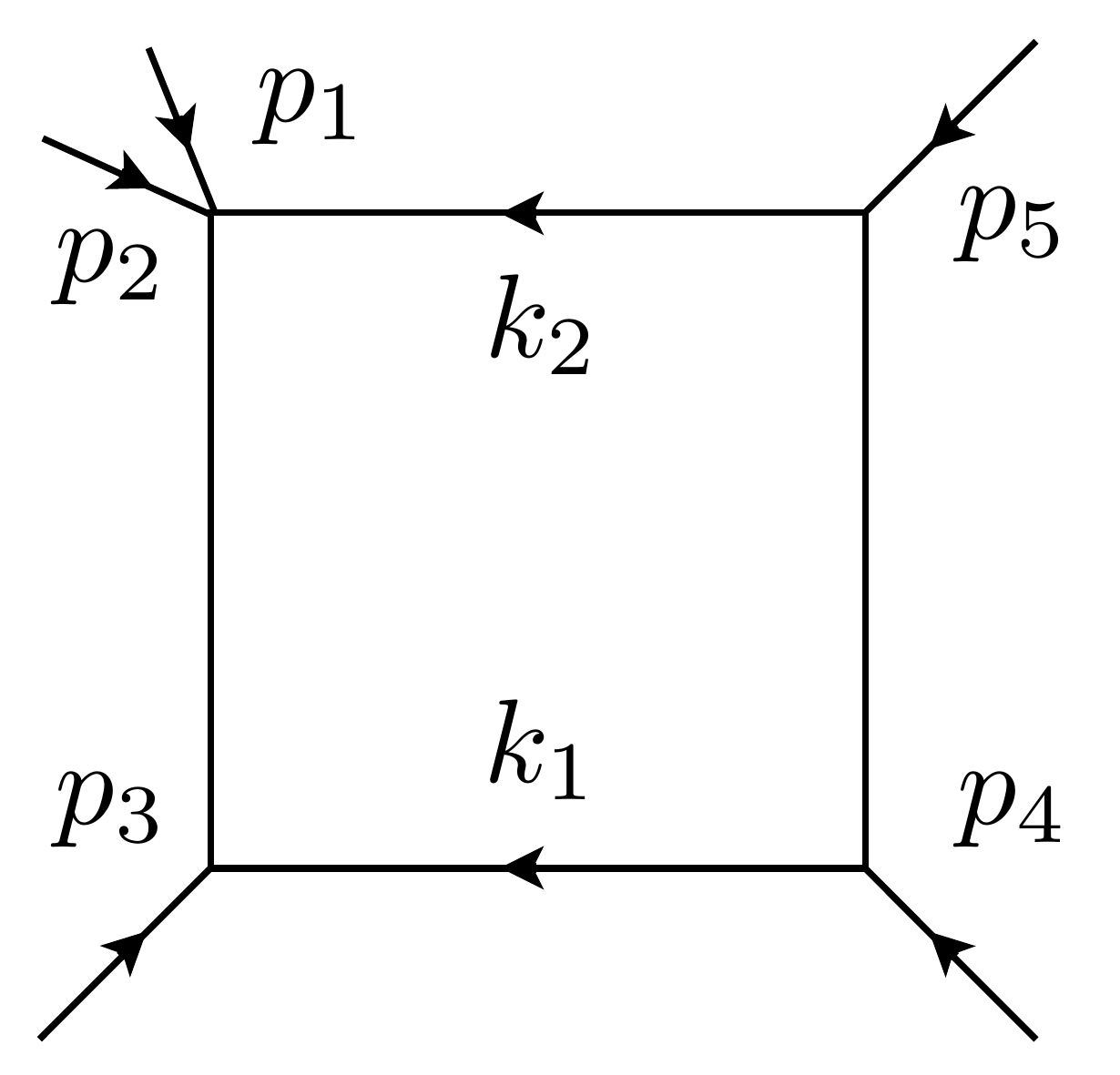}
}
\, ,
\end{align}
where we use the Mandelstam invariants $s_{ij} = (p_1 + p_j)^2$. The remaining two missing cyclically symmetric contributions
arise from the complementary $s_{234}$ cut. Its calculation is analogous to the above. Two-loop analysis along these lines
were presented Refs.\ \cite{Bern:1997it,Bern:2006vw,Cachazo:2008vp,Carrasco:2011mn}.

Notice that we have not attempted to simplify the algebra involved in intermediate calculations since the form of the MHV
amplitudes is quite concise to treat them in the most general form. This will not be quite the case in the analysis with the 
six-dimensional setup which will be performed in Sect.~\ref{6D5LegSection} in the main body of the paper. In the latter case, 
it will be convenient to choose a specific component of the five-leg amplitude, which will minimize necessary calculations. 
The component that does the job is the four-scalar--gluon amplitude
\begin{align}
\mathcal{A}_5
=
\eta^1_1 \eta^2_1 \eta^1_2 \eta^2_2 \eta^3_4 \eta^4_4 \eta^3_5 \eta^4_5
\vev{\phi^{12}_1 \phi^{12}_2 g^+_3 \phi^{34}_4 \phi^{34}_5} + \dots
\, ,
\end{align}
(where the ellipses stand for other contributions that will not be of our concern) with its tree-level expression being
\begin{align}
\label{4Dscalargluon}
\vev{\phi^{12}_1 \phi^{12}_2 g^+_3 \phi^{34}_4 \phi^{34}_5}^{(0)}
=
\frac{\vev{12}^2 \vev{45}^2}{\vev{12}\vev{23} \dots \vev{51}}
\, .
\end{align}
This particular choice will select only one contribution from superamplitudes in certain channels and is akin to the so-called singlet 
one used to evaluate loop integrand in massless gauge theories \cite{Bern:2007ct} where only gluon intermediate states survive. 
We will also dub our choice of the channel as the `singlet'.

\section{On-shell collinear limit}
\label{CollinearLimitAppendix}

Here, we recall results for the collinear behavior of individual one-loop integrals $I_i$ when the external lines are placed strictly on 
the mass shell 
\begin{align}
I_{0;i} = \lim_{m = 0} I_i
\, ,
\end{align}
and the invariant mass $s_{45} \to 0$
\begin{align}
\mathcal{J}_{0;i} = \mbox{col-lim}_{45} [I_{0;i}]
\, .
\end{align}
This will explicitly demonstrate what leading regions of the loop momentum define splitting amplitudes, which will be reviewed in the 
subsequent Appendix \ref{4DSplittingAmplitudeAppendix}.

Starting from the integral $I_{0;1}$ and applying MofR, we find that it possesses one hard and one collinear region. They read
\begin{align}
\label{I1MasslessS45Lim}
\mathcal{J}_{0;1}^{\rm h} = {\rm Box}_0 [s_{12}, s_{23}] =
- \frac{c_\varepsilon}{s_{12} s_{23}} V^{(1)}_4 + O (\varepsilon)
\, , \qquad
\mathcal{J}_{0;1}^{\rm c45}
&
= -
\frac{c_\varepsilon}{s_{12} s_{23}} \left( - \frac{\mu^2}{s_{45}} \right)^\varepsilon\frac{2}{\varepsilon^2}
\, .
\end{align}
The latter arises due to the possibility for the loop momentum to become collinear\footnote{Notice that MofR does not detect the other
three collinear regions when the loop momentum is collinear to either $p_1$, $p_2$ or $p_3$. The reason for this is that MofR
operates on dimensionally-regularized integrals, and these, in the strictly massless limit, are given by scaleless integrals which vanish.
Notrivial contributions from these arise solely from counterterms, as explained in Ref.\ \cite{Aybat:2006mz}.} with the external near-null 
momentum $(p_4 + p_5)$. Here, $V^{(1)}_4$ is determined by the massless box function and defines, in the nomenclature of Ref.\ 
\cite{Bern:1994zx}, the four-leg on-shell amplitude, see Eq.\ \re{V4}.

The $I_{0;2}$ graph is solely determined by the collinear region when the loop momentum is collinear to $p_4$,
\begin{align}
\mathcal{I}_{0;2}^{\rm c4}
=
\frac{c_\varepsilon}{s_{34} s_{45}} \left( - \frac{\mu^2}{z s_{45}} \right)^\varepsilon
\left[
\frac{2}{\varepsilon^2} + 2 {\rm Li}_2 (\bar{z}) + O (\varepsilon)
\right]
\, ,
\end{align}
and so does the $I_{0;3}$ when $k^\mu \sim p_5^\mu$
\begin{align}
\mathcal{I}_{0;3}^{\rm c5}
=
\frac{c_\varepsilon}{s_{45} s_{51}} \left( - \frac{\mu^2}{\bar{z} s_{45}} \right)^\varepsilon
\left[
\frac{2}{\varepsilon^2} + 2 {\rm Li}_2 (z) + O (\varepsilon)
\right]
\, .
\end{align}
Here and below, we do not impose collinear kinematics on the overall Mandelstam denominators since they cancel in the amplitude 
\re{RatioM5} with corresponding accompanying numerators.

The remaining two graphs $I_{0;4}$ and $I_{0;5}$ are `nonsingular' in the $s_{45} \to 0$ limit are contain only one hard region each.
The latter can be expressed in terms of the one-mass box. They possess only one hard region. The latter is given by the one-mass 
box that already made its stage appearance in Eq.\ \re{1LoopHard},
\begin{align}
\mathcal{I}^{\rm h}_{0;4} = {\rm Box}_1 [s_{51}, s_{12}; s_{34}] 
\, , \qquad
\mathcal{I}^{\rm h}_{0;5} = {\rm Box}_1 [s_{23}, s_{34}; s_{51}] 
\, .
\end{align}
Then, we find the collinear limit of the last two graphs gives the massless box
\begin{align}
\bar{z}
{\rm Box}_1 [\bar{z} s_{23}, s_{12}; z s_{12}] 
+
z
{\rm Box}_1 [s_{23}, z s_{12}; \bar{z} s_{23}] 
=
{\rm Box}_0 [s_{12}, s_{23}]
\, .
\end{align}

The sum of all hard regions gives twice the four-leg one-loop amplitude
\begin{align}
s_{12} s_{23} \mathcal{I}^{\rm h}_{0;1} 
+ 
s_{51} s_{12} \mathcal{I}^{\rm h}_{0;4} 
+ 
s_{23} s_{34} \mathcal{I}^{\rm h}_{0;5} = - 2 c_\varepsilon V^{(1)}_4
\, , 
\end{align}
while the collinear regions define the massless splitting amplitude
\begin{align}
s_{12} s_{23}  \mathcal{I}_{0;1}^{\rm c45}
+
s_{34} s_{45} \mathcal{I}_{0;2}^{\rm c4}
+
s_{45} s_{51} \mathcal{I}_{0;3}^{\rm c5}
= -  2 c_\varepsilon r^{(1)}_{\rm h} (s_{45}, z)
\, ,
\end{align}
where the one-loop massless splitting amplitude ratio is \cite{Bern:1994zx,Kosower:1999xi,Kosower:1999rx}
\begin{align}
\label{Massless1loopSplittingAmplitude}
r^{(1)}_{\rm h}  (s_{45}, z)
=
- \frac{1}{\varepsilon^2} \left( - \frac{\mu^2}{z \bar{z} s_{45}} \right)^\varepsilon + 2 \log z \log\bar{z} - \zeta_2
\, .
\end{align}
Notice that the $s_{45}$-collinear divergence stems from the hard regions of the original five-leg graphs for generic values of
the Maldestam invariants. There are no further sources for them.

\section{Four-dimensional gluon splitting amplitude}
\label{4DSplittingAmplitudeAppendix}

Let us review here a setup for direct calculations of on-shell splitting amplitudes. It was first introduced by Kosower a quarter
of a century ago in Ref.\ \cite{Kosower:1999xi} and applied to a one-loop calculation in \cite{Kosower:1999rx}. However, in 
the latter, the authors used a hybrid approach with some states treated covariantly (internal) and some using spinor-helicity 
(external) to be able to extract $D$-dimensional integrands. Presently, we will do it in a uniform manner by solely using 
four-dimensional spinor-helicity since it suffices in $\mathcal{N} = 4$ sYM: rational terms are not expected to appear. This will 
be the `jump board' for the subsequent six-dimensional extension, addressed in Sect.\ \ref{6DSplitSection}. Since all amplitudes 
in the $\mathcal{N} = 4$ sYM are related by supersymmetry, it will suffice for our goal to consider merely a single component of 
the supermultiplet. We will choose the transition $g^\pm \to g^\mp g^\mp$ for this purpose due to the simplicity of its unitarity cuts.

All of the ingredients for the unitarity reconstruction of integrands have been worked out in the previous sections, we  just need 
one more ingredient, the vertex. Let us use the nonlinear Gervais-Neveu gauge \cite{Gervais:1972tr}, where the three-gluon amplitude 
contracted with polarization vectors $\varepsilon_j$ of external gluons reads
\begin{align}
\label{4Dvertex}
V_{123} 
= 
(\varepsilon_1 \cdot \varepsilon_2) (\varepsilon_3 \cdot p_1)
+
(\varepsilon_2 \cdot \varepsilon_3) (\varepsilon_1 \cdot p_2)
+
(\varepsilon_3 \cdot \varepsilon_1) (\varepsilon_2 \cdot p_3)
\, .
\end{align}
It enjoys only cyclic rather than permutation symmetry. All momenta are incoming and obey the conservation law $p_1 + p_2 + p_3 = 0$. 
The legs 1 and 2 will be treated as on-shell. The polarization vectors for positive and negative helicities are given by
\begin{align}
\label{OnShellPolar}
\varepsilon_j^+ \equiv \varepsilon^+ (p_j) = \frac{| q \rangle [ j |}{\langle q j \rangle}
\, , \qquad
\varepsilon_j^- \equiv \varepsilon^- (p_j) = \frac{| q ] \langle j |}{[q j]}
\, ,
\end{align}
with the help of an auxiliary light-like vector $q = | q ] \langle q |$. For simplicity of our analysis, we chose the same 
gauge-fixing vector $q$ for all gluons. 

The tree-level $g^+ (-p_3) \to g^- (p_1) g^- (p_2)$ splitting amplitude then reads
\begin{align}
\label{4DTreeSplit}
{\rm Split}^{(0)}_{+ ; --} (p_3; p_1, p_2)
\equiv \frac{2  V_{--+}}{s_{12}} 
= 
\frac{(p_3 \cdot q)}{(p_1 \cdot q) (p_2 \cdot q)} \frac{\langle 2 | q \varepsilon_3^+ | 1 \rangle}{[21]}
=
\frac{1}{[12] \sqrt{z \bar z}}
\, , \nonumber
\end{align} 
where we used the near-collinear approximation 
\begin{align}
p_1 \to - z p_3
\, , \qquad
p_2 \to - \bar{z} p_3 
\, .
\end{align}
This is a well-known expression \cite{Parke:1986gb,Mangano:1990by,Dixon:1996wi}. In the last step in Eq.\ \re{4DTreeSplit},
we ignored the negligible effect of the off-shellness of the third leg and used the on-shell expression for its polarization vector.

Let us proceed to one loop. We will determine the integrand from the two-particle unitarity cut shown in Fig.\ \ref{1LoopSplitPic}.
It reads
\begin{align}
\mathcal{S}^{(1)}_{+ ; --} (p_3; p_1, p_2)|_{s_{12}-{\rm cut}}
=
{\rm Split}^{(0)}_{+ ; --} (p_3; - k_2, - k_1)
\widehat{A}_{--++}^{(0)\rm MHV} (p_1, p_2, k_1, k_2) 
\, ,
\end{align} 
with the 4-gluon Parke-Taylor MHV amplitude
\begin{align}
\widehat{A}_{--++}^{(0)\rm MHV} (p_1, p_2, k_1, k_2) 
=
\frac{[k_1 k_2]^3}{[12][2 k_1][k_2 1]}
\, .
\end{align}
Converting products of spinor brackets into strings of Dirac matrices, we find
\begin{align}
\mathcal{S}^{(1)}_{+ ; --} (p_3; p_1, p_2)|_{s_{12}-{\rm cut}}
=
\frac{(p_3 \cdot q)}{8 (k_1 \cdot q) (k_2 \cdot q) (k_1 \cdot p_2) (k_2 \cdot p_1)} \frac{\langle 2 | k_1 k_2 \varepsilon_3^+ q k_1 k_2 | 1 \rangle}{[21]}
\, ,
\end{align}
Finally, to extract the integrand, we need to project out its overlap on the spinor structure of the tree amplitude, i.e., 
\begin{align}
\langle 2 | k_1 k_2 \varepsilon_3^+ q k_1 k_2 | 1 \rangle
=
\langle 2 | q \varepsilon_3^+ | 1 \rangle
\rho
\, .
\end{align}
To this end, we can use the on-shell form of the polarization vector $\varepsilon_3$ \re{OnShellPolar}, since the difference is
not sufficiently singular to induce a non-vanishing effect in the splitting amplitude \cite{Bern:2004cz}, such that
\begin{align}
\rho
&
=
\frac{\tr_2 [p_1 q k_1 k_2] \tr_2 [q p_2 k_1 k_2]}{\tr_2 [p_1 q] \tr_2 [p_2 q]}
\nonumber\\
&
=
\frac{(k_2 + p_1)^2 (p_3 \cdot q)}{2 (p_1 \cdot q) (p_2 \cdot q)}
\left[
(p_3 \cdot q) (k_2 + p_1)^2 + (p_3 \cdot q) s_{12} - (k_2 \cdot q) s_{12} z - (k_1 \cdot q) s_{12} \bar{z}
\right]
\, .
\end{align}
Here, in the last step, we used Eq.\ \re{2Dtraces}, which can be readily automated with {\tt FeynCalc} by converting all two-dimensional
traces to chiral four-dimensional ones. The odd-parity structure proper was ignored in the above result since there is not a sufficient 
number of four-vectors to warrant a non-trivial effect after the loop-momentum integration. However, the product of parity-odd 
contributions stemming from the two traces in the numerator is crucial to getting the above (correct) result. 

Restoring the cut propagators, the integrand for the splitting amplitude in question is
\begin{align}
\label{4D1loopSplitIntegrand}
\mathcal{S}^{(1)}_{+ ; --} (p_3; p_1, p_2)
&=
{\rm Split}^{(0)}_{+ ; --}
\left[
\frac{1}{(k_1 \cdot q)} + \frac{1}{(k_2 \cdot q)}
\right]
\nonumber\\
&\times
\frac{
(p_3 \cdot q) (k_2 + p_1)^2 + (p_3 \cdot q) s_{12} - (k_2 \cdot q) s_{12} z - (k_1 \cdot q) s_{12} \bar{z}
}{2 k_1^2 k_2^2 (k_2 + p_1)^2}
\, ,
\end{align}
and the one-loop splitting amplitude reads
\begin{align}
\label{1loopSplitting}
{\rm Split}^{(1)}_{+ ; --}
=
-
\frac{1}{2}
{\rm Split}^{(0)}_{+ ; --}
\left[
J_{0;1} + z s_{12} J_{0;2} + (\bar z - z) s_{12} J_{0;3} + ({z \leftrightarrow \bar{z}})
\right]
\, ,
\end{align}
in terms of the Feynman integrals (use momentum conservation to rewrite $- k_1 = k_2 + p_1 + p_2$)
\begin{align}
\label{MasslessIntSplitting}
J_{0;1} 
&= \int_{k_2} \frac{(p_3 \cdot q)}{k_1^2 k_2^2 (k_2 \cdot q)}
\, ,\\
J_{0;2}
&= \int_{k_2} \frac{(p_3 \cdot q)}{k_1^2 k_2^2 (k_2 + p_1)^2(k_2 \cdot q)}
\, ,\\
J_{0;3} 
&= \int_{k_2} \frac{1}{k_1^2 k_2^2 (k_2 + p_1)^2}
\, .
\end{align}
The first two integrals contribute in the same fashion as in the result from Ref.\ \cite{Kosower:1999rx} for $g^+ \to g^+ g^-$ amplitude. 
However, we did not need to perform a Passarino-Veltman reduction since we did not encounter any tensor integrals. The contribution 
of the $z$-independent triangle $J_{0;3}$ cancels in the final expression due to the antisymmetry of the accompanying coefficient.
These integrals do not contain any soft scales and are all hard in MofR terminology. Their calculation reproduces Eq.\ 
\re{Massless1loopSplittingAmplitude}. This concludes our re-analysis of the on-shell case.

\section{Six-dimensional gluon splitting amplitude}
\label{6DSplitComponentsAppendix}

In this Appendix, we present a component derivation for the one-loop integrand of the gluon splitting amplitude in six dimensions.
The $S_{12}$-channel unitarity cut, shown in Fig.\ \ref{1LoopSplitPic}, yields for the one-loop integrand
\begin{align}
\mathcal{S}^{(1)} (P_3; P_1, P_2)|_{S_{12}-{\rm cut}}
=
{\rm Split}^{(0)} (P_3; - K_2, - K_1)
\widehat{A}_4^{(0)} (P_1, P_2, K_1, K_2) 
\, ,
\end{align} 
with the amputated 4-gluon amplitude
\begin{align}
\widehat{A}_4^{(0)} (P_1, P_2, K_1, K_2) 
=
\frac{\vev{1_a 2_b K_{1,c} K_{2,d}} [1_{\dot{a}} 2_{\dot{b}} K_{1,\dot{c}} K_{2,\dot{d}}]}{(P_1 \cdot P_2)(P_1 \cdot K_2)}
\, ,
\end{align}
which is defined in terms of the chiral and antichiral four-brackets \cite{Cheung:2009dc}
\begin{align}
\vev{1_a 2_b 3_c 4_d} 
\equiv 
\varepsilon_{ABCD} \Lambda^A_{1, a} \Lambda^B_{2, b} \Lambda^C_{3, c} \Lambda^D_{4, d}
\, , \qquad
[1_{\dot{a}} 2_{\dot{b}} 3_{\dot{c}} 4_{\dot{d}}] 
\equiv 
\varepsilon^{ABCD} \bar\Lambda_{1, A, \dot{a}}  \bar\Lambda_{2, B, \dot{b}}  \bar\Lambda_{3, C, \dot{c}}  \bar\Lambda_{4, D, \dot{d}} 
\, .
\end{align}

The six-dimensional algebra is strikingly different from its four-dimensional counterpart and it is, therefore, instructive to
provide an example. Notice that the superspace derivation given in the main text was far less laborious! Let us perform 
a detailed calculation for the second term in the square brackets of the vertex \re{6Dvertex}. Coupling it with the amplitude 
$\widehat{A}_4^{(0)}$, we can re-arrange this expression using the momentum conservation as
\begin{align}
T_2 \equiv
\langle K_1^c | \bar{\mathcal{E}}_3 \mathcal{Q} | K_1^{\dot{c}}] [K_2^{\dot{d}} | \mathcal{Q} \bar{K}_1 | K_2^{d} \rangle 
&
\vev{1_a 2_b K_{1,c} K_{2,d}} [1_{\dot{a}} 2_{\dot{b}} K_{1,\dot{c}} K_{2,\dot{d}}]
\\
=
&
- \varepsilon_{ABCD} \Lambda^A_{1, a} \Lambda^B_{2, b}  (P_1+P_2)^{DD'} (\bar{K}_1 \mathcal{Q} \bar{K}_2)_{D'H}
\nonumber\\
&
\times
\varepsilon^{EFGH} \bar\Lambda_{1, A, \dot{a}}  \bar\Lambda_{2, B, \dot{b}} (K_1 \bar{\mathcal{E}} \mathcal{Q} \bar{K}_1)^C{}_G
\, . \nonumber
\end{align}
The Levi-Civita tensor in the first line after the equality sign can be eliminated by making use of the duality condition \re{6Dduality} for 
the six-dimensional Dirac matrices such that
\begin{align}
\varepsilon_{ABCD} \langle 1_a|^A \langle 2_b|^B  (P_1+P_2)^{DD'}
=
\langle 1_a| \bar{P}_2)_C \langle 2_b|^{D'}
-
\langle 2_b| \bar{P}_1)_C \langle 1_a|^{D'}
\, .
\end{align}
This provides the intermediate result
\begin{align}
T_2
&
=
\varepsilon^{EFGH} \bar\Lambda_{1, A, \dot{a}}  \bar\Lambda_{2, B, \dot{b}} \bar{K}_{1,GG'}\bar{K}_{2,HH'}
\nonumber\\
&
\times
\left[
\langle 2_b| \bar{P}_1 K_1 \bar{\mathcal{E}}_3 \mathcal{Q} )^{G'} \langle 1_a| \bar{K}_1 \mathcal{Q})^{H'}
-
\langle 1_a| \bar{P}_2 K_1 \bar{\mathcal{E}}_3 \mathcal{Q} )^{G'} \langle 2_b| \bar{K}_1 \mathcal{Q})^{H'}
\right]
\, . 
\end{align}
The elimination of the second Levi-Civita tensor can be performed in two complementary ways, either solving $K_1$ or
$K_2$ via the momentum conservation condition. We will choose the former since it will generate expressions with
external momentum vectors adjacent to the spinors of external states. A short calculation produces
\begin{align}
\varepsilon^{EFGH} | 1_{\dot{a}}]_A [ 2_{\dot{b}} |_B \bar{K}_{1,GG'}\bar{K}_{2,HH'}
=
[ 2_{\dot{b}} | K_2 | 1_{\dot{a}} ] \bar{K}_{2, G'H'}
-
[2_{\dot{b}} | P_1 \bar{K}_2)_{H'} | 1_{\dot{a}}]_{G'}
+
[1_{\dot{a}} | P_2 \bar{K}_2)_{H'} | 2_{\dot{b}}]_{G'}
\, .
\end{align}
Assembling everything together, we decompose $T_2$ as
\begin{align}
T_2
= T_{2}^{\circ} + T_{2}^{\bullet}
\end{align}
and obtain the following two structures 
\begin{align}
T_{2}^{\circ} 
&
= [1_{\dot{a}} | K_2 | 2_{\dot{b}}]
\langle 1_{a} | \bar{P}_2 K_1 \bar{\mathcal{E}}_3 \mathcal{Q} \bar{K}_2 \mathcal{Q} \bar{K}_1 | 2_{b} \rangle
+ (1\leftrightarrow 2)
\, , \\
T_{2}^{\bullet} 
&=
[2_{\dot{b}} | (P_1 + P_2) \bar{K}_2 \mathcal{Q} \bar{K}_1
\big[ |2_b\rangle \langle 1_a| - |1_a\rangle \langle 2_b| \big]
(\bar{P}_1 + \bar{P}_2) K_1 \bar{\mathcal{E}}_3 \mathcal{Q} |1_{\dot{a}}]
+ (1\leftrightarrow 2)
\, ,
\end{align}
where the exchange  $(1\leftrightarrow 2)$ also implies the interchange of the accompanying little group indices of legs
with momenta $P_1$ and $P_2$. The empty and full circle superscripts indicate contributions with odd and even
numbers of Dirac matrices sandwiched between spinors of the first and second gluon. Its significance will become 
transparent momentarily.

The contribution from the last term in Eq.\ \re{6Dvertex} is obtained by the exchange of the lines with $K_1$ and $K_2$
momenta such that (the accompanying minus sign is included)
\begin{align}
T_3 = - T_2|_{K_1 \leftrightarrow K_2}
\, .
\end{align}
Finally, the first term is given by a factorized product of chiral and antichiral Dirac strings
\begin{align}
T_1 \equiv T_1^{\circ}
= \tr_4 [ \bar{\mathcal{E}}_3 P_1]
&
\langle 1_a | \bar{\mathcal{Q}} K_2 \bar{P}_1 + \bar{P}_2 K_2 \bar{\mathcal{Q}} - \bar{K}_2 \tr_4[\bar{\mathcal{Q}} K_2] | 2_b \rangle
\\
\times
&
[ 1_{\dot{a}} | {\mathcal{Q}} \bar{K}_2 {P}_1 + {P}_2 \bar{K}_2 {\mathcal{Q}} - {K}_2 \tr_4[{\mathcal{Q}} \bar{K}_2] | 2_{\dot{b}} ]
\, . \nonumber
\end{align}

Now, we are in a position to construct the integrand for the off-shell decay $+ \to --$. In fact, we will deduce it for its parity 
conjugate counterpart. The positive helicity W-bosons from the four-dimensional perspective corresponds to little group 
indices $1\dot{1} \to +$, while the negative one to $2\dot{2} \to -$. Then from Eq.\ \re{LambdaFromlambdas} and the 
off-diagonal-block nature of the six-dimensional Dirac matrices, one immediately observes that all contributions $T^\circ_i$ 
are proportional to positive powers of the off-shellness $w$ since they have an odd number of gamma matrices between 
states of the same helicity. In order to enforce it exactly (i.e., being exactly zero), we can without loss of generality from the 
four-dimensional perspective, choose the null six-vector $\mathcal{Q}^M$ to reside in the four-dimensional 
subspace\footnote{More on this choice, see Sect.~\ref{QChoiceSection}.} $\mathcal{Q}^M = (q^\mu, 0,0)$ and moreover to 
identify it with the null vector $q = |q]\langle q|$ which is used to define the virtuality portion of four-momenta of massive states 
\re{OffShellMomentum}. This condition then results in identities like
\begin{align}
\langle 1_1 | \bar{\mathcal{Q}} |2_{1} \rangle = 0
\, .
\end{align}
In this manner, we find in terms of four-dimensional spinor-helicity
\begin{align}
T_2^{\bullet}
=
[21] [1| k_2 k_1 \varepsilon_3 q (p_1 + p_2) k_2 q k_1 |2] + (1 \leftrightarrow 2)
\, .
\end{align}
Projecting this down on the tree-level structure
\begin{align}
T_2^{\bullet}
=
[1 | \varepsilon_3 q |2] R_2
\end{align}
for negative-helicity third W-boson \re{OnShellPolar}, we find
\begin{align}
R_2
&
= 
[21]
\frac{\tr_2 [p_1 k_2 k_1 q] \tr_2 [q (p_1 + p_2) k_2 q k_1 p_2]
-
\tr_2 [p_2 k_2 k_1 q] \tr_2 [q (p_1 + p_2) k_2 q k_1 p_1]}{\tr_2 [p_1 q] \tr_2 [p_2 q]}
\\
&
=
[21]
\frac{(K_1 \cdot \mathcal{Q}) (P_3 \cdot \mathcal{Q})}{(P_1 \cdot \mathcal{Q}) (P_2 \cdot \mathcal{Q})} S_{12}
\left[
(P_3 \cdot \mathcal{Q}) (K_2 + P_1)^2 + (K_2 \cdot \mathcal{Q})  S_{12} (\bar{z} - z) + (P_3 \cdot \mathcal{Q})  S_{12} z
\right]
\, .
\end{align}
Adding to this $T_3^\bullet$, we get the numerator of the integrand
\begin{align}
\label{SumGluonComponentSplit6D}
T_2^{\bullet}
+
T_3^{\bullet}
&
=
[21]
\frac{(P_3 \cdot \mathcal{Q})^2}{(P_1 \cdot \mathcal{Q}) (P_2 \cdot \mathcal{Q})} S_{12}
\\
&\times
\left[
(P_3 \cdot \mathcal{Q}) (K_2 + P_1)^2 + (P_3 \cdot \mathcal{Q}) S_{12} - (K_2 \cdot \mathcal{Q}) S_{12} z - (K_1 \cdot \mathcal{Q}) S_{12} \bar{z}
\right]
. \nonumber
\end{align}
Multiplying this by the six-dimensional propagators and dimensionally reducing it down to four dimensions in a manner by
treating the out-of-four-dimensional components as masses, we conclude that the integrand is the same as in the purely 
massless case but now with external legs possessing nonvanishing virtuality.


\begin{thebibliography}{100}

\bibitem{Collins:1989gx}
J.~C. Collins, D.~E. Soper and G.~F. Sterman, \emph{{Factorization of Hard
  Processes in QCD}},
  \href{http://dx.doi.org/10.1142/9789814503266_0001}{\emph{Adv. Ser. Direct.
  High Energy Phys.} {\bf 5} (1989) 1--91},
  [\href{https://arxiv.org/abs/hep-ph/0409313}{{\tt hep-ph/0409313}}].

\bibitem{Agarwal:2021ais}
N.~Agarwal, L.~Magnea, C.~Signorile-Signorile and A.~Tripathi, \emph{{The
  infrared structure of perturbative gauge theories}},
  \href{http://dx.doi.org/10.1016/j.physrep.2022.10.001}{\emph{Phys. Rept.}
  {\bf 994} (2023) 1--120}, [\href{https://arxiv.org/abs/2112.07099}{{\tt
  2112.07099}}].

\bibitem{Parke:1986gb}
S.~J. Parke and T.~R. Taylor, \emph{{An Amplitude for $n$ Gluon Scattering}},
  \href{http://dx.doi.org/10.1103/PhysRevLett.56.2459}{\emph{Phys. Rev. Lett.}
  {\bf 56} (1986) 2459}.

\bibitem{Berends:1987me}
F.~A. Berends and W.~T. Giele, \emph{{Recursive Calculations for Processes with
  n Gluons}}, \href{http://dx.doi.org/10.1016/0550-3213(88)90442-7}{\emph{Nucl.
  Phys. B} {\bf 306} (1988) 759--808}.

\bibitem{Mangano:1990by}
M.~L. Mangano and S.~J. Parke, \emph{{Multiparton amplitudes in gauge
  theories}}, \href{http://dx.doi.org/10.1016/0370-1573(91)90091-Y}{\emph{Phys.
  Rept.} {\bf 200} (1991) 301--367},
  [\href{https://arxiv.org/abs/hep-th/0509223}{{\tt hep-th/0509223}}].

\bibitem{Bern:1994zx}
Z.~Bern, L.~J. Dixon, D.~C. Dunbar and D.~A. Kosower, \emph{{One loop n point
  gauge theory amplitudes, unitarity and collinear limits}},
  \href{http://dx.doi.org/10.1016/0550-3213(94)90179-1}{\emph{Nucl. Phys. B}
  {\bf 425} (1994) 217--260}, [\href{https://arxiv.org/abs/hep-ph/9403226}{{\tt
  hep-ph/9403226}}].

\bibitem{Kosower:1999xi}
D.~A. Kosower, \emph{{All order collinear behavior in gauge theories}},
  \href{http://dx.doi.org/10.1016/S0550-3213(99)00251-5}{\emph{Nucl. Phys. B}
  {\bf 552} (1999) 319--336}, [\href{https://arxiv.org/abs/hep-ph/9901201}{{\tt
  hep-ph/9901201}}].

\bibitem{Kosower:1999rx}
D.~A. Kosower and P.~Uwer, \emph{{One loop splitting amplitudes in gauge
  theory}}, \href{http://dx.doi.org/10.1016/S0550-3213(99)00583-0}{\emph{Nucl.
  Phys. B} {\bf 563} (1999) 477--505},
  [\href{https://arxiv.org/abs/hep-ph/9903515}{{\tt hep-ph/9903515}}].

\bibitem{Bern:2004cz}
Z.~Bern, L.~J. Dixon and D.~A. Kosower, \emph{{Two-loop g ---\ensuremath{>} gg
  splitting amplitudes in QCD}},
  \href{http://dx.doi.org/10.1088/1126-6708/2004/08/012}{\emph{JHEP} {\bf 08}
  (2004) 012}, [\href{https://arxiv.org/abs/hep-ph/0404293}{{\tt
  hep-ph/0404293}}].

\bibitem{Badger:2004uk}
S.~D. Badger and E.~W.~N. Glover, \emph{{Two loop splitting functions in QCD}},
  \href{http://dx.doi.org/10.1088/1126-6708/2004/07/040}{\emph{JHEP} {\bf 07}
  (2004) 040}, [\href{https://arxiv.org/abs/hep-ph/0405236}{{\tt
  hep-ph/0405236}}].

\bibitem{Guan:2024hlf}
X.~Guan, F.~Herzog, Y.~Ma, B.~Mistlberger and A.~Suresh, \emph{{Splitting
  amplitudes at N$^3$LO in QCD}},  \href{https://arxiv.org/abs/2408.03019}{{\tt
  2408.03019}}.

\bibitem{Dixon:1996wi}
L.~J. Dixon, \emph{{Calculating scattering amplitudes efficiently}},  in
  \emph{{Theoretical Advanced Study Institute in Elementary Particle Physics
  (TASI 95): QCD and Beyond}}, pp.~539--584, 1, 1996.
\newblock \href{https://arxiv.org/abs/hep-ph/9601359}{{\tt hep-ph/9601359}}.

\bibitem{Bern:2007dw}
Z.~Bern, L.~J. Dixon and D.~A. Kosower, \emph{{On-Shell Methods in Perturbative
  QCD}}, \href{http://dx.doi.org/10.1016/j.aop.2007.04.014}{\emph{Annals Phys.}
  {\bf 322} (2007) 1587--1634}, [\href{https://arxiv.org/abs/0704.2798}{{\tt
  0704.2798}}].

\bibitem{Sen:1982bt}
A.~Sen, \emph{{Asymptotic Behavior of the Wide Angle On-Shell Quark Scattering
  Amplitudes in Nonabelian Gauge Theories}},
  \href{http://dx.doi.org/10.1103/PhysRevD.28.860}{\emph{Phys. Rev. D} {\bf 28}
  (1983) 860}.

\bibitem{Catani:1998bh}
S.~Catani, \emph{{The Singular behavior of QCD amplitudes at two loop order}},
  \href{http://dx.doi.org/10.1016/S0370-2693(98)00332-3}{\emph{Phys. Lett. B}
  {\bf 427} (1998) 161--171}, [\href{https://arxiv.org/abs/hep-ph/9802439}{{\tt
  hep-ph/9802439}}].

\bibitem{Sterman:2002qn}
G.~F. Sterman and M.~E. Tejeda-Yeomans, \emph{{Multiloop amplitudes and
  resummation}},
  \href{http://dx.doi.org/10.1016/S0370-2693(02)03100-3}{\emph{Phys. Lett. B}
  {\bf 552} (2003) 48--56}, [\href{https://arxiv.org/abs/hep-ph/0210130}{{\tt
  hep-ph/0210130}}].

\bibitem{Aybat:2006mz}
S.~M. Aybat, L.~J. Dixon and G.~F. Sterman, \emph{{The Two-loop soft anomalous
  dimension matrix and resummation at next-to-next-to leading pole}},
  \href{http://dx.doi.org/10.1103/PhysRevD.74.074004}{\emph{Phys. Rev. D} {\bf
  74} (2006) 074004}, [\href{https://arxiv.org/abs/hep-ph/0607309}{{\tt
  hep-ph/0607309}}].

\bibitem{Dixon:2008gr}
L.~J. Dixon, L.~Magnea and G.~F. Sterman, \emph{{Universal structure of
  subleading infrared poles in gauge theory amplitudes}},
  \href{http://dx.doi.org/10.1088/1126-6708/2008/08/022}{\emph{JHEP} {\bf 08}
  (2008) 022}, [\href{https://arxiv.org/abs/0805.3515}{{\tt 0805.3515}}].

\bibitem{Gardi:2009qi}
E.~Gardi and L.~Magnea, \emph{{Factorization constraints for soft anomalous
  dimensions in QCD scattering amplitudes}},
  \href{http://dx.doi.org/10.1088/1126-6708/2009/03/079}{\emph{JHEP} {\bf 03}
  (2009) 079}, [\href{https://arxiv.org/abs/0901.1091}{{\tt 0901.1091}}].

\bibitem{Becher:2009qa}
T.~Becher and M.~Neubert, \emph{{On the Structure of Infrared Singularities of
  Gauge-Theory Amplitudes}},
  \href{http://dx.doi.org/10.1088/1126-6708/2009/06/081}{\emph{JHEP} {\bf 06}
  (2009) 081}, [\href{https://arxiv.org/abs/0903.1126}{{\tt 0903.1126}}].

\bibitem{Basso:2018tif}
B.~Basso and A.~V. Belitsky, \emph{{ABJM flux-tube and scattering amplitudes}},
  \href{http://dx.doi.org/10.1007/JHEP09(2019)116}{\emph{JHEP} {\bf 09} (2019)
  116}, [\href{https://arxiv.org/abs/1811.09839}{{\tt 1811.09839}}].

\bibitem{Alday:2007hr}
L.~F. Alday and J.~M. Maldacena, \emph{{Gluon scattering amplitudes at strong
  coupling}},
  \href{http://dx.doi.org/10.1088/1126-6708/2007/06/064}{\emph{JHEP} {\bf 06}
  (2007) 064}, [\href{https://arxiv.org/abs/0705.0303}{{\tt 0705.0303}}].

\bibitem{Drummond:2007cf}
J.~M. Drummond, J.~Henn, G.~P. Korchemsky and E.~Sokatchev, \emph{{On planar
  gluon amplitudes/Wilson loops duality}},
  \href{http://dx.doi.org/10.1016/j.nuclphysb.2007.11.007}{\emph{Nucl. Phys. B}
  {\bf 795} (2008) 52--68}, [\href{https://arxiv.org/abs/0709.2368}{{\tt
  0709.2368}}].

\bibitem{Brandhuber:2007yx}
A.~Brandhuber, P.~Heslop and G.~Travaglini, \emph{{MHV amplitudes in N=4 super
  Yang-Mills and Wilson loops}},
  \href{http://dx.doi.org/10.1016/j.nuclphysb.2007.11.002}{\emph{Nucl. Phys. B}
  {\bf 794} (2008) 231--243}, [\href{https://arxiv.org/abs/0707.1153}{{\tt
  0707.1153}}].

\bibitem{Corrigan:1978zg}
E.~Corrigan and B.~Hasslacher, \emph{{A Functional Equation for Exponential
  Loop Integrals in Gauge Theories}},
  \href{http://dx.doi.org/10.1016/0370-2693(79)90518-5}{\emph{Phys. Lett. B}
  {\bf 81} (1979) 181--184}.

\bibitem{Durand:1979sw}
L.~Durand and E.~Mendel, \emph{{Functional Equations for Path Dependent Phase
  Factors in {Yang-Mills} Theories}},
  \href{http://dx.doi.org/10.1016/0370-2693(79)90588-4}{\emph{Phys. Lett. B}
  {\bf 85} (1979) 241--245}.

\bibitem{Alday:2010ku}
L.~F. Alday, D.~Gaiotto, J.~Maldacena, A.~Sever and P.~Vieira, \emph{{An
  Operator Product Expansion for Polygonal null Wilson Loops}},
  \href{http://dx.doi.org/10.1007/JHEP04(2011)088}{\emph{JHEP} {\bf 04} (2011)
  088}, [\href{https://arxiv.org/abs/1006.2788}{{\tt 1006.2788}}].

\bibitem{Belitsky:2011nn}
A.~V. Belitsky, \emph{{OPE for null Wilson loops and open spin chains}},
  \href{http://dx.doi.org/10.1016/j.physletb.2012.02.027}{\emph{Phys. Lett. B}
  {\bf 709} (2012) 280--284}, [\href{https://arxiv.org/abs/1110.1063}{{\tt
  1110.1063}}].

\bibitem{Sever:2011da}
A.~Sever, P.~Vieira and T.~Wang, \emph{{OPE for Super Loops}},
  \href{http://dx.doi.org/10.1007/JHEP11(2011)051}{\emph{JHEP} {\bf 11} (2011)
  051}, [\href{https://arxiv.org/abs/1108.1575}{{\tt 1108.1575}}].

\bibitem{Basso:2013vsa}
B.~Basso, A.~Sever and P.~Vieira, \emph{{Spacetime and Flux Tube S-Matrices at
  Finite Coupling for N=4 Supersymmetric Yang-Mills Theory}},
  \href{http://dx.doi.org/10.1103/PhysRevLett.111.091602}{\emph{Phys. Rev.
  Lett.} {\bf 111} (2013) 091602}, [\href{https://arxiv.org/abs/1303.1396}{{\tt
  1303.1396}}].

\bibitem{Abbott:1980hw}
L.~F. Abbott, \emph{{The Background Field Method Beyond One Loop}},
  \href{http://dx.doi.org/10.1016/0550-3213(81)90371-0}{\emph{Nucl. Phys. B}
  {\bf 185} (1981) 189--203}.

\bibitem{Abbott:1981ke}
L.~F. Abbott, \emph{{Introduction to the Background Field Method}}, {\emph{Acta
  Phys. Polon. B} {\bf 13} (1982) 33}.

\bibitem{Gates:1983nr}
S.~J. Gates, M.~T. Grisaru, M.~Rocek and W.~Siegel, \emph{{Superspace Or One
  Thousand and One Lessons in Supersymmetry}},
  \href{https://arxiv.org/abs/hep-th/0108200}{{\tt hep-th/0108200}}.

\bibitem{Caron-Huot:2021usw}
S.~Caron-Huot and F.~Coronado, \emph{{Ten dimensional symmetry of $ \mathcal{N}
  $ = 4 SYM correlators}},
  \href{http://dx.doi.org/10.1007/JHEP03(2022)151}{\emph{JHEP} {\bf 03} (2022)
  151}, [\href{https://arxiv.org/abs/2106.03892}{{\tt 2106.03892}}].

\bibitem{Selivanov:1999ie}
K.~G. Selivanov, \emph{{An Infinite set of tree amplitudes in
  Higgs-Yang-Mills}},
  \href{http://dx.doi.org/10.1016/S0370-2693(99)00760-1}{\emph{Phys. Lett. B}
  {\bf 460} (1999) 116--118}, [\href{https://arxiv.org/abs/hep-th/9906001}{{\tt
  hep-th/9906001}}].

\bibitem{Boels:2010mj}
R.~H. Boels, \emph{{No triangles on the moduli space of maximally
  supersymmetric gauge theory}},
  \href{http://dx.doi.org/10.1007/JHEP05(2010)046}{\emph{JHEP} {\bf 05} (2010)
  046}, [\href{https://arxiv.org/abs/1003.2989}{{\tt 1003.2989}}].

\bibitem{Bork:2022vat}
L.~V. Bork, N.~B. Muzhichkov and E.~S. Sozinov, \emph{{Infrared properties of
  five-point massive amplitudes in $ \mathcal{N} $ = 4 SYM on the Coulomb
  branch}}, \href{http://dx.doi.org/10.1007/JHEP08(2022)173}{\emph{JHEP} {\bf
  08} (2022) 173}, [\href{https://arxiv.org/abs/2201.08762}{{\tt 2201.08762}}].

\bibitem{Belitsky:2022itf}
A.~V. Belitsky, L.~V. Bork, A.~F. Pikelner and V.~A. Smirnov, \emph{{Exact Off
  Shell Sudakov Form Factor in N=4 Supersymmetric Yang-Mills Theory}},
  \href{http://dx.doi.org/10.1103/PhysRevLett.130.091605}{\emph{Phys. Rev.
  Lett.} {\bf 130} (2023) 091605},
  [\href{https://arxiv.org/abs/2209.09263}{{\tt 2209.09263}}].

\bibitem{Belitsky:2023ssv}
A.~V. Belitsky, L.~V. Bork and V.~A. Smirnov, \emph{{Off-shell form factor in $
  \mathcal{N} $=4 sYM at three loops}},
  \href{http://dx.doi.org/10.1007/JHEP11(2023)111}{\emph{JHEP} {\bf 11} (2023)
  111}, [\href{https://arxiv.org/abs/2306.16859}{{\tt 2306.16859}}].

\bibitem{Belitsky:2024agy}
A.~V. Belitsky, L.~V. Bork, J.~M. Grumski-Flores and V.~A. Smirnov,
  \emph{{Three-leg form factor on Coulomb branch}},
  \href{http://dx.doi.org/10.1007/JHEP11(2024)169}{\emph{JHEP} {\bf 11} (2024)
  169}, [\href{https://arxiv.org/abs/2402.18475}{{\tt 2402.18475}}].

\bibitem{Belitsky:2024dcf}
A.~V. Belitsky and L.~V. Bork, \emph{{Off-shell minimal form factors}},
  \href{https://arxiv.org/abs/2411.16941}{{\tt 2411.16941}}.

\bibitem{Cheung:2009dc}
C.~Cheung and D.~O'Connell, \emph{{Amplitudes and Spinor-Helicity in Six
  Dimensions}},
  \href{http://dx.doi.org/10.1088/1126-6708/2009/07/075}{\emph{JHEP} {\bf 07}
  (2009) 075}, [\href{https://arxiv.org/abs/0902.0981}{{\tt 0902.0981}}].

\bibitem{Dennen:2009vk}
T.~Dennen, Y.-t. Huang and W.~Siegel, \emph{{Supertwistor space for 6D maximal
  super Yang-Mills}},
  \href{http://dx.doi.org/10.1007/JHEP04(2010)127}{\emph{JHEP} {\bf 04} (2010)
  127}, [\href{https://arxiv.org/abs/0910.2688}{{\tt 0910.2688}}].

\bibitem{Caron-Huot:2010nes}
S.~Caron-Huot and D.~O'Connell, \emph{{Spinor Helicity and Dual Conformal
  Symmetry in Ten Dimensions}},
  \href{http://dx.doi.org/10.1007/JHEP08(2011)014}{\emph{JHEP} {\bf 08} (2011)
  014}, [\href{https://arxiv.org/abs/1010.5487}{{\tt 1010.5487}}].

\bibitem{Bern:1994cg}
Z.~Bern, L.~J. Dixon, D.~C. Dunbar and D.~A. Kosower, \emph{{Fusing gauge
  theory tree amplitudes into loop amplitudes}},
  \href{http://dx.doi.org/10.1016/0550-3213(94)00488-Z}{\emph{Nucl. Phys. B}
  {\bf 435} (1995) 59--101}, [\href{https://arxiv.org/abs/hep-ph/9409265}{{\tt
  hep-ph/9409265}}].

\bibitem{Bern:2010qa}
Z.~Bern, J.~J. Carrasco, T.~Dennen, Y.-t. Huang and H.~Ita, \emph{{Generalized
  Unitarity and Six-Dimensional Helicity}},
  \href{http://dx.doi.org/10.1103/PhysRevD.83.085022}{\emph{Phys. Rev. D} {\bf
  83} (2011) 085022}, [\href{https://arxiv.org/abs/1010.0494}{{\tt
  1010.0494}}].

\bibitem{Alday:2009zm}
L.~F. Alday, J.~M. Henn, J.~Plefka and T.~Schuster, \emph{{Scattering into the
  fifth dimension of N=4 super Yang-Mills}},
  \href{http://dx.doi.org/10.1007/JHEP01(2010)077}{\emph{JHEP} {\bf 01} (2010)
  077}, [\href{https://arxiv.org/abs/0908.0684}{{\tt 0908.0684}}].

\bibitem{Coronado:2018cxj}
F.~Coronado, \emph{{Bootstrapping the Simplest Correlator in Planar $\mathcal N
  = 4$ Supersymmetric Yang-Mills Theory to All Loops}},
  \href{http://dx.doi.org/10.1103/PhysRevLett.124.171601}{\emph{Phys. Rev.
  Lett.} {\bf 124} (2020) 171601},
  [\href{https://arxiv.org/abs/1811.03282}{{\tt 1811.03282}}].

\bibitem{Belitsky:2019fan}
A.~V. Belitsky and G.~P. Korchemsky, \emph{{Exact null octagon}},
  \href{http://dx.doi.org/10.1007/JHEP05(2020)070}{\emph{JHEP} {\bf 05} (2020)
  070}, [\href{https://arxiv.org/abs/1907.13131}{{\tt 1907.13131}}].

\bibitem{Belitsky:2020qzm}
A.~V. Belitsky, \emph{{Null octagon from Deift-Zhou steepest descent}},
  \href{http://dx.doi.org/10.1016/j.nuclphysb.2022.115844}{\emph{Nucl. Phys. B}
  {\bf 980} (2022) 115844}, [\href{https://arxiv.org/abs/2012.10446}{{\tt
  2012.10446}}].

\bibitem{Polyakov:1980ca}
A.~M. Polyakov, \emph{{Gauge Fields as Rings of Glue}},
  \href{http://dx.doi.org/10.1016/0550-3213(80)90507-6}{\emph{Nucl. Phys. B}
  {\bf 164} (1980) 171--188}.

\bibitem{Korchemsky:1987wg}
G.~P. Korchemsky and A.~V. Radyushkin, \emph{{Renormalization of the Wilson
  Loops Beyond the Leading Order}},
  \href{http://dx.doi.org/10.1016/0550-3213(87)90277-X}{\emph{Nucl. Phys. B}
  {\bf 283} (1987) 342--364}.

\bibitem{Fishbane:1971jz}
P.~M. Fishbane and J.~D. Sullivan, \emph{{Asymptotic behavior of the vertex
  function in quantum electrodynamics}},
  \href{http://dx.doi.org/10.1103/PhysRevD.4.458}{\emph{Phys. Rev. D} {\bf 4}
  (1971) 458--475}.

\bibitem{Mueller:1981sg}
A.~H. Mueller, \emph{{Perturbative QCD at High-Energies}},
  \href{http://dx.doi.org/10.1016/0370-1573(81)90030-2}{\emph{Phys. Rept.} {\bf
  73} (1981) 237}.

\bibitem{Korchemsky:1988hd}
G.~P. Korchemsky, \emph{{Sudakov Form-factor in {QCD}}},
  \href{http://dx.doi.org/10.1016/0370-2693(89)90799-5}{\emph{Phys. Lett. B}
  {\bf 220} (1989) 629--634}.

\bibitem{Sudakov:1954sw}
V.~V. Sudakov, \emph{{Vertex parts at very high-energies in quantum
  electrodynamics}}, {\emph{Sov. Phys. JETP} {\bf 3} (1956) 65--71}.

\bibitem{Belitsky:2024yag}
A.~V. Belitsky, L.~V. Bork and V.~A. Smirnov, \emph{{Pinching Sudakov}},
  \href{https://arxiv.org/abs/2409.05945}{{\tt 2409.05945}}.

\bibitem{Beneke:1997zp}
M.~Beneke and V.~A. Smirnov, \emph{{Asymptotic expansion of Feynman integrals
  near threshold}},
  \href{http://dx.doi.org/10.1016/S0550-3213(98)00138-2}{\emph{Nucl. Phys. B}
  {\bf 522} (1998) 321--344}, [\href{https://arxiv.org/abs/hep-ph/9711391}{{\tt
  hep-ph/9711391}}].

\bibitem{Nair:1988bq}
V.~P. Nair, \emph{{A Current Algebra for Some Gauge Theory Amplitudes}},
  \href{http://dx.doi.org/10.1016/0370-2693(88)91471-2}{\emph{Phys. Lett. B}
  {\bf 214} (1988) 215--218}.

\bibitem{Galperin:2001seg}
A.~S. Galperin, E.~A. Ivanov, V.~I. Ogievetsky and E.~S. Sokatchev,
  \emph{{Harmonic superspace}}.
\newblock Cambridge Monographs on Mathematical Physics. Cambridge University
  Press, 2007,
  \href{http://dx.doi.org/10.1017/CBO9780511535109}{10.1017/CBO9780511535109}.

\bibitem{Bern:1997sc}
Z.~Bern, L.~J. Dixon and D.~A. Kosower, \emph{{One loop amplitudes for e+ e- to
  four partons}},
  \href{http://dx.doi.org/10.1016/S0550-3213(97)00703-7}{\emph{Nucl. Phys. B}
  {\bf 513} (1998) 3--86}, [\href{https://arxiv.org/abs/hep-ph/9708239}{{\tt
  hep-ph/9708239}}].

\bibitem{Britto:2004nc}
R.~Britto, F.~Cachazo and B.~Feng, \emph{{Generalized unitarity and one-loop
  amplitudes in N=4 super-Yang-Mills}},
  \href{http://dx.doi.org/10.1016/j.nuclphysb.2005.07.014}{\emph{Nucl. Phys. B}
  {\bf 725} (2005) 275--305}, [\href{https://arxiv.org/abs/hep-th/0412103}{{\tt
  hep-th/0412103}}].

\bibitem{Cachazo:2008vp}
F.~Cachazo, \emph{{Sharpening The Leading Singularity}},
  \href{https://arxiv.org/abs/0803.1988}{{\tt 0803.1988}}.

\bibitem{Bern:2011qt}
Z.~Bern and Y.-t. Huang, \emph{{Basics of Generalized Unitarity}},
  \href{http://dx.doi.org/10.1088/1751-8113/44/45/454003}{\emph{J. Phys. A}
  {\bf 44} (2011) 454003}, [\href{https://arxiv.org/abs/1103.1869}{{\tt
  1103.1869}}].

\bibitem{Carrasco:2011hw}
J.~J.~M. Carrasco and H.~Johansson, \emph{{Generic multiloop methods and
  application to N=4 super-Yang-Mills}},
  \href{http://dx.doi.org/10.1088/1751-8113/44/45/454004}{\emph{J. Phys. A}
  {\bf 44} (2011) 454004}, [\href{https://arxiv.org/abs/1103.3298}{{\tt
  1103.3298}}].

\bibitem{Brandhuber:2010mm}
A.~Brandhuber, D.~Korres, D.~Koschade and G.~Travaglini, \emph{{One-loop
  Amplitudes in Six-Dimensional (1,1) Theories from Generalised Unitarity}},
  \href{http://dx.doi.org/10.1007/JHEP02(2011)077}{\emph{JHEP} {\bf 02} (2011)
  077}, [\href{https://arxiv.org/abs/1010.1515}{{\tt 1010.1515}}].

\bibitem{Plefka:2014fta}
J.~Plefka, T.~Schuster and V.~Verschinin, \emph{{From Six to Four and More:
  Massless and Massive Maximal Super Yang-Mills Amplitudes in 6d and 4d and
  their Hidden Symmetries}},
  \href{http://dx.doi.org/10.1007/JHEP01(2015)098}{\emph{JHEP} {\bf 01} (2015)
  098}, [\href{https://arxiv.org/abs/1405.7248}{{\tt 1405.7248}}].

\bibitem{Mertig:1990an}
R.~Mertig, M.~Bohm and A.~Denner, \emph{{FEYN CALC: Computer algebraic
  calculation of Feynman amplitudes}},
  \href{http://dx.doi.org/10.1016/0010-4655(91)90130-D}{\emph{Comput. Phys.
  Commun.} {\bf 64} (1991) 345--359}.

\bibitem{Shtabovenko:2023idz}
V.~Shtabovenko, R.~Mertig and F.~Orellana, \emph{{FeynCalc 10: Do multiloop
  integrals dream of computer codes?}},
  \href{http://dx.doi.org/10.1016/j.cpc.2024.109357}{\emph{Comput. Phys.
  Commun.} {\bf 306} (2025) 109357},
  [\href{https://arxiv.org/abs/2312.14089}{{\tt 2312.14089}}].

\bibitem{Bern:2007ct}
Z.~Bern, J.~J.~M. Carrasco, H.~Johansson and D.~A. Kosower, \emph{{Maximally
  supersymmetric planar Yang-Mills amplitudes at five loops}},
  \href{http://dx.doi.org/10.1103/PhysRevD.76.125020}{\emph{Phys. Rev. D} {\bf
  76} (2007) 125020}, [\href{https://arxiv.org/abs/0705.1864}{{\tt
  0705.1864}}].

\bibitem{Huang:2011um}
Y.-t. Huang, \emph{{Non-Chiral S-Matrix of N=4 Super Yang-Mills}},
  \href{https://arxiv.org/abs/1104.2021}{{\tt 1104.2021}}.

\bibitem{Bern:1997it}
Z.~Bern, J.~Rozowsky and B.~Yan, \emph{{Two loop N=4 supersymmetric amplitudes
  and QCD}}, \href{http://dx.doi.org/10.1063/1.53686}{\emph{AIP Conf. Proc.}
  {\bf 407} (1997) 908}, [\href{https://arxiv.org/abs/hep-ph/9706392}{{\tt
  hep-ph/9706392}}].

\bibitem{Bern:2006vw}
Z.~Bern, M.~Czakon, D.~A. Kosower, R.~Roiban and V.~A. Smirnov, \emph{{Two-loop
  iteration of five-point N=4 super-Yang-Mills amplitudes}},
  \href{http://dx.doi.org/10.1103/PhysRevLett.97.181601}{\emph{Phys. Rev.
  Lett.} {\bf 97} (2006) 181601},
  [\href{https://arxiv.org/abs/hep-th/0604074}{{\tt hep-th/0604074}}].

\bibitem{Carrasco:2011mn}
J.~J.~M. Carrasco and H.~Johansson, \emph{{Five-Point Amplitudes in N=4
  Super-Yang-Mills Theory and N=8 Supergravity}},
  \href{http://dx.doi.org/10.1103/PhysRevD.85.025006}{\emph{Phys. Rev. D} {\bf
  85} (2012) 025006}, [\href{https://arxiv.org/abs/1106.4711}{{\tt
  1106.4711}}].

\bibitem{Henn:2010ir}
J.~M. Henn, S.~G. Naculich, H.~J. Schnitzer and M.~Spradlin, \emph{{More loops
  and legs in Higgs-regulated N=4 SYM amplitudes}},
  \href{http://dx.doi.org/10.1007/JHEP08(2010)002}{\emph{JHEP} {\bf 08} (2010)
  002}, [\href{https://arxiv.org/abs/1004.5381}{{\tt 1004.5381}}].

\bibitem{Smirnov:2021rhf}
A.~V. Smirnov, N.~D. Shapurov and L.~I. Vysotsky, \emph{{FIESTA5: Numerical
  high-performance Feynman integral evaluation}},
  \href{http://dx.doi.org/10.1016/j.cpc.2022.108386}{\emph{Comput. Phys.
  Commun.} {\bf 277} (2022) 108386},
  [\href{https://arxiv.org/abs/2110.11660}{{\tt 2110.11660}}].

\bibitem{Fabricius:1979tb}
K.~Fabricius and I.~Schmitt, \emph{{Calculation of dimensionally regularized
  box graphs in the zero mass case}},
  \href{http://dx.doi.org/10.1007/BF01577398}{\emph{Z. Phys. C} {\bf 3} (1979)
  51--53}.

\bibitem{Papadopoulos:1981ju}
S.~Papadopoulos, A.~P. Contogouris and J.~Ralston, \emph{{Calculation of Box
  Graph With Lightlike Particles}},
  \href{http://dx.doi.org/10.1103/PhysRevD.25.2218}{\emph{Phys. Rev. D} {\bf
  25} (1982) 2218}.

\bibitem{Bern:1993kr}
Z.~Bern, L.~J. Dixon and D.~A. Kosower, \emph{{Dimensionally regulated pentagon
  integrals}},
  \href{http://dx.doi.org/10.1016/0550-3213(94)90398-0}{\emph{Nucl. Phys. B}
  {\bf 412} (1994) 751--816}, [\href{https://arxiv.org/abs/hep-ph/9306240}{{\tt
  hep-ph/9306240}}].

\bibitem{Bern:1993mq}
Z.~Bern, L.~J. Dixon and D.~A. Kosower, \emph{{One loop corrections to five
  gluon amplitudes}},
  \href{http://dx.doi.org/10.1103/PhysRevLett.70.2677}{\emph{Phys. Rev. Lett.}
  {\bf 70} (1993) 2677--2680},
  [\href{https://arxiv.org/abs/hep-ph/9302280}{{\tt hep-ph/9302280}}].

\bibitem{Almelid:2015jia}
O.~Almelid, C.~Duhr and E.~Gardi, \emph{{Three-loop corrections to the soft
  anomalous dimension in multileg scattering}},
  \href{http://dx.doi.org/10.1103/PhysRevLett.117.172002}{\emph{Phys. Rev.
  Lett.} {\bf 117} (2016) 172002},
  [\href{https://arxiv.org/abs/1507.00047}{{\tt 1507.00047}}].

\bibitem{Bern:2005iz}
Z.~Bern, L.~J. Dixon and V.~A. Smirnov, \emph{{Iteration of planar amplitudes
  in maximally supersymmetric Yang-Mills theory at three loops and beyond}},
  \href{http://dx.doi.org/10.1103/PhysRevD.72.085001}{\emph{Phys. Rev. D} {\bf
  72} (2005) 085001}, [\href{https://arxiv.org/abs/hep-th/0505205}{{\tt
  hep-th/0505205}}].

\bibitem{Bern:1991aq}
Z.~Bern and D.~A. Kosower, \emph{{The Computation of loop amplitudes in gauge
  theories}}, \href{http://dx.doi.org/10.1016/0550-3213(92)90134-W}{\emph{Nucl.
  Phys. B} {\bf 379} (1992) 451--561}.

\bibitem{Kunszt:1993sd}
Z.~Kunszt, A.~Signer and Z.~Trocsanyi, \emph{{One loop helicity amplitudes for
  all 2 ---\ensuremath{>} 2 processes in QCD and N=1 supersymmetric Yang-Mills
  theory}}, \href{http://dx.doi.org/10.1016/0550-3213(94)90456-1}{\emph{Nucl.
  Phys. B} {\bf 411} (1994) 397--442},
  [\href{https://arxiv.org/abs/hep-ph/9305239}{{\tt hep-ph/9305239}}].

\bibitem{Gervais:1972tr}
J.-L. Gervais and A.~Neveu, \emph{{Feynman rules for massive gauge fields with
  dual diagram topology}},
  \href{http://dx.doi.org/10.1016/0550-3213(72)90071-5}{\emph{Nucl. Phys. B}
  {\bf 46} (1972) 381--401}.

\bibitem{Srednicki:2007qs}
M.~Srednicki, \emph{{Quantum field theory}}.
\newblock Cambridge University Press, 1, 2007,
  \href{http://dx.doi.org/10.1017/CBO9780511813917}{10.1017/CBO9780511813917}.

\bibitem{Brandhuber:2012vm}
A.~Brandhuber, G.~Travaglini and G.~Yang, \emph{{Analytic two-loop form factors
  in N=4 SYM}}, \href{http://dx.doi.org/10.1007/JHEP05(2012)082}{\emph{JHEP}
  {\bf 05} (2012) 082}, [\href{https://arxiv.org/abs/1201.4170}{{\tt
  1201.4170}}].

\bibitem{Catani:2011st}
S.~Catani, D.~de~Florian and G.~Rodrigo, \emph{{Space-like (versus time-like)
  collinear limits in QCD: Is factorization violated?}},
  \href{http://dx.doi.org/10.1007/JHEP07(2012)026}{\emph{JHEP} {\bf 07} (2012)
  026}, [\href{https://arxiv.org/abs/1112.4405}{{\tt 1112.4405}}].

\bibitem{DelDuca:1999rs}
V.~Del~Duca, L.~J. Dixon and F.~Maltoni, \emph{{New color decompositions for
  gauge amplitudes at tree and loop level}},
  \href{http://dx.doi.org/10.1016/S0550-3213(99)00809-3}{\emph{Nucl. Phys. B}
  {\bf 571} (2000) 51--70}, [\href{https://arxiv.org/abs/hep-ph/9910563}{{\tt
  hep-ph/9910563}}].

\bibitem{DelDuca:1999iql}
V.~Del~Duca, A.~Frizzo and F.~Maltoni, \emph{{Factorization of tree QCD
  amplitudes in the high-energy limit and in the collinear limit}},
  \href{http://dx.doi.org/10.1016/S0550-3213(99)00657-4}{\emph{Nucl. Phys. B}
  {\bf 568} (2000) 211--262}, [\href{https://arxiv.org/abs/hep-ph/9909464}{{\tt
  hep-ph/9909464}}].

\bibitem{Belitsky:2021huz}
A.~V. Belitsky and V.~A. Smirnov, \emph{{An off-shell Wilson loop}},
  \href{http://dx.doi.org/10.1007/JHEP04(2023)071}{\emph{JHEP} {\bf 04} (2023)
  071}, [\href{https://arxiv.org/abs/2110.13206}{{\tt 2110.13206}}].

\bibitem{Belitsky:2003sh}
A.~V. Belitsky, S.~E. Derkachov, G.~P. Korchemsky and A.~N. Manashov,
  \emph{{Superconformal operators in N=4 superYang-Mills theory}},
  \href{http://dx.doi.org/10.1103/PhysRevD.70.045021}{\emph{Phys. Rev. D} {\bf
  70} (2004) 045021}, [\href{https://arxiv.org/abs/hep-th/0311104}{{\tt
  hep-th/0311104}}].

\bibitem{Sohnius:1985qm}
M.~F. Sohnius, \emph{{Introducing Supersymmetry}},
  \href{http://dx.doi.org/10.1016/0370-1573(85)90023-7}{\emph{Phys. Rept.} {\bf
  128} (1985) 39--204}.

\bibitem{tHooft:1976snw}
G.~'t~Hooft, \emph{{Computation of the Quantum Effects Due to a
  Four-Dimensional Pseudoparticle}},
  \href{http://dx.doi.org/10.1103/PhysRevD.14.3432}{\emph{Phys. Rev. D} {\bf
  14} (1976) 3432--3450}.

\bibitem{Belitsky:2000ii}
A.~V. Belitsky, S.~Vandoren and P.~van Nieuwenhuizen, \emph{{Instantons,
  Euclidean supersymmetry and Wick rotations}},
  \href{http://dx.doi.org/10.1016/S0370-2693(00)00183-0}{\emph{Phys. Lett. B}
  {\bf 477} (2000) 335--340}, [\href{https://arxiv.org/abs/hep-th/0001010}{{\tt
  hep-th/0001010}}].

\bibitem{Gastmans:1990xh}
R.~Gastmans and T.~T. Wu, \emph{{The Ubiquitous photon: Helicity method for QED
  and QCD}}, vol.~80.
\newblock 1990.

\bibitem{Xu:1986xb}
Z.~Xu, D.-H. Zhang and L.~Chang, \emph{{Helicity Amplitudes for Multiple
  Bremsstrahlung in Massless Nonabelian Gauge Theories}},
  \href{http://dx.doi.org/10.1016/0550-3213(87)90479-2}{\emph{Nucl. Phys. B}
  {\bf 291} (1987) 392--428}.

\bibitem{Dixon:2013uaa}
L.~J. Dixon, \emph{{A brief introduction to modern amplitude methods}},  in
  \emph{{Theoretical Advanced Study Institute in Elementary Particle Physics}:
  {Particle Physics: The Higgs Boson and Beyond}}, pp.~31--67, 2014.
\newblock \href{https://arxiv.org/abs/1310.5353}{{\tt 1310.5353}}.
\newblock \href{http://dx.doi.org/10.5170/CERN-2014-008.31}{DOI}.

\bibitem{Boels:2009bv}
R.~Boels, \emph{{Covariant representation theory of the Poincare algebra and
  some of its extensions}},
  \href{http://dx.doi.org/10.1007/JHEP01(2010)010}{\emph{JHEP} {\bf 01} (2010)
  010}, [\href{https://arxiv.org/abs/0908.0738}{{\tt 0908.0738}}].

\bibitem{Drummond:2008bq}
J.~M. Drummond, J.~Henn, G.~P. Korchemsky and E.~Sokatchev, \emph{{Generalized
  unitarity for N=4 super-amplitudes}},
  \href{http://dx.doi.org/10.1016/j.nuclphysb.2012.12.009}{\emph{Nucl. Phys. B}
  {\bf 869} (2013) 452--492}, [\href{https://arxiv.org/abs/0808.0491}{{\tt
  0808.0491}}].

\bibitem{Passarino:1978jh}
G.~Passarino and M.~J.~G. Veltman, \emph{{One Loop Corrections for e+ e-
  Annihilation Into mu+ mu- in the Weinberg Model}},
  \href{http://dx.doi.org/10.1016/0550-3213(79)90234-7}{\emph{Nucl. Phys. B}
  {\bf 160} (1979) 151--207}.

\bibitem{Bern:1995db}
Z.~Bern and A.~G. Morgan, \emph{{Massive loop amplitudes from unitarity}},
  \href{http://dx.doi.org/10.1016/0550-3213(96)00078-8}{\emph{Nucl. Phys. B}
  {\bf 467} (1996) 479--509}, [\href{https://arxiv.org/abs/hep-ph/9511336}{{\tt
  hep-ph/9511336}}].

\end{thebibliography}
\end{document}